\pdfoutput=1
\documentclass[12pt,a4paper]{article}%

\usepackage{adjustbox}
\usepackage{lmodern}
\usepackage[T1]{fontenc}
\usepackage[utf8]{inputenc}
\usepackage{parskip}
\usepackage{amssymb}
\usepackage{amsfonts}
\usepackage{xfrac}
\usepackage{amsmath}
\usepackage{mathtools}
\usepackage{enumitem}
\usepackage[UKenglish]{isodate}
\usepackage{cases}
\usepackage{diagbox}
\usepackage{rotating}

\usepackage{geometry}
\usepackage[authoryear]{natbib}
\usepackage{setspace}
\usepackage{booktabs}
\usepackage[table,usenames,dvipsnames]{xcolor}

\usepackage{graphicx}
\usepackage{tikz}
\usepackage{pgfplots}
\usepackage[font=small,labelfont=sf]{caption}

\usepackage{array}
\newcolumntype{L}{>{$}l<{$}}

\usepackage[colorlinks=false,
            urlcolor=blue,
            citecolor=blue,
            linkcolor=blue,
            pdfauthor={Alexander Mayer, Universita Ca' Foscari, Venice, Italy, and Michael Massmann, WHU, Vallendar, Germany},
            pdftitle={Least squares estimation in nonlinear cohort panels with learning from experience},
            bookmarks,
            bookmarksopen=true,
            bookmarksnumbered=true,
            pdfstartview={FitH},
            ]{hyperref}

\newcommand{\bk}{\color{black}}

\newcommand{\rd}{\color{red}}

\newcommand{\Tr}{{\textnormal{\textsf{T}}}}
\newcommand{\Ex}{\textnormal{\textsf{E}}}

\newcommand{\var}{\textnormal{\textsf{var}}}
\newcommand{\cov}{\textnormal{\textsf{cov}}}

\newcommand{\st}{\star}
\newcommand{\inn}{\, \in \,}
\newcommand{\eps}{\varepsilon}

\DeclareMathOperator*{\ssup}{\textnormal{\textsf{sup}}}
\DeclareMathOperator*{\iinf}{\textnormal{\textsf{inf}}}
\DeclareMathOperator*{\llim}{\textnormal{\textsf{lim}}}

\DeclareMathOperator*{\lln}{\textnormal{\textsf{ln}}}
\DeclareMathOperator*{\mmax}{\textnormal{\textsf{max}}}
\DeclareMathOperator*{\mmin}{\textnormal{\textsf{min}}}
\DeclarePairedDelimiter\floor{\lfloor}{\rfloor}

\usepackage{accents}
\newcommand\ubar[1]{%
  \underaccent{\bar}{#1}}
\newtheorem{remark}{Remark}
\newtheorem{assumption}{Assumption}
\newtheorem{lemma}{Lemma}
\newtheorem{proposition}{Proposition}
\newtheorem{corollary}{Corollary}
 \newtheorem{example}{Example}

\usepackage{xkvltxp} 
\usepackage{fixme} 
\fxsetup{author=,marginclue,marginface=\footnotesize\itshape\color{red}}
\FXRegisterAuthor{mm}{mhfm}{MM} 
\FXRegisterAuthor{am}{asm}{AM} 

\makeatletter
\renewcommand*\FXLayoutMarginClue[3]{\marginpar[{\raggedleft\@fxuseface{margin}\ignorespaces#3 \fxnotename{#1}!}]{\raggedright\@fxuseface{margin}\ignorespaces#3 \fxnotename{#1}}}
\makeatother

\usepackage[page]{appendix} 
\renewcommand{\appendixpagename}{\centering \quad Supplementary material to \newline\noindent {\it `Least squares estimation in nonlinear cohort panels with learning from experience'}} 

\begin{document}

\title{\vspace*{-2cm}Least squares estimation in nonstationary nonlinear cohort panels with learning from experience\thanks{We would like to thank Eric Beutner, Chico Blasques, Jörg Breitung, Paolo Gorgi, Andre Lucas, Sven Otto, Mario Padula, Dario Palumbo, Nic Schaub, and
    Dominik Wied for helpful discussions. The first author is grateful for the support he received from Monica Billio and Davide Raggi, the second author is indebted to Ben Litzinger for his excellent research assistance. Highly
    appreciated were the suggestions and comments made at the NBER-NSF Time Series meeting 2023 in Montreal, and at the research seminars at WHU--Otto Beisheim School of
    Management, Universitat de les Illes Balears, Aix-Marseille School of Economics, Universität zu Köln, Università Ca' Foscari in Venice, and Vrije Universiteit Amsterdam.}}

\author{Alexander Mayer\footnote{{\it Corresponding author}, email: \href{mailto:alexandersimon.mayer@unive.it}{alexandersimon.mayer@unive.it}}\\
{\it Università Ca' Foscari}\\
Venezia, Italy\\\vspace{-1em}
  \and
Michael Massmann\\
{\it WHU -- Otto Beisheim School of Management}\\
Vallendar, Germany\\
and\\
{\it Vrije Universiteit}\\
Amsterdam, The Netherlands}
\date{\today}
\maketitle
\thispagestyle{empty}
\vspace*{-.5cm}
\begin{abstract} \noindent We discuss techniques of estimation and inference for nonstationary nonlinear cohort panels with learning from experience, showing, inter alia, the consistency and asymptotic normality of the nonlinear
  least squares estimator used in empirical practice. Potential pitfalls for hypothesis testing are identified and solutions proposed. Monte Carlo simulations verify the properties of the estimator and corresponding test
  statistics in finite samples, while an application to a panel of survey expectations demonstrates the usefulness of the theory developed.

  \vspace{1em}\noindent{\bf Keywords:} adaptive learning, inflation expectations, nonlinear least squares with nonsmooth objective function, cohort panel data, asymptotic theory, nuisance parameters  
\end{abstract}

\newpage

\section{Introduction}

Following \cite{Sargent93,Sargent99}, the literature has seen a renewed interest in how economic agents form expectations. In particular, researchers and policy makers alike increasingly question the orthodox framework of viewing
agents as forming full-information rational expectations, see e.g.\ \cite{evans:01}, \cite{MankiwReis02}, and \cite{Bernanke07}. 
Especially expectations about future inflation are relevant for understanding economic outcomes. Recent empirical work on the formation process of inflation expectations includes
\cite{BachmannBergSims15} and \cite{CoibionGorodnichenkoRopele20} who emphasise the importance of expected inflation for consumption and investment decisions, respectively, and \cite{coibion:20} who analyse how inflation
expectations can be used as a policy tool by monetary authorities.

A concomitant development is the increasing recognition that representative agent theory, the predominant approach to modelling in economics, may be insufficient for explaining economic fluctuations and that the heterogeneity
between agents needs to be taken account of, see e.g.\ \cite{HeathcoteStoreslettenViolante09} and \cite{KaplanViolante18} for surveys and \cite{Yellen16} for the view of a policy maker. One of the driving forces behind this
development has been the growing availability and analysis of surveys of both households' and firms' beliefs, see for instance \cite{WeberD’AcuntoGorodnichenkoCoibion22} and \cite{DAcuntoMalmendierWeber23} for overviews, and
\cite{link:23} for a recent investigation into the heterogeneity of housholds' and firms' expectations.  The Michigan Survey of Consumers (MSC) is one of the longest-running surveys that contains
information on agents' inflation expectations. Early work on the MSC concentrated on analysing aggregates of the data, see e.g.\ \cite{MankiwReisWolfers04}, \cite{Branch04}, and \cite{CoibionGorodnichenko12}. Recently, however,
the focus has shifted to taking full advantage of the entire panel of survey respondents, as do, for instance, \cite{BachmannBergSims15}, \cite{mn:2016}, and \cite{meekmonti:23}.

At the confluence of these two strands of the literature stand \cite{CoibionGorodnichenkoKamdar18} who forcefully argue for ``a careful (re-)consideration of the expectations formation process and a more systematic inclusion of
real-time expectations through survey data'' (p.\ 1447). One recent line of such research is on so-called `experience effects', stipulating that exposure to personal or public economic or political outcomes tends to shape agents'
behaviour, see \cite{Malmendier21} for a current survey. In particular, in their seminal paper on how inflation expectations are determined by individual experiences, \cite{mn:2016} (MN, henceforth) depart from the rational
expectations paradigm by making use of an adaptive learning framework in which agents entertain their own --potentially mis-specified-- model of how inflation is determined and estimate it recursively to form their
expectations. Similarly, in their empirical analysis, MN make full use of the MSC, in both the cross-sectional and the time dimensions.  The two main findings of MN are ($i$) substantial heterogeneity between individuals of
different age and ($ii$) what MN call recency bias, a concept related to the availability heuristic by \cite{TverskyKahneman74}. The heterogeneity is manifest in the weight that individuals give to new data as they update
their inflation forecasts and that depends on their age.
The recency bias is captured by the magnitude  of the so-called `gain parameter' $\gamma$ in the estimated updating equation and
indicates that individuals' recent experiences have a stronger impact on their expectations than distant ones.

These results have spawned a string of papers on `learning-from-experience' that either  re-use MN's parameter estimates in similar models for different empirical  applications  or that extend MN's specification 
for describing the MSC data. Examples of the former category are \cite{naknun:15}  on  stock prices and dividends, and \cite{Acedanski17} on  the wealth distribution; both papers calibrate their models with
 MN's estimated gain parameter of $\hat \gamma = 3.044$. In the latter category fall \cite{MadeiraZafar15} who extend MN's analysis of the MSC dataset by allowing for the heterogeneous use of private information, and
\cite{Gwak22} who builds a model similar to MN's yet includes in the specification a Markov-switching component to distinguish between learning-from-experience in high and low volatile inflation regimes. Recently,
\cite{MalmendierNagelYan21} analyse the voting behaviour of the Fed's FOMC members using individuals' learnt-from-experience inflation expectations as given input variable, and \cite{nagel:24} explores the implications on real interest rates of learning from experience. 

What all these papers have in common is that the models they consider are highly complex and that neither their microfoundation nor their econometrics is yet fully understood. Indeed, from an economic theory point of view,
\cite{DuffyShin23} take a step back and use the concept of learning-from-experience in a microfounded demography-based model to rationalise constant gain learning. In the present paper, we follow their example, go back to square
one, and derive the econometric theory of a learning-from-experience model. In fact, the full complexity of the empirical models estimated by \cite{MadeiraZafar15}, \cite{mn:2016}, \cite{MalmendierNagelYan21}, and \cite{Gwak22} is
beyond the scope of the present paper. Instead, we consider a special case of theirs that is analytically tractable. It nevertheless allows us to estimate a plausible learning-from-experience model empirically and engage in
statistically well-founded inference. Doing so, we make progress on two fronts: Empirically, we shed new light on the question of heterogeneous inflation expectations and recency bias, obtaining conclusions that
are in line with the aforementioned empirical papers on the MSC. Theoretically, we establish novel econometric results for the analysis of learning-from-experience models that set the scene for future work analysing the
econometrics of even more complex models.

The econometric specification we adopt in the present paper is a special case of the model used in MN and can be viewed as nonstationary nonlinear cohort panel data model with time fixed effects.  The nonlinearity in the
regression function stems from the recursively generated expectations, while the nonstationarity arises due to a stochastic evaporating trend component, reminiscent of the linear regressions with deterministic evaporating trends
studied by \cite{phillips:07}. Related research demonstrates that the statistical analysis of estimators in macroeconomic models with similar adaptive learning schemes provides a challenging task, see e.g. \cite{chev:10},
\cite{chev:17}, \cite{chrismass:18,chrismass:19}, \cite{mayer:22,mayer:23}, or \cite{christiano:24}. This literature shows that estimation of and inference in models with adaptive learning is far from standard and often marred by weak-identification,
asymptotic collinearity, or non-standard convergence rates. As will be discussed below, one of MN's contributions is to sidestep these issues to some extent by exploiting the cross-sectional variation across individuals. However,
as mentioned earlier, the theoretical properties of the nonlinear econometric methods employed in the empirical learning-from-experience papers have not been examined to date, and implicit or explicit claims that the model
parameters are identified or that certain statistics have some given asymptotic distribution call for verification.

The aim of the present paper is thus to bridge the existing gap between econometric theory and empirical practice. Our contributions are twofold: {\it First}, we derive new asymptotic results for point estimation and inference in
a nonlinear cohort panel data model with learning from experience, thereby extending the established econometric results in the literature in terms of ($i$) a panel dimension, ($ii$) a heterogeneous gain sequence, and ($iii$)
multivariate estimation by nonlinear least squares (NLS, henceforth). As a {\it second} contribution, we apply our results to an empirical model of the MSC dataset. Our model is akin to the baseline specification employed by
\cite{MadeiraZafar15}, \cite{mn:2016}, \cite{MalmendierNagelYan21}, \cite{Gwak22} and \cite{nagel:24}. The present paper is therefore in the tradition of \cite{milani:07}, \cite{chev:10}, \cite{adam:16}, and \cite{hommes:23} who derive rigorous econometric results
for the modelling of substantive empirical problems in the economics of adaptive learning. Yet while all three papers use constant gain learning specifications, estimated by Bayesian methods in \cite{milani:07} and by
continuously updated GMM  in \cite{chev:10} and by the method of simulated moments in \cite{adam:16}, our paper is the first to provide well-founded econometric insights for a decreasing gain model specification.

In the theory part of the paper, we consider different asymptotic regimes that depend on whether the number of cohorts is fixed or not.  One important conclusion for point estimation is that, albeit consistent in all scenarios, the
NLS estimator might not be asymptotically normal if the number of cohorts is fixed due to  an objective function that is not differentiable everywhere. As argued below, this problem can be overcome if the number of cohorts diverges. However, in this case, we are confronted with the additional challenge of asymptotic collinear regressors. This technical hurdle notwithstanding, asymptotic normality, albeit at a nonstandard convergence rate, is established by combining results from analytical number theory with seminal results for extremum estimators with nonsmooth objective function, see e.g.\ \citet[Section 7]{newmc:94}.

A further focus is placed on hypothesis testing. The inference conducted in the aforementioned empirical learning-from-experience papers is, by and large, classical. Yet our asymptotic analysis identifies potential pitfalls
due to slow convergence and parameter non-identification. To address the latter issue, we propose a solution that builds on the results of \cite{hansen:1996} regarding hypotheses that involve non-identified nuisance parameters. We
investigate the properties of the NLS estimator and the test statistics in finite samples by use of Monte Carlo simulations. In the empirical part of the paper we revisit the MSC dataset.
Using the estimation and inference procedures developed in the theory part of our paper we confirm, on the whole, the findings of previous empirical papers on the MSC. Yet we note conflicting
evidence on the weight given by agents to private experiences when they form expectations about future inflation.

The remainder of the paper is organised as follows. Sections \ref{sec:model} and \ref{sec:estimation} introduce the model and the NLS estimator, respectively. We lay out assumptions, discuss consistency and asymptotic normality of
the estimator in Section \ref{sec:asymptotics}. Standard errors and inferential methods are addressed in Section \ref{sec:inf}. Section \ref{sec:MC} contains a Monte Carlo study, while the empirical application is presented in
Section \ref{sec:emp}. Additional results are relegated to the Supplementary Material.

\section{Model}\label{sec:model}

Consider the following model that relates observed survey expectations, denoted by $z_{t,s}$, to the learnt expectation about future macro-level inflation $y_{t+1}$, denoted by $a_{t,s}$, via
\begin{equation} \label{surveyexp}
  z_{t,s} = \alpha_t + \beta a_{t,s} + \eps_{t,s}, 
\end{equation}
with $t$ and $s$ indexing the time and birth period, respectively. Here, \(\alpha_t\) represents a time fixed effect capturing a component in the expectations formation process that is common to all cohorts, while $\beta$ is
the weight given to private, i.e.\ cohort-specific, experiences. The term \(\eps_{t,s}\) is some error term further specified below. The model in Eq.\ \eqref{surveyexp} is supplemented by an equation specifying the way in which different cohorts update their beliefs over time. In particular, it will be assumed that, at the end of period $t$, individuals born in period
\(s\) form expectations about future aggregate inflation based on the available inflation history according to an adaptive learning rule
\begin{equation}\label{recursion} 
a_{t,s} = a_{t-1,s} + \gamma_{t,s}(y_t-a_{t-1,s}), \quad a_{0,s} \coloneqq \mathfrak{a}_s,
\end{equation}
where \(\mathfrak{a}_s \in \mathbb R\) is some initial value. The updating scheme in Eq.\ \eqref{recursion} is a stochastic approximation algorithm that can be viewed as a generalisation of recursive least squares (see, e.g., \citealp{ben:90}), where the so-called `gain sequence' \(\gamma_{t,s}\) measures the responsiveness to previous prediction mistakes.

Implicit in Eq.\ \eqref{recursion} is the assumption that individuals use a constant  level  model as their so-called perceived law of motion (PLM) for prediction. Within the macroeconomic learning literature, the PLM is the model individuals use to forecast $y_{t+1}$ and which in general does not coincide with the true data generating process of $y_t$. Note that the model in Eqs.\ (\ref{surveyexp}) and (\ref{recursion}) is not of the self-referential type found in  the classical adaptive learning literature surveyed by \cite{evans:01} since the dependent variable $z_{t,s}$ is different from the covariate $y_t$ upon which the learning recursion is based. 

 Following the empirical learning-from-experience literature, we assume that the weight individuals attach to previous observations depends on their age \(t-s\) 
 according to
\(\gamma_{t,s} \coloneqq \gamma_{t,s}(\gamma)\) such that
 \begin{equation}\label{gain_mn}
 \gamma_{t,s}(\gamma) = \begin{cases}
\displaystyle\frac{\gamma}{t-s}  & \text{ if }\,t-s > \gamma\\
 1 & \text{ otherwise}, \end{cases}
 \end{equation}
 where \(\gamma>0\) is some unknown gain parameter. This specification creates heterogeneity in the way different cohorts form their expectations. The economic interpretation of \(\gamma\) is that of a `forgetting factor', where
 \(\gamma > 1\) (\(\gamma < 1\)) means that agents attach more (less) weight to recent revisions of the data while \(\gamma=1\) results in ordinary least squares learning with equally weighted observations. The choice of gain sequence $\gamma_{t,s}$  in Eq.\ (\ref{gain_mn}) extends the classical least-squares recursion of, e.g., \cite{marsar:89} by making the updating weight a function of age $t-s$ rather than merely time $t$.

It follows from Eqs.\ \eqref{recursion} and \eqref{gain_mn} that the forecast of an individual of age $t-s$ is simply a weighted average of past and present information. That is, $a_{t,s} \coloneqq a_{t,s}(\gamma)$ where, for any $s>\gamma > 0$,
\begin{align}\label{amean}
a_{t,t-s}(\gamma) =  \sum_{j= \floor{\gamma}}^{s}\kappa_{j,s}(\gamma)y_{t-s+j}, \quad \kappa_{j,s}(\gamma) \coloneqq \displaystyle\frac{\gamma}{j}\prod_{i = j+1}^{s}\left(1-\frac{\gamma}{i}\right),
\end{align} 
 see Lemma \ref{lem:A0} of the appendix for details. The `floor' function is defined as \(\floor{x} = \{m \in \mathbb{Z}: m \leq x\}\), and we use the conventions $\prod_{i = s+1}^{s}(1-\gamma/i) \coloneqq 1$, and 
$\kappa_{\floor{\gamma},s}(\gamma) \coloneqq \prod_{i = \floor{\gamma}+1}^{s}(1-\gamma/i).$ Importantly, all information before birth \textit{plus} \(\gamma\) is discarded, thereby ensuring an economically reasonable sequence of
non-negative weights $\kappa_{j,s}(\gamma)$ .

The complete model is thus given by Eqs.\ \eqref{surveyexp}, \eqref{recursion}, and \eqref{gain_mn}. It is a nonlinear cohort panel data model with time fixed effects.  The age effect is captured by the nonlinear
age-dependent belief updating mechanism $a_{t,s}(\gamma)$, while the time effects enter the model linearly.  The parameters to be estimated are thus $\theta \coloneqq (\beta,\gamma)^\Tr$.  This model is akin to MN's baseline specification; see also Eq.\ (6) in MN. Note that Eq.\
\eqref{recursion} is a special case of the updating scheme that MN equip individuals with, in that they consider an AR(1) as PLM.  Agents in our setup are therefore assumed to be less sophisticated when compared to their
counterparts in MN, yet given the prominence of `simple' forecasting rule for inflation in publications by the Federal Reserve Banks\footnote{See \cite{AtkesonOhania01}, \cite{PasaogullariMeyer2010}, and \cite{BauerMcCarthy15} for
  examples.}  a more restricted perception of how inflation is generated can arguably be seen as more realistic for boundedly rational agents. A comparison of both specifications in terms of a Monte Carlo study and an extended empirical application  is included in the Supplementary Material; a theoretical treatment is, however,  beyond the scope of the paper.

\section{Estimation}\label{sec:estimation}

The parameters in the model given by Eqs.\ \eqref{surveyexp}, \eqref{recursion}, and \eqref{gain_mn} will be estimated by nonlinear least squares (NLS). Before proceeding with a discussion of the estimator, some additional notation is needed. In particular, we denote by $l$ and $u$ the first and the last age group to be considered, and by \(n\) the last time period. Consequently, the time index $t$
and the birth year $s$ take on values in
\begin{equation*}
  t \in \{u+1,u+2,\dots,n\} \quad \text{and} \quad s \in \{t-u, t-u+1,\dots, t-l\}, 
\end{equation*}
respectively. Defining
\[
  m \coloneqq u-l+1, \quad 1 \leq l < u < n,
\]
to be the number of cohorts, the pooled data set is seen to consist of a total of \[N \coloneqq (n-u)m\] observations. The structure of the dataset, illustrated in Table \ref{tab:indices}, is similar to age-period-cohort panels covered elsewhere in the literature (e.g. \citealp{HarnauNielsen18} or \citealp{FannonNielsen19}). Our specification is, however, fundamentally different from the aforementioned literature due to the nature of the nonlinearly and recursively generated cohort effects (see also the discussion in \citealp[p. 20]{maletal:21}). 

\colorlet{myred}{white} 
\begin{table}[htbp]
  \small
  \begin{center}
    \begin{tabular}{r|ccccccccccc}
      \diagbox{\rule{0.2em}{0em}$t$\rule{1.2em}{0em}}{\vspace{-1.2em}$s$} & 1  & 2  & 3  & $\cdots$ & 50 & 51 & 52 & $\cdots$ &$n -u$& $\cdots$ &$n-l$\\ \hline
      $u+1 = 75$    & \cellcolor{blue!10} 74 & \cellcolor{blue!10} 73 &\cellcolor{blue!10} 72 & \cellcolor{blue!10} $\cdots$ & \cellcolor{blue!10} 25 & \cellcolor{myred}   &  \cellcolor{myred}  &   \cellcolor{myred}   & \cellcolor{myred}&\cellcolor{myred} &\cellcolor{myred}\\
      $u+2 = 76$    &  \cellcolor{myred}  & \cellcolor{blue!10}  74 & \cellcolor{blue!10}  73 &\cellcolor{blue!10}   $\cdots$ & \cellcolor{blue!10}  26 & \cellcolor{blue!10}  25 & \cellcolor{myred}   &  \cellcolor{myred}  &\cellcolor{myred} &\cellcolor{myred}&  \cellcolor{myred} \\
      $u+3 = 77$    &  \cellcolor{myred}  &  \cellcolor{myred}  &\cellcolor{blue!10} 74 & \cellcolor{blue!10}$\cdots$ & \cellcolor{blue!10} 27 & \cellcolor{blue!10} 26 & \cellcolor{blue!10} 25 &   \cellcolor{myred}    &\cellcolor{myred}&\cellcolor{myred}& \cellcolor{myred}\\
      $\vdots$\;\;\;\;\;\;&&&&&&&& $\ddots$ &&&\\
      $n = 150$ &\cellcolor{myred}&\cellcolor{myred}&\cellcolor{myred}&\cellcolor{myred}&\cellcolor{myred}&\cellcolor{myred}&\cellcolor{myred}&\cellcolor{myred} &\cellcolor{blue!10}74&\cellcolor{blue!10}\dots&\cellcolor{blue!10}25\\
    \end{tabular}  
  \end{center}
  \caption{Data structure when the final time period is $n = 150$, the minimum age $l = 25$ and the maximum age $u = 74$. The blue cells indicate the age ($t-s$) groups included in the sample.}\label{tab:indices}
\end{table}

We are now ready to introduce the NLS estimator of the true parameter vector \(\theta_0 \coloneqq (\beta_0,\gamma_0)^\Tr\). In a first step, the time fixed effects are eliminated by subtracting
from Eq.\ \eqref{surveyexp} cohort-means, yielding
\begin{equation}
\tilde z_{t,s} = \beta_0 \tilde a_{t,s}(\gamma_0) + \tilde \eps_{t,s},
\end{equation}
where a tilde indicates deviations from cohort-means:
\begin{equation}
\tilde{e}_{t,s} \coloneqq e_{t,s} - \bar{e}_{t\cdot} \quad \text{ and } \quad	 \bar{e}_{t\cdot} \coloneqq \frac{1}{m}\sum_{s = t-u}^{t-l}e_{t,s}, \quad e \in \{a,z,\eps\}.
\end{equation}
For economy of notation, the dependence of \(\tilde e_{t,s}\) on the number of cohorts \(m\) is implicitly understood. In terms of the data structure illustrated in Table \ref{tab:indices}, the cohort-means are given by row-wise
averages. The NLS estimator \(\theta_n \coloneqq (\theta_{\beta,n},\theta_{\gamma,n})^\Tr\), say, then minimises the objective 
\begin{equation}\label{Qn}
  Q_n(\theta) \coloneqq \sum_{t = u+1}^{n} \sum_{s = t-u}^{t-l}(\tilde{z}_{t,s} - \beta \tilde{a}_{t,s}(\gamma))^2 = \sum_{t = u+1}^{n} \sum_{s = l}^{u}(\tilde{z}_{t,t-s} - \beta \tilde{a}_{t,t-s}(\gamma))^2,
\end{equation}
over \(\theta = (\beta, \gamma)^\Tr \in {\mathbb R} \times \Gamma \), $\Gamma \coloneqq [\ubar\gamma,\bar\gamma]$ for $0<\ubar\gamma<\bar\gamma<\infty$ specified below. Note that since
\(\tilde{a}_{t,s}(\gamma) \coloneqq a_{t,s}(\gamma)-\bar a_t(\gamma)\), \(Q_n(\theta)\) depends on \(\gamma\) through both  \(a_{t,s}(\gamma)\) and its cohort-mean. Clearly, the objective is highly nonlinear in $\gamma$,
necessitating the use of numerical routines for estimation. However, we note that the computational burden of numerical optimisation can be reduced significantly by profiling $Q_n (\theta)$ in Eq.\ \eqref{Qn} further w.r.t.\ $\beta$. Put differently, upon exploiting that the model is linear in \(\beta\) we get $(\beta_n,\gamma_n)^\Tr$ where
$\beta_n \coloneqq  \beta_n(\gamma_n)$, and $\gamma_n$ minimises  
  \begin{align*}
    Q^\st_n(\gamma) \coloneqq \sum_{t = u+1}^{n} \sum_{s = t-u}^{t-l}(\tilde{z}_{t,s} - \beta_n(\gamma) \tilde{a}_{t,s}(\gamma))^2, \quad
    \beta_n(\gamma) \coloneqq \frac{ \sum_{t = u+1}^{n} \sum_{s = t-u}^{t-l} \tilde a_{t,s}(\gamma) \tilde z_{t,s}}{\sum_{t = u+1}^{n} \sum_{s = t-u}^{t-l} \tilde a^2_{t,s}(\gamma)}.
  \end{align*}
The same approach is commonly taken in threshold models (see, e.g.  \citealp[Section 3]{hansen:17}).

\section{Assumptions, consistency, and asymptotic normality}\label{sec:asymptotics}

The statistical analysis of the NLS estimator requires some care. Upon inspecting Eq.\ \eqref{amean}, it is evident that the sample objective function $Q_n(\theta)$ is continuous in $\theta$ but has ``kinks'', i.e.\ lacks
differentiability, if $\gamma \in \mathbb{Z}$. Importantly, even its suitably standardised population counterpart might not be differentiable everywhere, a crucial
prerequisite for the asymptotic normality of $\theta_n$ as $n \rightarrow \infty$, see e.g. \citet[Section 7]{newmc:94}. To ensure differentiability --at least in the limit-- we must allow for the number of cohorts $m$ to diverge. However, the resulting setting poses two further challenges:
First, contraction (or invertibility) arguments, that turn out to be very convenient when establishing the required uniform results in similar settings like GARCH models, score driven models, or, more generally, nonlinear time
series models (e.g.\ \citealp{jensen:04}, \citealp{strau:06}, \citealp{blasques:18}, or \citealp[Section 6]{potscher:13}), fall short in our setting. In view of Eq.\ \eqref{Qn}, this follows readily by recognising that the mapping
$f_{t,s}(\cdot)$ defined via $a_{t,s} = f_{t,s}(a_{t-1,s},y_t;\gamma)$, $$f_{t,s}(a,y;\gamma) = (1-\gamma_{t,s}(\gamma))a+\gamma_{t,s}(\gamma)y $$ 
is not contracting in its first
argument because
$$\ssup\limits_{l \leq t-s \leq u, \gamma \in \Gamma}\lln\left|\frac{\partial f_{t,s}(a,y;\gamma)}{\partial a}\right| = 0$$
if $m$ (and thus $u$) diverges.   Second, the nonlinear regression function degenerates if the number of cohorts diverges. To see this, it is helpful to recognise that under certain regularity conditions laid down below
\begin{align}\label{phi}
  (t-s) \, \var[a_{t,s}(\gamma)]
  \rightarrow \omega^2\varphi(\gamma,\gamma),
\end{align}
as age grows large (i.e \(t-s \rightarrow \infty\)), with   
\begin{align}  \nonumber
  \omega^2 \coloneqq \sum_{j=-\infty}^\infty   \cov[y_0,y_{j}]\quad \varphi(\gamma_1,\gamma_2) \coloneqq \frac{\gamma_1\gamma_2}{\gamma_1+\gamma_2-1}.
\end{align}

Put differently, our setting can be described as a nonlinear panel specification with stochastic, potentially evaporating trend component. The analysis
of linear models with deterministic (e.g. \citealp{phillips:07}) and stochastic (e.g., \citealp{chrismass:19} or \citealp{mayer:22}) evaporating trends leads us to expect nonstandard limiting behaviour also in the present case. More specifically, the evaporating component of the recursively generated expectations causes time-varying moments (e.g. $\var[a_{t,s}] = O(t-s)$ cf. Eq.\ \eqref{phi}) and thus nonstationarity. 

 To summarize, we cannot derive the properties of $\theta_n$ from standard asymptotic theory. Instead, we must derive them from first principles. In particular, the recursive solution in Eq.\ \eqref{amean} enables us, by leveraging insights from analytical number theory, to directly apply seminal results from $M$-estimation. This approach allows us to establish, among other things, the consistency, convergence rates, limiting distribution, and inferential methods for $\theta_n$. In doing so, we impose the following assumptions.

\subsection{Assumptions}

 First, we distinguish between two asymptotic regimes: 
 
\renewcommand{\theassumption}{A}
\begin{assumption}\label{ass:A1}
\(u\) and \(l\) are fixed constants independent of \(n\).
\end{assumption}

\renewcommand{\theassumption}{A'}
\begin{assumption}\label{ass:A2}\(u \coloneqq u(n) \rightarrow \infty\) as \(n \rightarrow \infty\) such that \(m  = u-l+1\rightarrow \infty\), where \(l\) is either fixed or \(l \coloneqq l(n) \rightarrow \infty\), with
  \[
    (\lln(u)-\lln(l))/\lln(n) \rightarrow \lambda_{1} \in (0,\infty), \quad u/n \rightarrow \lambda_2 \in [0,1),
  \]
\end{assumption}

Assumption \ref{ass:A1} refers to a fixed cohort length (\(m\)), while Assumption \ref{ass:A2} allows \(m = m(n)\) to diverge pathwise as a function of the sample size \(n\). Note, that the lower bound \(l\) can be fixed, while \(u = u(n) \rightarrow \infty\) ensures that \(m\) diverges with $n$.\footnote{The assumption allows the (finite) limit of the ratio $m/n$ to be either strictly positive or zero. For
  example, to see that the latter case is covered, suppose that $l$ is fixed and $u \sim n^{\kappa}$, $\kappa \in (0,1)$, so that $\lln(u)/\lln(n) \sim \kappa > 0$ but $m/n \rightarrow 0$.} Intuitively, under Assumption
\ref{ass:A2}, we are able to consistently estimate the fixed effects so that (asymptotically) the additional estimation error due to cohort-demeaning --present under Assumption \ref{ass:A1}-- vanishes. More importantly, however,
$m = m(n) \rightarrow \infty$ ensures a sufficiently smooth population objective such that we can hope to derive the limiting distribution of $\theta_n$.

The next assumption specifies the distributional characteristics of $y_t$ and the error term:

	\renewcommand{\theassumption}{B}
\begin{assumption}\label{ass:B}\textcolor[rgb]{1,1,1}{.}
\begin{enumerate}[label=\textnormal{\textbf{B.\arabic*}},ref=B.\arabic*]
\item\label{ass:B1} \(\{y_t\}_t\) is fourth-order stationary with  continuous spectral density bounded away from zero, autocovariance function $c(\cdot)$ such that $\ssup_{\tau \geq 1}(1+|\tau|)^2|c(\tau)| \leq \infty$, and absolutely summable cumulants up to order four. 
\item\label{ass:B2} For each \(s\leq t\),  \(\{\eps_{t,s}\}_{t}\)  form martingale difference sequences with respect to \(\mathcal{F}_{t} \coloneqq \sigma(\{y_{i+1},\eps_{i,j}: i \leq t, j\leq i\})\) such that \(\Ex[\eps_{t,s}^r \mid \mathcal{F}_{t-1}]\), \(r \in \{2,3,4\}\) are finite constants a.s. so that \(\Ex[\eps_{t,s}\eps_{t,k} \mid \mathcal{F}_{t-1}] = \sigma^21\{k=s\}\) a.s..
\end{enumerate}
\end{assumption}

Assumption \ref{ass:B} places some structure on the dependence of the process \(\{y_t\}_t\) using a fourth-order cumulant condition. Any stationary Gaussian process with $\ssup_{\tau \geq 1}(1+|\tau|)^2|c(\tau)| \leq \infty$ satisfies Assumption \ref{ass:B1} because higher order cumulants are zero in this case. More generally, stationary processes under (strong) mixing conditions (see, e.g., \citealp{doukhan:89}) as well as linear processes with absolutely summable Wold coefficients and \textsf{IID} innovations that have finite fourth moments (see, e.g., \citealp{hannan:70}) can be shown to have absolutely summable fourth cumulants, i.e.
$\sum_{i,j,k=-\infty}^\infty |c(i,j,k)| < \infty,$ $c(i,j,k) \coloneqq \textsf{cum}[y_t,y_{t+i},y_{t+j},y_{t+k}].$
These summability conditions restrict the memory of \(\{y_t\}_t\) to be short and allow us to evaluate higher order moments of the recursion in Eq.\ \eqref{recursion} based on arguments borrowed from \cite{dem:2008}. Assumption \ref{ass:B2} assumes that the error term is a homoskedastic martingale difference sequence with finite homokurtosis; importantly, the assumption rules out serial correlation among time (\(t\)) and birth period (\(s\)). We leave any weakening of Assumption \ref{ass:B2} for future research, but return to this issue briefly as part of a Monte Carlo study in the Supplementary Material. 

Finally, Assumption \ref{ass:C} restricts the parameter space:

	\renewcommand{\theassumption}{C}
 \begin{assumption}\label{ass:C} \(\theta_0 \in \textnormal{\sf int}(\Theta)\), where \(\Theta \coloneqq \Xi \times \Gamma\), $\Xi \coloneqq [\ubar\beta,\bar\beta]$, $\Gamma \coloneqq [\ubar\gamma,\bar\gamma]$ for $-\infty<\ubar\beta\leq \bar\beta < \infty$ and \(\frac2{3} < \ubar\gamma < \bar\gamma <\infty\). \end{assumption}

 Assuming a compact and convex parameter space is a standard assumption for nonlinear regression (e.g. \citealp{jennrich:1969} or \citealp{chan:15}). In particular, we impose compactness on the parameter space of $\beta$. We follow \citet[Assumption 2.2]{hansen:17}, who argues that, in principle, the restriction could be relaxed such that $\Xi = \mathbb{R}$ at the expense of more technical detail, as, for instance, discussed in \citet[Section 2.6]{newmc:94}. Whenever interest lies in identifying jointly the parameter vector $\theta$, we impose the additional identification restriction \(\beta_0 \neq 0\). Fortunately, as shown in Section \ref{sec:inf}, it is still possible to draw statistical inferences involving the hypothesis \(\beta = 0\), provided a suitable test statistic is used. Although it seems possible to relax the constraint \(\ubar\gamma > 2/3\) and allow for \(\gamma \in (1/2,2/3]\), this comes with a substantial increase in additional technicalities and is thus left for future research. The boundary point $\gamma = 1/2$, in particular, presents several difficulties as already discussed in \cite{chrismass:18} for a linear regression model. Importantly, Assumption \ref{ass:C} allows for ``recency bias'' ($\gamma > 1$) as well as updating schemes where distant data points are weighted more heavily than recent ones ($\gamma < 1$). As we will see in Section \ref{sec:emp}, this allows us to empirically test the hypothesis of ``recency bias'' put forward by MN.

\subsection{Consistency}

Inspired by the analysis of the NLS estimator with trending data by \cite{park:01},  we make use of the following seminal result of \cite{jennrich:1969}: If the `identification criterion' 
$
D_n(\theta) \coloneqq Q_n(\theta)-Q_n(\theta_0),
$
scaled suitably by some sequence \(\nu_n \rightarrow \infty\), converges uniformly in probability to a continuous (deterministic) function that is uniquely minimised at \(\theta = \theta_0\), then, \(\theta_n  \rightarrow_p \theta_0\). This allows us to establish the consistency of the NLS estimator as summarized below:

\renewcommand{\theproposition}{1}
\begin{proposition}\label{prop:1}{\textcolor{white}{.}} 
\begin{enumerate} 
\item If Assumptions \textnormal{\ref{ass:A1}, \ref{ass:B}}, and \textnormal{\ref{ass:C}} are satisfied, then 
$ \ssup_{\theta \in \Theta}|\frac1{n}D_n(\theta) - {\sf D}_{m}(\theta)| \rightarrow_p 0,$
where ${\sf D}_{m}(\theta) \coloneqq   \sum_{s,k=l}^u\left[1\{s=k\}-\frac{1}{m}\right]{\sf D}_{s,k,m}(\theta)$, with 
\begin{align*}    \normalfont
\textsf{D}_{s,k,m}(\theta) \coloneqq  &\ \beta^2\sum_{i=\floor{\gamma}}^s\sum_{j=\floor{\gamma}}^k\kappa_{i,s}(\gamma)\kappa_{j,k}(\gamma)c(k-s+i-j)\\
&+\beta_0^2\sum_{i=\floor{\gamma_0}}^s\sum_{j=\floor{\gamma_0}}^k\kappa_{i,s}(\gamma_0)\kappa_{j,k}(\gamma_0)c(k-s+i-j)\\
&-2\beta\beta_0\sum_{i=\floor{\gamma_0}}^s\sum_{j=\floor{\gamma}}^k\kappa_{i,s}(\gamma_0)\kappa_{j,k}(\gamma)c(k-s+i-j).
\end{align*}
\item If Assumptions \textnormal{\ref{ass:A2}, \ref{ass:B}}, and \textnormal{\ref{ass:C}} are satisfied, then
$
\ssup_{\theta \in \Theta}|\frac1{n\lln(n)}D_n(\theta) - {\sf D}(\theta)| \rightarrow_p 0,
$
where \({\sf D}(\theta) \coloneqq (\lambda\omega)^2(\beta^2\varphi(\gamma,\gamma)+\beta_0^2\varphi(\gamma_0,\gamma_0)-2\beta\beta_0\varphi(\gamma,\gamma_0))\), with  $\omega^2$ and \(\varphi(\cdot,\cdot)\) defined in Eq.\ \eqref{phi} while
\(
\lambda^2 \coloneqq  \lambda_1(1-\lambda_2).
\)
\end{enumerate}
If, in addition, \(\beta_0 \neq 0\), then \(\normalfont\theta \mapsto \textsf{D}_{m}(\theta)\) and \(\normalfont\theta \mapsto \textsf{D}(\theta)\) are uniquely minimised at \(\theta = \theta_0\). Consequently,  \(\theta_n \rightarrow_p \theta_0\).
\end{proposition}

Under Assumption \ref{ass:A1}, the population criterion function \(\theta \mapsto \textsf{D}_m(\theta)\) can be viewed as a quadratic form of the (Toeplitz) covariance matrix \(\{c(|j-i|)\}_{0\leq i \leq j \leq u}\), that is positive definite under Assumption \ref{ass:B1}. Note how the population objective function for fixed $m$ is smooth in $\beta$ but viewed as a function of the gain parameter $\gamma \mapsto {\sf D}_m(\beta,\gamma)$ lacks differentiability.
Also under the asymptotic regime of Assumption \ref{ass:A2}, the map \(\theta \mapsto \textsf{D}(\theta)\) is non-negative as inspection of the function \(\varphi(\gamma_1,\gamma_2)\) reveals. Moreover, \(\textsf{D}(\cdot)\) can be viewed as the smooth limit of the rescaled \(\textsf{D}_m(\cdot)\), i.e. \({\lln}^{-1}(n)\textsf{D}_m(\theta) \rightarrow \textsf{D}(\theta)/(1-\lambda_2)\) for \(m \rightarrow \infty\) as \(n \rightarrow \infty\). Finally, the factor $\lambda^2 =  \llim\limits_{n \rightarrow \infty} \lln(u/l)/\lln(n)(1-u/n)$ in Proposition \ref{prop:1}, Part 2, can be interpreted as the asymptotic relative proportion of cohorts to time periods.

An intriguing aspect of Proposition \ref{prop:1} is the different scaling of the objective function. Specifically, the scaling depends on whether $m$ is fixed or $m$ diverges. In the former case, the scaling is $\nu_n = n$, while in the latter, it is $\nu_n = n\lln(n)$. The following example is intended to provide further intuition on this point:

\begin{example}\label{example}
Consider the case of the scalar (OLS) estimator $\theta_n$ of $\theta = \beta$ in case $\gamma = 1$ is known. Here, it is known that the convergence rate of the estimator is determined by the scaling $\nu_n$, say, needed to stabilize the regresser second sample-moment, which, when scaled by $\nu_n$, is given by $H_n \coloneqq \frac1{\nu_n}\sum_{t=u+1}^n\sum_{s=l}^u \tilde a_{t,t-s}^2$. Assuming $\var[\eps] = 1$ and $c(\tau) = 1\{\tau = 0\}$, then we get under Assumption \ref{ass:A2} with $\nu_n = n\lln(n)$
\begin{align}\label{eq:example}
\Ex[H_n] = \,&  \left(1-\frac{u}{n}\right)\frac1{\lln(n)} \bigg[\left(1-\frac1{m}\right)(\psi(u+1)-\psi(l)) - 2\left(1-\frac1{m}\right)\\
\,& \qquad \qquad \quad + \frac{2l}{m}(\psi(u+1)-\psi(l+1))\bigg] = \left(1-\frac{u}{n}\right)\frac{\lln(u/l)}{\lln(n)}+O\left(\frac1{\lln(n)}\right) \nonumber,
\end{align}
for the digamma function $\psi(\cdot)$ (see the appendix for details). That is, the scaling by $\nu_n = n\lln(n)$ ensures that the expected regressor second-moment stabilizes and converges to the asymptotic relative sample-size $\lambda^2 = \llim\limits_{n \rightarrow \infty} \lln(u/l)/\lln(n)(1-u/n)$ as $m=m(n) \rightarrow \infty$ with $n \rightarrow \infty$. If, on the other hand, under Assumption \ref{ass:A1} only $n$ diverges and $m$ is fixed, then inspection of Eq. \eqref{eq:example} reveals that scaling by $\nu_n = n$ suffices to obtain a nondegenerate limit. 

\end{example}

Beyond the special case treated in the preceding example, one might want to make a more general statement about the convergence rate of the $\theta_n$. While the convergence rate of the estimator can be directly deduced from the limiting distribution derived in the next section under the asymptotic regime \ref{ass:A2}, the same approach cannot be taken if $m$ is fixed. The reason is that under asymptotic regime \ref{ass:A1}, the {\it population} objective is not differentiable, which, however, is an indispensable requirement to derive the limiting distribution. Instead, to derive the rate of convergence under Assumption \ref{ass:A1}, we make use of \citet[Theorem 3.2.5]{van:1996}, exploiting a stochastic Lipschitz bound on $\gamma \mapsto a_{t,t-s}(\gamma)$. Due to the non-differentiability of $\gamma \mapsto {\sf D}_m(\beta,\gamma)$ the remaining difficulty lies in verifying
$-{\sf D}_m(\theta) \leq - C\Vert \theta-\theta_0 \Vert^2$ for all $\theta$ in a neighbourhood of $\theta_0$ and some finite $C>0$. Because a standard (Taylor) expansion approach fails, our argument instead rests on deriving the subgradient. In doing so, as we saw in Proposition \ref{prop:1} already, we have to exclude the case $\beta_0 = 0$ to {\it jointly} identify $\beta$ and $\gamma$. However\footnote{We are grateful to a reviewer for pointing this out.}, {\it individually}, the first element $\theta_{\beta,n}$ of $\theta_n \coloneqq (\theta_{\beta,n},\theta_{\gamma,n})^\Tr$, still estimates $\beta_0 = 0$ consistently. In showing this, we adapt the discussion in \citet[Section 5.3]{saikkonen:95} and \citet[Theorem 1]{seo:11} that both build on \citet[Lemma 1]{wu:81}. The preceding discussion can be summarized as follows:

\renewcommand{\thecorollary}{1}
\begin{corollary}
    \label{cor:0} Under the conditions of Proposition  \textnormal{\ref{prop:1}}, $\sqrt{\nu_n}\Vert \theta_n-\theta_0\Vert = O_p(1)$, while, if $\beta_0 = 0$, $\sqrt{\nu_n}|\theta_{n,\beta}| = O_p(1)$, where $\nu_n = n$ or $\nu_n = n\lln(n)$ depending on whether Assumption \textnormal{\ref{ass:A1}} or \textnormal{\ref{ass:A2}} holds, respectively. 
\end{corollary}

 When comparing the convergence rates under both asymptotic regimes, we note that the rather slow additional \(\lln(n)\)-factor of the convergence rate \(\nu_n = n\lln(n)\) under Assumption \ref{ass:A2} arises due to the trending behaviour of the data mentioned earlier and captures the variation as cohorts grow older (\(t-s\)); this is similar in nature to the convergence rates featuring in earlier related work in models with macroeconomic time series (see, e.g., \citealp{chrismass:18} or \citealp{mayer:22}). It is thus only the variation across cohorts (\(s\)) that leads to the
 convergence factor \(n\), thereby making the NLS estimator practically appealing. Or, in the words of \citet[p. 63]{mn:2016} the ``cross-sectional heterogeneity\ ...\ provides a new source of identification''. Our results therefore provide a rigorous justification for this assertion.

\subsection{Asymptotic normality}

 As discussed before, the (centred) objective $D_n(\cdot)$ is not differentiable on the set of ``kink points'' where the gain is integer-valued. This complicates the proof of asymptotic normality. However, as discussed in \citet[Section 7]{newmc:94}, a less restrictive notion of smoothness called \textit{stochastic differentiability} can bypass the common requirement that the sample objective is differentiable twice, provided the population objective is sufficiently smooth (see e.g. \citealp{srisuma2013supplement}, \citealp{oh2013simulated}, or \citealp{mayerwied23} for similar arguments). As Proposition \ref{prop:1} reveals, this requires that $m \rightarrow \infty$, because even the population objective ${\sf D}_m(\cdot)$, that obtains for $m$ fixed, is not differentiable when viewed as a function of $\gamma.$

 To that end, we approximate in a first step $D_n(\cdot)$ with a smooth counterpart $D^\dagger_n(\cdot)$, say. More specifically, under the asymptotic regime of Assumption \ref{ass:A2}, results from analytical number theory can be used to obtain a smooth approximation:
 \[
 \ssup\limits_{\theta\in \Theta}\nu_n^{-1}|D_n(\theta)-D_n^\dagger(\theta)| = o_p(1), \quad \nu_n =  n \lln(n),
 \]
 where $D^\dagger_n(\cdot)$, satisfying the stochasticly differentiability mentioned above and whose probability limit coincides with the smooth function $\theta \mapsto \textsf{D}(\theta)$, is given by  
\begin{align}\label{eq:Ddagger}
D_n^\dagger(\theta) \coloneqq \sum_{t=u+1}^n\sum_{s=l}^u(\beta_0\tilde r_{t,t-s}(\gamma_0) -\beta \tilde r_{t,t-s}(\gamma))(\beta_0\tilde r_{t,t-s}(\gamma_0) -\beta \tilde r_{t,t-s}(\gamma)-2\tilde\eps_{t,t-s}),  
\end{align}
with $r_{t,t-s}(\gamma) \coloneqq \sum_{i=1}^s h_{i,s}(\gamma)y_{t-s+i}$, $h_{i,s}(\gamma) \coloneqq \frac{\gamma}{s^\gamma}i^{\gamma-1}$. Note that $D_n(\beta,1) = D_n^\dagger(\beta,1)$; see Lemma \ref{lem:A4} of the appendix for details.  Based on seminal results for extremum estimators with non-smooth objective function collected in \citet[Thm. 7.1]{newmc:94}, we can then derive the following proposition.

 \renewcommand{\theproposition}{2}
\begin{proposition}\label{prop:2} If Assumptions \textnormal{\ref{ass:A2}, \ref{ass:B}} and \textnormal{\ref{ass:C}} are satisfied and $\beta_0 \neq 0$, then 
\[\normalfont
\sqrt{\nu_n}(\theta_n-\theta_0) = -\left(\frac1{2}\frac{\partial^2}{\partial\theta\partial\theta^\Tr}\textsf{D}(\theta)\bigg\rvert_{ \theta=\theta_{\scalebox{.45}{0}}}\right)^{-1} \frac1{\sqrt{4\nu_n}}\frac{\partial}{\partial\theta}  D^\dagger_n(\theta)\bigg\rvert_{ \theta=\theta_{\scalebox{.45}{0}}} + o_p(1),
\]
where
\[\normalfont
\frac1{\sqrt{4\nu_n}}\frac{\partial}{\partial\theta}  D^\dagger_n(\theta)\bigg\rvert_{ \theta=\theta_{\scalebox{.45}{0}}} \rightarrow_d \mathcal{N}_2(0,(\sigma \omega\lambda)^2{\sf H}), \quad \normalfont\textsf{H} \coloneqq \textsf{H}(\theta_0),
\]
 with $\normalfont\frac1{2}\frac{\partial^2}{\partial\theta\partial\theta^\Tr}\textsf{D}(\theta) = (\omega\lambda)^2{\sf H}(\theta)$,
\[\normalfont
\normalfont\sqrt{\nu_n}(\theta_n-\theta_0) \rightarrow_d \mathcal{N}_2\left(0,\left(\frac{\sigma}{\omega\lambda}\right)^2\textsf{H}^{-1}\right), \quad \textsf{H}(\theta)\coloneqq \varphi(\gamma,\gamma) \begin{bmatrix} 1 &  \frac{\beta}{\gamma}\frac{\gamma-1}{(2\gamma-1)}\\    \frac{\beta}{\gamma}\frac{\gamma-1}{(2\gamma-1)} &  \frac{\beta^2}{\gamma}\frac{1/\gamma+2(\gamma-1)}{(2\gamma-1)^2}\end{bmatrix}.
\]
 \end{proposition}

In general, the variance-covariance matrix depends on the $2 \times 1$ parameter vector of interest $\theta$ through ${\sf H}(\theta)$, which might lead to non-similar inference (see, e.g. \citealp{nankervis:85}). Indeed, as discussed in the following section, the local power of a $t$-test for $H_0:$ $\gamma = \gamma_0$ can be arbitrarily close to the nominal significance level for values of $\gamma_0$ found in empirical studies. If, however, $\gamma_0 = 1$, then ${\sf H}(\theta)$ is a diagonal matrix\footnote{We are grateful to a reviewer for pointing this out.} so that the limiting marginal distributions of the elements of $\theta_n = (\theta_{n,\beta},\theta_{n,\gamma})^\Tr$ are independent of each other and free of the respective parameters itself, i.e.  $\nu_n \var[\theta_{\beta,n}] \rightarrow (\sigma/\gamma_0)^2/(\omega\lambda)^{2}$ and $\nu_n\var[\theta_{\gamma,n}]\rightarrow (\sigma/\beta_0)^2/(\omega\lambda)^{2}$. Intuitively, $\gamma_0 = 1$ reduces the weighted least-squares recursion with data-dependent weights in Eq.\ \eqref{recursion} to an on-line ordinary least-squares estimator free of the nuisance parameter $\gamma$. Finally, we note that the relative asymptotic sample size $\lambda^2$, featuring before in Proposition \ref{prop:1} and Example \ref{prop:1}, enters inversely the limiting variance-covariance matrix. We could thus also directly scale the estimator with the relative sample size to get a limiting distribution free of $\lambda^2$, i.e.
$\sqrt{(n-u)\lln(u/l)}(\theta_n-\theta_0) \rightarrow_d \mathcal{N}_2(0,(\sigma/\omega)^2{\sf H}^{-1}),$ see also the discussion of Example \ref{example}.

\section{Standard errors and inference}\label{sec:inf}

Standard errors require consistent estimators of the error variance and the Hessian. While the former is consistently estimated using  
$s_n^2 \coloneqq s_n^2(\theta_n)$, $s^2_n(\theta) \coloneqq \frac1{N}Q_n(\theta)$, estimators of the latter can alternatively be based on any of the following three expressions:
\begin{align}
H_n(\theta) \coloneqq \frac1{\nu_n}\sum_{t=u+1}^n\sum_{s=l}^u (\tilde r_{t,t-s}(\gamma),  \beta\tilde r^{(1)}_{t,t-s}(\gamma))^\Tr(\tilde r_{t,t-s}(\gamma),  \beta\tilde r^{(1)}_{t,t-s}(\gamma)), \label{eq:obsHess} 
\end{align}
with, see Eq. \eqref{eq:Ddagger} above, $r^{(1)}_{t,t-s}(\gamma) \coloneqq \sum_{j=1}^sh^{(1)}_{j,s}(\gamma)$, $h^{(1)}_{j,s}(\gamma) \coloneqq h_{j,s}(\gamma)(\lln(j/s)+1/\gamma)$; or
\begin{align}\label{eq:expHess}
\Ex[H_n(\theta)] =\,&  \frac{1-u/n}{\lln(n)}\sum_{s,k=l}^u\left[1\{s=k\}-\frac{1}{m}\right]\nonumber\\
\,& \qquad\qquad\qquad \times \sum_{i=1}^s\sum_{j=1}^k c(k-s+i-j) e_{i,s}(\theta)e_{j,k}(\theta)^\Tr,   
\end{align}
with $e_{j,s}(\theta) \coloneqq (h_{j,s}(\gamma),\beta h_{j,s}^{(1)}(\gamma))^\Tr$; or
\begin{align}\label{eq:limexpHess}
 p\llim_{n \rightarrow \infty} H_n(\theta) = \llim_{n \rightarrow \infty} \Ex[H_n(\theta)] = (\omega\lambda)^2{\sf H}(\theta), 
\end{align}

where ${\sf H}(\theta_0)$ is given in Proposition \ref{prop:2}. Akin to the discussion of the relative accuracy of observed and expected Fisher information (see e.g. \citealp{efron:78} or \citealp{lind:97}), we may refer to \eqref{eq:obsHess} and \eqref{eq:expHess} as `observed' Hessian and `expected' Hessian, respectively,  while \eqref{eq:limexpHess} is  referred to as the `asymptotic' Hessian. We find that, although computationally attractive, an estimator based on the asymptotic Hessian is in finite samples inferior. It is instructive to illustrate these quantities by returning to our discussion of the OLS estimator in  Example \ref{example}: Here, the observed Hessian is $H_n(\theta) = H_n = \frac1{\nu_n}\sum_{t=u+1}^n\sum_{s=l}^u \tilde a_{t,t-s}^2$. An analytical expression of the
expected Hessian $\Ex[H_n]$ is given by Eq. \eqref{eq:example}, which shows that the asymptotic Hessian --given here by ${\sf H}(\theta) = \lambda^2$-- differs from $\Ex[H_n]$ by an order of magnitude of ${\lln}^{-1}(n)$. Therefore, the use of the latter provides even in large samples only a poor approximation of the finite sample variance $\nu_n\var[\theta_n]$.

Turning back to the general case, we can readily construct estimators from the three different Hessians \eqref{eq:obsHess}, \eqref{eq:expHess}, and \eqref{eq:limexpHess}  using appropriate sample counterparts. We call theses estimators $H_{j,n}$, $j \in \{1,2,3\},$ respectively. Specifically, one obvious estimator is the observed Hessian in  \eqref{eq:obsHess}  evaluated at $\theta_n$, i.e. $H_{1,n} = H_n(\theta_n)$. From the expected Hessian in \eqref{eq:expHess} an estimator $H_{2,n} \coloneqq H_{2,n}(\theta_n)$ obtains by replacing the unknown quantities $(\theta_0,\{c(\tau)\}_{\tau=0}^u)$ entering $\Ex[H_n(\theta_0)]$ with the sample counterparts $(\theta_n,\{c_n(\tau)\}_{\tau=0}^u)$, $c_n(\tau) \coloneqq \frac{1}{n}\sum_{t=1}^{n-\tau}(y_t-\bar y)(y_{t+\tau}-\bar y)$, i.e.
\begin{equation}\label{eq:H2n}  
\begin{split}
\hspace*{-.2cm}H_{2,n}(\theta) \coloneqq \,&   \frac{1-u/n}{\lln(n)}\sum_{s,k=l}^u\left(1\{s=k\}-\frac{1}{m}\right)  \sum_{i=1}^s\sum_{j=1}^k c_n(k-s+i-j) e_{i,s}(\theta)e_{j,k},(\theta)^\Tr.    \end{split}    
\end{equation}  
Similarly, we can make \eqref{eq:limexpHess} operational via $H_{3,n} \coloneqq H_{3,n}(\theta_n)$, $H_{3,n}(\theta) \coloneqq   (\omega_n\lambda_n)^2 {\sf H}(\theta)$, where $\lambda_n^2 \coloneqq \lln(u/l)/\lln(n)(1-u/n)$ and $\omega_n^2$ is an estimator of the long-run variance $\omega^2$.

Finally, we propose an additional estimator $H_{4,n} \coloneqq H_{4,n}(\theta_n)$, say, that does not exploit the analytical expressions of the Hessian. In particular, because $Q_n(\cdot)$ is not differentiable, this estimator is simply based on a second-order numerical derivative of the objective function  with $i$,$j$-th element ($1\leq i,j \leq 2$) given by
\begin{equation}
\begin{split}\label{eq:H4n}
[H_{4,n}(\theta)]_{i,j} \coloneqq \,& \frac1{2\nu_n}(Q_n(\theta_n+e_i\ell_n+e_j\ell_n)-Q_n(\theta_n-e_i\ell_n+e_j\ell_n) \\
&\quad\qquad-Q_n(\theta_n+e_i\ell_n-e_j\ell_n)+Q_n(\theta_n-e_i\ell_n-e_j\ell_n))/(2\ell_n)^2
\end{split}
\end{equation}
for \(\ell_n \searrow 0\) and $e_i$, $i \in \{1,2\}$, denoting some sequence of step-sizes and the $2 \times 1$ unit vector, respectively. 

\renewcommand{\thecorollary}{2}
\begin{corollary}
    \label{cor:1} Suppose the assumptions of Proposition \textnormal{\ref{prop:2}} hold, then \textnormal{($i$)} $H_{1,n} \rightarrow_p (\omega\lambda)^2{\sf H}$, \textnormal{($ii$)} $H_{2,n} \rightarrow_p (\omega\lambda)^2{\sf H}$ if  $\mmax\limits_{1 \leq  \tau \leq u}   \sqrt{n}|c_n(\tau)-c(\tau)|= O_p(1)$ and $m = o({\lln}(n)\sqrt{n})$, \textnormal{($iii$)} $H_{3,n} \rightarrow_p (\omega\lambda)^2{\sf H}$ if $\omega_n^2 \rightarrow_p \omega^2$, and \textnormal{($iv$)} $H_{4,n} \rightarrow_p (\omega\lambda)^2{\sf H}$ if $\sqrt{\nu_n}\ell_n\rightarrow \infty$, where convergence in probability holds elementwise.
\end{corollary}

Two comments seem warranted: First, as discussed in \cite{xiao:14}, the condition in ($ii$) holds under $u = O(n^{\iota})$, $\iota \in (0,1)$, for a wide range of stationary processes $\{y_t\}_t$; the condition in ($iii$) on the long-run variance estimator can be verified for various candidates of $\omega_n^2$ (see, e.g. \citealp{andrews:91}); the condition in ($iv$) on the step-size is common in the literature (see, e.g. \citealp[Theorem 7.4]{newmc:94} or \citealp[Section 2.4]{oh2013simulated}). Second, we expect $H_{3,n}$, i.e. the estimator based on the asymptotic Hessian, to perform worst among the four estimators in finite, medium, and even large samples. As already suggested by Example \ref{example}, the reason is that ${H}_{3,n}$ provides only a poor approximation of the finite sample Hessian due to a bias of order $O(\lln^{-1}(n))$.

As the following corollary reveals, hypotheses of the form \(H_0\): \(R\theta_0 = \rho_0\), for some \(q\times 2\), \(q \leq 2\), restriction matrix \(R\) and \(\rho_0 \in \mathbb{R}^q\), can be tested using the Wald statistic 
\begin{align}\label{eq:Wald}
\textsf{W}_{j,n} \coloneqq \nu_n\frac{(R\theta_n-\rho_0)^\Tr(R H^{-1}_{j,n} R^\Tr)^{-1} (R\theta_n-\rho_0)}{s_n^2}, \quad j \in \{1,2,3,4\}.
\end{align}

\renewcommand{\thecorollary}{3}
\begin{corollary}\label{cor:2} Under the assumptions of Corollary \ref{cor:1}, \(\normalfont\textsf{W}_{j,n} \rightarrow_d \chi^2(q)\), $j \in \{1,2,3,4\},$ given $H_0$ is true. 
\end{corollary}

Under a sequence $\theta_{0,n} \coloneqq \theta_0 + \Delta/\sqrt{\nu_n}$, $\Delta \coloneqq (\Delta_\beta,\Delta_\gamma)^\Tr \in \mathbb{R}^2$, of so-called `Pitman drifts' the limiting distribution is $\chi^2(q)$ with non-centrality parameter $\kappa \coloneqq (\kappa_\beta,\kappa_\gamma)^\Tr =  (\omega\lambda/\sigma)^{2}\Delta^\Tr R^\Tr R {\sf H} R^\Tr R\Delta$, i.e. the test has non-trivial local power in a $\nu_n^{-1/2}$-vicinity around the null. It is instructive to consider the special case of the (squared) $t$-statistics for the hypotheses $H_0$: $\beta = \beta_0$ and $H_0$: $\gamma = \gamma_0$.  From the above we then get for the former $\kappa_\beta = (\omega\lambda\gamma\Delta_\beta/\sigma)^2/(2\gamma-1)$, which is independent of $\beta$, and, when viewed as function of $\gamma$, decreasing on $(1/2,1]$ but increasing on $[1,\infty)$. On the other hand, we obtain $\kappa_\gamma = (\omega\lambda\beta\Delta_\gamma/\sigma)^2(1+2\gamma(\gamma-1))/(2\gamma-1)^3$, which is an increasing function of $\beta$ but decreasing in $\gamma$ on the interval $(1/2,\infty)$.  Moreover, for $\gamma = 1$, $\kappa_\beta = (\Delta_\beta/\Delta_\gamma)^2\kappa_\gamma/\beta^2 = (\omega\lambda\Delta_\beta/\sigma)^2$. This is illustrated in Figure \ref{figpow}, depicting the power curves of the $t$-statistics  as a function of $\gamma$. 

\begin{figure}[h]
  \begin{minipage}[c]{0.55\textwidth}
       \begin{tikzpicture}
        \begin{axis}[
            xmin = .5, xmax = 2, ymin = 0, ymax = 1, xticklabel style={/pgf/number format/.cd,fixed},
        tick label style={font=\scriptsize},
		label style={font=\small},
		tick label style={font=\scriptsize},
		legend style={font=\scriptsize},
  width=1\textwidth,
    height=.2\textheight,
     			y label style={at={(axis description cs:0.0,0.5)},anchor=north},
  clip=false]
\draw[black, dotted] (axis cs:\pgfkeysvalueof{/pgfplots/xmin},0.05) -- (axis cs:\pgfkeysvalueof{/pgfplots/xmax},0.05);

\draw[gray!40] (axis cs:1,\pgfkeysvalueof{/pgfplots/ymin}) -- (axis cs:1,\pgfkeysvalueof{/pgfplots/ymax});
\addplot[black] table[x="gamma",y="pgamma"]{power.txt};
\addplot[black,dashed] table[x="gamma",y="pbeta"]{power.txt};
        \end{axis}
   \end{tikzpicture}
  \end{minipage}
  \begin{minipage}[c]{0.45\textwidth}
   \vspace*{.3cm} \caption{
    Theoretical asymptotic local power of the $t$-statistic for $H_0$: $\gamma = \gamma_0$ (solid line) and for $H_0$: $\beta = \beta_0$ (dashed line) as a function of $\gamma \in (1/2,3]$ for $\beta = \sigma = \omega = \lambda = \Delta_\beta = \Delta_\gamma =1$. The dotted horizontal line indicates the five per cent significance level.
    } \label{figpow}
  \end{minipage}
\end{figure}

Importantly, Corollary \ref{cor:2} does not allow for hypotheses containing the restriction \(\beta = 0\). Since this testing problem involves a nuisance parameter (viz. $\gamma$) that is not identified under the null, see e.g. \cite{andpol:1994} or \cite{hansen:1996} and the references therein. More specifically, we follow \cite{hansen:1996, hansen:17} and consider the following `supF' statistic
\begin{align}\label{supF}
{\it sup}\textsf{F} \coloneqq \ssup\limits_{\gamma \inn \Gamma} \frac{N(\tilde\sigma_n^2-\sigma^2_n(\gamma))}{\sigma^2_n(\gamma)} = \frac{N(\tilde\sigma_n^2-\sigma_n^2(\gamma_n))}{\sigma_n^2(\gamma_n)},
\end{align}
where $\tilde\sigma_n^2 \coloneqq \frac{1}{N}\sum_{t=u+1}^n\sum_{s=t-u}^{t-l} \tilde z_{t,s}^2$ and  $\sigma_n^2(\gamma) \coloneqq \frac{1}{N}Q_n^\st(\gamma)$,
with $Q^\st_n(\cdot)$ being the profiled objective defined at the end of Section \ref{sec:estimation}. The following corollary summarises the limiting behaviour of ${\it sup}{\sf F}$ under the null. 

\renewcommand{\thecorollary}{4}
\begin{corollary}\label{cor:3} Suppose Assumptions \textnormal{\ref{ass:B}}, \textnormal{\ref{ass:C}} are satisfied and \(H_0\)\textnormal{:} \(\beta = 0\). 
\begin{itemize}
    \item[\textnormal{($a$)}] If Assumption \textnormal{\ref{ass:A1}}, holds, then
\(\normalfont
{\it sup}\textsf{F} \rightarrow_d  T \coloneqq \ssup_{\gamma \inn \Gamma}\mathbb{S}_m^2(\gamma)/\varphi_m(\gamma,\gamma),
\)
where $\mathbb{S}_m(\cdot)$ is a Gaussian process with covariance kernel 
\[
\varphi_m(\gamma_1,\gamma_2) \coloneq \sum_{s,k=l}^u\left(1\{s=k\}-\frac{1}{m}\right)\sum_{i=\floor{\gamma_1}}^s\sum_{j=\floor{\gamma_2}}^kc(k-s+i-j) \kappa_{i,s}(\gamma_1)\kappa_{j,k}(\gamma_2).
\]
\item[\textnormal{($b$)}]
If Assumption \textnormal{\ref{ass:A2}}, holds, then
\(\normalfont
{\it sup}\textsf{F} \rightarrow_d  T \coloneqq \ssup_{\gamma \inn \Gamma}\mathbb{S}^2(\gamma)/\varphi(\gamma,\gamma),
\)
where $\mathbb{S}(\gamma)$ is a Gaussian process with covariance kernel 
 $\varphi(\cdot,\cdot)$ of Eq. \eqref{phi}.
 \end{itemize}
\end{corollary}

Three aspects of Corollary \ref{cor:3} are worth exploring. Firstly, as part ($a$) of Corollary \ref{cor:3} reveals, for this testing problem to be operational it is not necessary to require $m$ to diverge with $n$.  To see this, note that, under the null $\beta = 0$,
\[
\frac{N(\tilde\sigma_n^2-\sigma^2_n(\gamma))}{\sigma^2_n(\gamma)} = \frac{\sigma^2}{\sigma_n^2(\gamma)}\frac{(\frac1{\sigma\sqrt{\nu_n}}\sum_{t=u+1}^nS_t(\gamma))^2}{\frac1{\nu_n}\sum_{t=u+1}A_t(\gamma,\gamma)}, 
\]
where $S_t(\gamma) \coloneqq \sum_{s=l}^u \tilde a_{t,t-s}(\gamma)\eps_{t,t-s}$ and $A_{t}(\gamma_1,\gamma_2)\coloneqq \sum_{s=l}^u \tilde a_{t,t-s}(\gamma_1) \tilde a_{t,t-s}(\gamma_2)$.
Due to a stochastic Lipschitz condition verified in the appendix, we can deduce that (upon scaling by sample size and $\sigma$) $\sum_{t=u+1}^n S_t(\gamma)$ converges weakly to a Gaussian process with kernel coinciding with the limit of the process $\sum_{t=u+1}^n A_t(\gamma,\gamma)$, while, by the same arguments and the LLN, we get 
$
\sigma_n^2(\gamma) = \frac{1}{N}\sum_{t=u+1}^n\sum_{s=l}^u \eps_{t,t-s}^2(1+O_p(N^{-1})) \rightarrow_p \sigma^2
$ uniformly in $\gamma$. This yields the claim without the need of differentiability with respect to $\gamma$. 
 
Secondly, it follows readily that under a sequence of local alternatives $\beta_{0,n} = \Delta_\beta/\sqrt{\nu_n}$, $\Delta_\beta\in \mathbb R$, Corollary \ref{cor:3} holds with ${\mathbb S}_m(\cdot)$ and ${\mathbb S}(\cdot)$ replaced by ${\mathbb S}^\st_m(\cdot) = {\mathbb S}_m(\cdot)+\varphi_m(\gamma_0,\cdot)\Delta_\beta/\sigma$ and ${\mathbb S}^\st(\cdot) = {\mathbb S}(\cdot)+\varphi(\gamma_0,\cdot)\Delta_\beta/\sigma$, respectively; thus implying that tests based on ${\sf F}_n$ have non-trivial power in a $\nu_n^{-1/2}$-neighbourhood of the null. 

Thirdly, the process in ($b$) can be seen as the limit of that in ($a$) for $m = m(n) \rightarrow \infty$. However, akin to our discussion of standard errors, the convergence rate is slow, i.e. $|\lln^{-1}(n)\varphi_m(\gamma_1,\gamma_2)-\lambda_1\varphi(\gamma_1,\gamma_2)|=O(\lln^{-1}(n))$. This means that even in large samples it might not be a good idea to use critical values obtained by simulating the limiting process in ($b$). As an alternative, we could simulate the process in ($a$). However, its generation would depend on the autocovariances, implying that such a procedure becomes computationally highly expensive. Instead, to implement the test, we adopt a simple Gaussian multiplier bootstrap  proposed by \cite{hansen:1996}: For each $b \in \{1,\dots,B\}$, let ${\sf F}_{n,b}$ be the test statistic in \eqref{supF}, where $\tilde z_{t,s}$ is replaced by $z_{t,s,b}$,  with $\{z_{t,s,b}: u < t \leq n, t-u \leq s \leq t-l\}$ denoting a sample of  {\sf IID} standard normal variates. Next, define the bootstrap $p$-value $p_n \coloneqq 1- {\sf G}_n({\sf F}_n)$, where  ${\sf G}_n(t) \coloneqq  {\sf P}({\sf F}_{n,b} \leq t \mid {\mathcal S}_n)$ denotes the cumulative distribution function (cdf) of 
  ${\sf F}_{n,b}$,
  conditional on the data $\mathcal{S}_n \coloneqq \sigma(\{\{z_{t,s}\}_{s=t-u}^{t-l}\}_{t=u+1}^n, \{y_t\}_{t=1}^n)$. As revealed by the following corollary, this bootstrap replicates correctly the first-order asymptotic distribution of the test statistic.
  
\renewcommand{\thecorollary}{5}
\begin{corollary}\label{cor:5}
 Under the conditions of Corollary $\textnormal{\ref{cor:3}}$, $p_n = 1-{\sf G}(T)+o_p(1)$, where $ 1-{\sf G}(T)\sim {\sf Unif}[0,1]$, ${\sf G}(\cdot)$ is the cdf of the random variable $T$ defined in Corollary \ref{cor:3}.
  \end{corollary}
  
  Because ${\sf G}_n(\cdot)$ is unobservable, we simulate it via $G_{n,B}(t) \coloneqq \frac1{B} \sum_{b=1}^B1\{{\sf F}_{n,b} \leq t\}$ and define the simulated $p$-value $p_{n,B} \coloneqq 1-{G}_{n,B}({\sf F}_n).$ The approximation can be made arbitrarily accurate by letting $B \rightarrow \infty$. Thus, for a large value of $B$ and some predefined significance level $\alpha$, we reject the null $\beta = 0$ if $p_{n,B} \leq \alpha$.

\section{Monte Carlo simulation} \label{sec:MC}

In our simulation exercise we simulate from the model given by the three equations \ \eqref{surveyexp}, \eqref{recursion}, and \eqref{gain_mn}. In particular, the survey expectation at time $t$ formed by individuals born in period $s$,  that is, the dependent variable in the nonlinear regression model in \eqref{surveyexp}, is simulated according to  $z_{t,s} = \alpha_t + \beta a_{t,s}(\gamma) + \eps_{t,s}$, with error term $\eps_{t,s} \stackrel{\textsf{IID}}{\sim}  \mathcal{N}(0,1/2)$ and time-specific effect $\alpha_t = \xi_t+y_t/2$, $\xi_t  \stackrel{\textsf{IID}}{\sim} {\sf UNIF}[0,1].$ The recursion for the nonlinear regression function $a_{t,s}(\gamma)$ evolves according to Eq. \eqref{recursion} using \(\{y_t\}\), which, in turn, is generated as an AR(1) process
$y_t = \rho y_{t-1} + v_t,$ $v_t \stackrel{\textsf{IID}}{\sim} \mathcal{N}(0,1-\rho^2),$
for which we consider a mildly ($\rho=0.50$) and a highly ($\rho=0.99$) dependent scenario. We set the learning parameter in Eq. \eqref{gain_mn} to $\gamma \in \{0.8,3.0\}$ and distinguish between the case of identification ($\beta = 0.6$) and non-identification ($\beta = 0$) of the joint parameter vector $\theta$. The case $(\beta,\gamma) = (0.6, 3.0)$ corresponds to our empirical findings.  Simulations with more general specifications, including correlated $\eps$ and additional predetermined regressors, do not yield substantial differences when appropriate standard errors are used and are therefore relegated to the Supplementary Material. 

 Inspired by the empirical application and the analysis in MN, we consider three different sample sizes indexed by $k \in \{2,3,4\}$:
\[
n = k\times 150,\quad u = k\times 75, \quad l = 25.
\]
 Numerical optimisation over $\theta$ is based on the \textsf{optimize} routine of the statistical software \textsf{R} (\citealp{rcovre:21}). More specifically, we obtain \(\gamma_n\), the minimizer of the profiled NLS objective discussed in Section \ref{sec:estimation} on \(\Gamma = [2/3,10]\), which yields $\beta_n = \beta_n(\gamma_n)$.\footnote{The results numerically very close to jointly minimising $Q_n(\theta)$ based on the BFGS algorithm with $(\beta_n,\gamma_n)$ as starting  values. Also, extending \(\Gamma = [2/3,10]\) to \(\Gamma = [1/10,10]\) did not change the results substantially.} Using 1,000 Monte Carlo repetitions, we report the mean and variance of the estimator as well as rejection frequencies of two-sided \(t\)-tests for $H_0:$ \(\gamma = \gamma_0\) and for $H_0:$ \(\beta = \beta_0\) based on asymptotic critical values derived from Corollary \ref{cor:2}. The $t$-statistics, labelled $t_j$, $j \in \{1,2,3,4\}$, are equipped with the four different standard errors discussed in in Section \ref{sec:inf}, two of which require specification of additional nuisance parameters: First, in case of \eqref{eq:H2n}, the estimator of the long-run variance $\omega^2$ is chosen to be the estimator in \cite{newwest:94} with Bartlet kernel and automated bandwidth selection. Second, the numerical derivative in \eqref{eq:H4n} is calculated using a tuning parameter \(\ell_n = \delta_{n}(\gamma+\delta_n)\), where $\delta_{n} = \nu_n^{-2/5}$ in accordance with the requirement $\sqrt{\nu_n}\ell_n \rightarrow \infty$ of Corollary \ref{cor:1}. Note that $\delta_n$ is larger by at least one order of magnitude than step-sizes for numerical derivatives typically encountered in statistical software. We also report the empirical rejection frequencies of the {\it sup}{\sf F} statistic using $p$-values obtained from the Gaussian multiplier bootstrap with $B = 99$ (see Corollary \ref{cor:5}). All test decisions are executed at a nominal significance level of five per cent.

\setlength{\tabcolsep}{4.66pt}
\begin{table}[h]
\begin{center} 
\begin{adjustbox}{max width=\textwidth}
\begin{tabular}{lccccccccccccccccc} \toprule \\[-.55cm]  \hline\\[-.4cm]
&&&&&\multicolumn{4}{c}{$\gamma$} & &  \multicolumn{7}{c}{$\beta$} \\
 \cmidrule(l){5-10} \cmidrule(l){12-18}
$\beta$&$\gamma$&$\rho$&$k$& \sf mean  &   \sf var   &  \(t_1\)  & \(t_2\)  &\(t_3\)&\(t_4\)&&  \sf mean  &   \sf var   &  \(t_1\)  & \(t_2\)  &$t_3$ & $t_4$ & ${\it sup}{\sf F}$\\  \hline
0.6 & 3 & 0.50 &2 &	3.0064	&	0.0608	&	0.0330	&	0.0400	&	0.1900	&	0.0390	&&	0.6010	&	0.0032	&	0.0440	&	0.0590	&	0.3900	&	0.0420	&	1.0000		\\
&&& 3 &	3.0000	&	0.0278	&	0.0480	&	0.0480	&	0.1540	&	0.0530	&&	0.6032	&	0.0013	&	0.0440	&	0.0560	&	0.3560	&	0.0440	&	1.0000		\\
&&& 4 &	3.0028	&	0.0159	&	0.0300	&	0.0440	&	0.1460	&	0.0330	&&	0.6015	&	0.0007	&	0.0440	&	0.0560	&	0.2990	&	0.0440	&	1.0000		\\ 
  \cmidrule(l){3-18}																
 &  & 0.99 &2 &	3.0144	&	0.0661	&	0.0320	&	0.0810	&	0.6860	&	0.0320	&&	0.6024	&	0.0024	&	0.0550	&	0.0980	&	0.7670	&	0.0380	&	1.0000		\\
&&& 3 &	3.0162	&	0.0176	&	0.0430	&	0.0560	&	0.6860	&	0.0430	&&	0.6000	&	0.0006	&	0.0500	&	0.0730	&	0.7440	&	0.0400	&	1.0000		\\
  &&& 4 &	3.0026	&	0.0076	&	0.0500	&	0.0800	&	0.6440	&	0.0470	&&	0.5996	&	0.0002	&	0.0520	&	0.0770	&	0.7130	&	0.0490	&	1.0000		\\	
\cmidrule(l){2-18}		
&0.80& 0.50& 2 &	0.8008	&	0.0021	&	0.0110	&	0.0150	&	0.2840	&	0.0380	&&	0.6007	&	0.0026	&	0.0320	&	0.0440	&	0.3090	&	0.0340	&	1.0000	\\
 && & 3 &	0.8025	&	0.0009	&	0.0160	&	0.0160	&	0.2150	&	0.0440	&&	0.6027	&	0.0011	&	0.0340	&	0.0470	&	0.2700	&	0.0350	&	1.0000		\\
  &&& 4 &	0.8007	&	0.0005	&	0.0150	&	0.0200	&	0.2020	&	0.0410	&&	0.6009	&	0.0007	&	0.0510	&	0.0580	&	0.2640	&	0.0510	&	1.0000		\\	\cmidrule(l){3-18}																	
 &  & 0.99 &2 &	0.8012	&	0.0025	&	0.0140	&	0.0440	&	0.7520	&	0.0350	&&	0.5998	&	0.0011	&	0.0520	&	0.0740	&	0.6400	&	0.0430	&	1.0000		\\
&& & 3 &	0.7988	&	0.0007	&	0.0210	&	0.0430	&	0.7520	&	0.0350	&&	0.5984	&	0.0003	&	0.0520	&	0.0680	&	0.5860	&	0.0480	&	1.0000		\\
  &&& 4 &	0.7996	&	0.0003	&	0.0240	&	0.0490	&	0.7300	&	0.0480	&&	0.5995	&	0.0001	&	0.0440	&	0.0690	&	0.5730	&	0.0360	&	1.0000	 \\
  \hline
  0.0 & 3 & 0.50 &2 &	3.8511	&	12.0863	&	0.3470	&	0.3500	&	0.4310	&	0.3760	&&	-0.0021	&	0.0052	&	0.1130	&	0.1360	&	0.6700	&	0.1180	&	0.0560	\\
&&& 3	&3.8553	&	12.0547	&	0.3330	&	0.3290	&	0.4080	&	0.3550	&&	0.0014	&	0.0021	&	0.1170	&	0.1330	&	0.5950	&	0.1140	&	0.0510	\\
&&& 4	&3.6381	&	11.7419	&	0.3770	&	0.3730	&	0.4540	&	0.3940	&&	0.0015	&	0.0012	&	0.1240	&	0.1330	&	0.5480	&	0.1260	&	0.0570	\\							  \cmidrule(l){3-18}														
 &  & 0.99 &2 &	4.0306	&	12.3713	&	0.1630	&	0.2060	&	0.7540	&	0.2020	&&	0.0001	&	0.0045	&	0.1140	&	0.1950	&	0.9580	&	0.0720	&	0.0490	\\
&&& 3&	4.1486	&	12.9122	&	0.1780	&	0.2090	&	0.7450	&	0.2140	&&	-0.0002	&	0.0012	&	0.1240	&	0.1680	&	0.9600	&	0.0820	&	0.0500	\\
&&& 4&	4.2133	&	12.6796	&	0.1820	&	0.2050	&	0.7380	&	0.2120	&&	-0.0010	&	0.0005	&	0.1150	&	0.1760	&	0.9450	&	0.0910	&	0.0570	\\					\cmidrule(l){2-18}		
 & 0.8 & 0.50 &2 &	3.8511	&	12.0863	&	0.0310	&	0.0400	&	0.0610	&	0.0570	&&	-0.0021	&	0.0052	&	0.1130	&	0.1360	&	0.5000	&	0.1180	&	0.0560	\\
&&& 3	&3.8553	&	12.0547	&	0.0160	&	0.0260	&	0.0240	&	0.0290	&&	0.0014	&	0.0021	&	0.1170	&	0.1330	&	0.4170	&	0.1140	&	0.0510	\\
&&& 4&3.6381	&	11.7419	&	0.0200	&	0.0180	&	0.0180	&	0.0300	&&	0.0015	&	0.0012	&	0.1240	&	0.1330	&	0.3680	&	0.1260	&	0.0570	\\								\cmidrule(l){3-18}		
 &  & 0.99 &2 &	4.0306	&	12.3713	&	0.0200	&	0.0250	&	0.6330	&	0.0110	&&	0.0001	&	0.0045	&	0.1140	&	0.1950	&	0.9180	&	0.0720	&	0.0490	\\
&&& 3&	4.1486	&	12.9122	&	0.0200	&	0.0220	&	0.5920	&	0.0160	&&	-0.0002	&	0.0012	&	0.1240	&	0.1680	&	0.9150	&	0.0820	&	0.0500	\\
&&& 4	&4.2133	&	12.6796	&	0.0150	&	0.0230	&	0.5730	&	0.0170	&&	-0.0010	&	0.0005	&	0.1150	&	0.1760	&	0.9030	&	0.0910	&	0.0570	\\
\bottomrule
\end{tabular}
\end{adjustbox}
\caption{Simulation results based on 1,000 Monte Carlo repetitions. Rejection frequencies of two-sided $t$-tests based on Corollary \ref{cor:2}, where $t_j$ correspond to standard errors based on $H_{j,n}$, $j \in \{1,2,3,4\}$. The {\it sup}{\sf F} statistics are based on the wild multiplier in Corollary \ref{cor:5} using $B=99$.}\label{tab:MCnew}  
\end{center}
\end{table}

In line with Propositions \ref{prop:1} and \ref{prop:2}, Table \ref{tab:MCnew} shows that estimation precision increases with sample size. For $\beta = 0.6$, the empirical size of all $t$-tests but $t_3$ becomes reasonably close to the nominal size of five per cent. As expected, $t_3$ performs very poorly even in large samples. The ``supF'' test, using the Gaussian multiplier bootstrap, appears to consistently reject the alternative $\beta = 0.6$ of the null $\beta = 0$. Next, turn to the scenario under $\beta = 0$, where identification of $\gamma$ breaks down. This is reflected by the poor performance of $\gamma_n$.  However, the small sample evidence suggests that we can still consistently estimate $\beta = 0$, thereby corroborating the theoretical result from Corollary \ref{cor:1}. Moreover, we observe that the $t$-statistics for $\beta = 0$ are oversized because of the non-identified gain $\gamma$ under the null. This problem is solved by the use of ${\sf F}_n$.

\section{Empirical application}\label{sec:emp}

Reassured by our Monte Carlo evidence in the previous section, we now turn to the empirical analysis of the MSC dataset. The model we consider explains surveyed inflation expectations by age-specific inflation forecast that are
learnt from experience, and is given by Eqs.\ \eqref{surveyexp}, \eqref{recursion} and \eqref{gain_mn} in Section \ref{sec:model}, i.e.\
\begin{align}
  z_{t,s} &= \alpha_t + \beta a_{t,s} + \eps_{t,s}, \label{eq:strucmodel}\\
  a_{t,s} &= a_{t-1,s} + \gamma_{t,s} ( y_t - a_{t-1,s} ) ,\label{eq:a}
  \end{align}
  and 
  \begin{align}
  \gamma_{t,s} &= \begin{cases}
                   \displaystyle \frac{\gamma}{t-s} & \text{if } t-s > \gamma \\
                   1 & \text{otherwise.}
                  \end{cases} \label{eq:gamma}
\end{align}
The two observed variables in this model are ($i$) the survey expectation of next period's inflation, as recorded in the MSC, and ($ii$) the U.S.\ consumer price index (CPI). Inflation expectations formed by cohort $s$ in
time period $t$ are derived from the underlying raw data of the MSC. Numerical MSC micro data are available at a monthly frequency from 1978 onwards\footnote{See the MSC website at \url{https://data.sca.isr.umich.edu/}.}. Following
MN, we aggregate these data to quarterly frequency for cohorts that are between 25 and 74 years of age, yielding the cohort-specific inflation expectations $z_{t,s}$ used in Eq.\ \eqref{eq:strucmodel} and displayed in Figure
\ref{fig1}.

\begin{figure}[h]
\centering\medskip\medskip
\pgfplotsset{scaled y ticks=false}
   \begin{tikzpicture}
        \begin{axis}[ymin = -0.03,
                     xmin=0,xmax=183,
                     legend style={draw=none},
                     width=0.9\textwidth,
                     height=.28\textheight,
				     tick label style={font=\scriptsize},
		             label style={font=\small},
		             tick label style={font=\scriptsize},
		             legend style={font=\scriptsize},
			         ylabel style={at={(axis description cs:-0.02,.5)},anchor=north},
                     ytick={-0.02,-0.01,0,0.01,0.02},
                     yticklabels={-0.02,-0.01,0,0.01,0.02},
                     xtick={1,49,89,129,183},
                     xticklabels={1978Q1,1990Q1,2000Q1,2010Q1,2023Q3},
                     ylabel = {\sf survey inflation expectations},
		             x label style={at={(axis description cs:0.5,0)}, anchor=north},
		             xlabel = {\sf quarter},
                     clip=false]
\addplot[blue!70] table[x=X,y=a.6064]{figure1_new.txt}; 
\label{6064}
\addplot[blue!70,dashed] table[x=X,y=a.6569]{figure1_new.txt};
\label{6569}
\addplot[blue!70,dotted] table[x=X,y=a.7074]{figure1_new.txt};
\label{7074}
\addplot[blue!100,line width=2pt] table[x=X,y=Z]{figure1_new.txt};
\label{6074}

\node [draw=none,fill=white] at (rel axis cs: .9,0.15) {\shortstack[l]{
\ref{6064} \scriptsize \,60-64  \\
\ref{6569} \scriptsize \,65-69  \\
\ref{7074} \scriptsize \,70-74 \\
\ref{6074} \scriptsize 60-74  }};

\addplot[red!70] table[x=X,y=a.2529]{figure1_new.txt};
\label{2529}
\addplot[red!70,dashed] table[x=X,y=a.3034]{figure1_new.txt};
\label{3034}
\addplot[red!70,dotted] table[x=X,y=a.3539]{figure1_new.txt};
\label{3539}
\addplot[red,line width=2pt] table[x=X,y=Y]{figure1_new.txt};
\label{2539}

\node [draw=none,fill=white] at (rel axis cs: .9,0.85) {\shortstack[l]{
\ref{2529} \scriptsize \,25-29  \\
\ref{3034} \scriptsize \,30-34  \\
\ref{3539} \scriptsize \,35-39 \\
\ref{2539} \scriptsize 25-39}};


        \end{axis}
   \end{tikzpicture}
\caption{\it Four-quarter moving-averages of inflation expectations by age group relative to the cross-sectional mean using quarterly MSC data from 1978Q1 to 2023Q3.}\label{fig1}
\end{figure} 

The recursively generated quarterly age-specific inflation forecasts $a_{t,s}$ in Eq.\ \eqref{eq:a} are based, for a given value of $\gamma$, on quarterly U.S.\ CPI $y_t$ which, in turn, is derived from the monthly CPI series
originally published\footnote{See Robert Shiller's website at \url{http://www.econ.yale.edu//~shiller/data.htm}.}  by \cite{shiller:00}. The result is a sample of in total 8,800 pairs of quarterly observations $(z_{t,s},a_{t,s})$
between 1978Q1 and 2023Q3.

We also consider two variant datasets. One includes pre-1978 archive data of the MSC. These exist, however, only for some intermittent time periods and are not always in the form of quantitative inflation expectations, yet
\cite{Curtin96} suggests a procedure for making them comparable to post-1978 data. We use the archive data as made available by MN\footnote{See the homepage of Stefan Nagel:
  \url{https://voices.uchicago.edu/stefannagel/files/2021/06/InflExpCode.zip}.}  since they no longer seem to be available for download to the same extent at the MSC\footnote{Only 35 pre-1978 surveys are available for download,
  see the ICPSR website at \url{https://www.icpsr.umich.edu/web/ICPSR/series/54}.}.  Grafting MN's pre-1978 data on the publicly available post-1978 data yields a sample that stretches back to 1953Q4 and comprises 10,615
observation pairs $(z_{t,s}, a_{t,s})$. The second variant dataset, considered for reasons of comparison, is the one used in the analysis of MN, containing 8,215 observations between 1953Q4 and 2009Q4.

The two unknown parameters in model \eqref{eq:strucmodel}--\eqref{eq:gamma}, viz. $ \theta = (\beta,\gamma)^\Tr $ are estimated by NLS, as discussed in Section \ref{sec:estimation}. We use Ox version 9.3, see \cite{Doornik07}, for
the computations. The results for our prime sample from 1978 to 2023 are reported in the first two columns of Table\ \ref{tab:emp}.  The point estimate $\hat\gamma = 3.16$ of the gain parameter is of a similar order of magnitude
as the values found by \cite{MadeiraZafar15}, \cite{mn:2016}, \cite{Gwak22} and \cite{nagel:24}, and the 95\% confidence interval for the true gain effectively covers the competing estimates despite the different model specification:
$[2.74,3.57]$. However, the 95\% confidence interval for $\beta$ is $[0.76,0.91]$, comprising values that are statistically different from the estimates reported in MN, \cite{Gwak22}, and, in particular,
\cite{MadeiraZafar15}. This is evidence indicating that the r\^{o}le of personal experience in forecasting inflation may be higher than has been indicated by the literature so far.

\begin{table}
\begin{center} \small
\begin{tabular}{ r cc c cc c cc} \toprule \\[-.55cm]  \hline \\[-.4cm]
                        & \multicolumn{2}{c}{\it 1978--2023}                                    & & \multicolumn{2}{c}{\it 1953--2023}                    & & \multicolumn{2}{c}{\it 1953--2009}                                    \\ \cmidrule(l){2-3} \cmidrule(l){5-6} \cmidrule(l){8-9}
                        & $\beta$                                   & $\gamma$                  & & $\beta$                   & $\gamma$                  & & $\beta$                                   & $\gamma$                  \\ \hline \\[-5pt]
{\sf estimate}          & 0.8338                                    & 3.1551                    & & 0.6987                    & 2.8969                    & & 0.7311                                    & 2.7472                    \\[-2pt]
{\sf SE}                & {\footnotesize (0.0394) }                 & {\footnotesize (0.2115) } & & {\footnotesize (0.0343) } & {\footnotesize (0.2010) } & & {\footnotesize (0.0402) }                 & {\footnotesize (0.2173) } \\ \cmidrule(l){2-3} \cmidrule(l){5-6} \cmidrule(l){8-9}  
{\sf $\#$ obs}          & \multicolumn{2}{c}{8,800}                                             & & \multicolumn{2}{c}{10,615}                            & & \multicolumn{2}{c}{8,215}                                             \\  
{$R^2$}                 & \multicolumn{2}{c}{0.5612}                                            & & \multicolumn{2}{c}{0.6346}                            & & \multicolumn{2}{c}{0.6373}                                            \\ \hline\\[-12pt]
                        & \multicolumn{2}{c}{$H_0$: $\beta = 0$} 	                            & & \multicolumn{2}{c}{$H_0$: $\beta = 0$}                & & \multicolumn{2}{c}{$H_0$: $\beta = 0$} 	                            \\ \cmidrule(l){2-3} \cmidrule(l){5-6} \cmidrule(l){8-9}
${\it sup}{\sf F}$            & \multicolumn{2}{c}{476.82}                                            & & \multicolumn{2}{c}{425.60}                            & & \multicolumn{2}{c}{370.70}                                            \\[-2pt]
$p$                     & \multicolumn{2}{c}{\footnotesize (0.00) }	                            & & \multicolumn{2}{c}{\footnotesize (0.00) }             & & \multicolumn{2}{c}{\footnotesize (0.00) }                             \\ \hline\\[-.50cm] \bottomrule
\end{tabular}
\caption{NLS estimation results for the sample currently available at MSC (1978Q1--2023Q3), for the current sample extended by MN's preparation of the MSC archive data (1953Q4--2023Q3), and for the sample used by MN
  (1953Q4-2009Q4). Note that the archive data contains missings. Standard errors of the  coefficient estimates and $p$-values of the {\it sup}{\sf F}-statistics, based on Corollaries \ref{cor:2} and \ref{cor:3}, are given in parentheses underneath the estimates and test statistics, respectively.}\label{tab:emp}
\end{center}  
\end{table}

Recall from the discussion in Section \ref{sec:inf} that caution needs to be exercised when testing the null hypothesis $H_0 \colon \beta = 0$. In particular, we argued that the nuisance parameter $\gamma$ is not identified under
the null, implying that the size of the usual Wald test is not controlled, see $\textsf W_n$ in Eq.\ \eqref{eq:Wald} and the Monte Carlo evidence in Section \ref{sec:MC}. Instead, we use the `supF'-test by \cite{hansen:1996}
discussed in Corollary \ref{cor:3} for testing $H_0$, yielding ${\it sup}{\sf F}n = 476.82$. The distribution of {\it sup}{\sf F} is again approximated by the Gaussian multiplier discussed before, based on 99 bootstrap replications,
resulting in a $p$-value of 0.00. This corroborates, for our model, the statement by \citet[p.\ 67]{mn:2016} that ``$\beta$ is significantly different from zero''.

Given the joint asymptotic normality of the NLS estimator that we establish in Proposition \ref{prop:2}, it is now also possible to put the hypothesis of `no recency bias' to a test. As explained in Section \ref{sec:estimation}, this
hypothesis can be parameterised as
\begin{align*}
\begin{array}{rl}
    H_0 \colon &  \gamma \leq 1 \\
    H_1 \colon &  \gamma > 1.
\end{array}    
\end{align*}
Computing the corresponding $t$-statistic yields a one-sided $p$-value of $0.00$. Thus, in our model, there is strong empirical evidence to reject the null in favour of MN's conjecture that economic agents weight more recent
observations more heavily than distant ones when forecasting inflation.

Estimating the model using the extended dataset from 1953 to 2023 yields 95\% confidence intervals for $\beta$ and $\gamma$ comprising parameter values that are considerably lower than for our prime sample, viz.\ $[0.63,0.77]$ and
$[2.49, 3.31]$, respectively. Note that the interval for $\beta$ effectively does not overlap with that based on our prime dataset. The intervals based on the MN dataset are similar. It appears that these results are driven by the
pre-1978 data, yet we leave it to future research to look in detail at issues such as structural change. Importantly, the tests of $H_0 \colon \beta=0$ and $H_0 \colon \gamma \leq 1$ continue to be soundly rejected when the two
variant datasets are used.

Several lessons can thus be learnt from the empirical application: First, using our model to describe post-1978 MSC data, we obtain a confidence interval for the gain parameter $\gamma$, viz.\ $[2.74, 3.57]$, that
comprises most parameter estimates found in the literature. This is re-assuring, given that MN's parameter estimate $\hat \gamma= 3.044$ has become something like a yardstick for calibrating `learning from experience' models.
Secondly, the hypothesis that there is no recency bias is rejected in our model, which lends support to MN's theory that recent experiences weigh more heavily when agents forecast the future. Thirdly, in line with the thrust of
the `learning from experience' literature, our model does not support the conjecture that private experiences do not matter in forecasting inflation. Our empirical evidence indicates that private experiences carry
statistically significantly more weight than has been previously reported in the literature by MN, \cite{Gwak22}, and, in particular, \cite{MadeiraZafar15}. Finally, it appears as if the aforementioned conclusions are
sensitive to the inclusion of pre-1978 MSC archive data. The question of whether this is data issue or a model issue is, however, left to future work.

\section{Concluding remarks}

This paper contributes to the burgeoning literature on analysing the heterogeneity in the expectations formation process. In particular, we establish the econometric theory for NLS estimation and inference in nonlinear panels
with learning from experience.  We show that the  estimator is consistent  and derive its rate of convergence. However,  we  find that asymptotic normality may not be obtained when the number of cohorts is small. If, on the other hand,  the number of cohorts diverges, we prove that the NLS estimator is asymptotically normal, albeit at a nonstandard convergence rate, using  seminal results on extremum estimation with nonsmooth objective functions.  Building on this rigorous econometric foundation, we apply our findings  to an empirical model of the Michigan Survey of Consumers (MSC) data and confirm conjectures made in the learning-from-experience literature on the gain parameter as well as on the contribution of private experiences.

Our analysis can be seen as a starting point for future extensions. One such research avenue would be to consider the econometric theory of more elaborate forms of belief updating  as in the empirical applications of \cite{mn:2016}, \cite{Acedanski17}, \cite{Gwak22}, and \cite{nagel:24}. For example, as mentioned above, \cite{mn:2016} 
assume that agents use a more general PLM for forecasting inflation, in the sense that the information contained in an additional regressor $x_t$ is taken into account. Specifically, an agent born in period \(s\) estimates in each period \(t\) the parameter of a linear regression $\phi$, say, according to the general  stochastic recursive algorithm:
\begin{align*}
 r_{t,s}                    & = r_{t-1,s} + \gamma_{t,s} (x_t x_t^\Tr - r_{t-1,s}) \\
  \phi_{t,s}               & = \phi_{t-1,s} + \gamma_{t,s} r^{-1}_{t,s} x_t (y_t - \phi_{t-1,s}^\Tr x_t), 
\end{align*}
with \(\gamma_{t,s} = \gamma_{t,s}(\gamma_0)\) as in Eq.\ \eqref{gain_mn}.  Similar to ordinary least squares with stochastic regressors, this updating scheme is a Newton-type algorithm that utilizes information on second moments through $r_{t,s}$. Clearly, it is a multivariate generalisation of our setup  in that our learning rule in Eq.\ \eqref{recursion} obtains with \(x_t=1\). Given the recursive estimate of $\phi_{t-1,s}$, the learnt expectation is defined as
$ a_{t,s} \coloneqq \phi_{t-1,s}^\Tr x_t$ so that the data generating process of the dependent variable obtains as $z_{t,s} = \alpha_t + \beta a_{t,s} + \eps_{t,s}$. 
 A technical treatment of the NLS estimator of $\beta$ and $\gamma$ would involve analysing the counterpart of the expressions in  Eq.\ \eqref{amean}, with the crucial difference that the weights  $\kappa_{j,s}$ are now ($i$) stochastic and ($ii$) dependent on the recursion of $r_{t,s}$. The analytical examination of this generalised model is, however, non-trivial. Some preliminary results are contained in the Supplementary Material, where we investigate the small sample behaviour of the NLS estimator empirically and by simulation. Several of the results we established above seem to carry over.

\singlespacing 
\addcontentsline{toc}{section}{References}
\bibliography{bibl}

\newpage

\appendix
\renewcommand{\theequation}{\Alph{section}.\arabic{equation}}
\setcounter{equation}{0}
\section{Proofs}

\subsection{Auxiliary lemmata}

The proofs of the following auxiliary lemmata are delegated to the Supplementary Material (SM). 

\renewcommand{\thelemma}{A.0}
\begin{lemma}\textcolor{white}{.}\label{lem:A0}
\begin{enumerate}
    \item[\textnormal{($i.$)}] $a_{t,t-s}(\gamma) = \sum_{j= \floor{\gamma}}^{s}\kappa_{j,s}(\gamma)y_{t-s+j}$
    \item[\textnormal{($ii.$)}] $\gamma \mapsto a_{t,t-s}(\gamma)$ is continuous  
    \item[\textnormal{($iii.$)}]  For any $\gamma_1,\gamma_2 \in \Gamma$, 
$|a_{t,t-s}(\gamma_1)-a_{t,t-s}(\gamma_2)|\leq c \sum_{i = 1}^si|y_{t-i}| |\gamma_1-\gamma_2|$
\item[\textnormal{($iv.$)}] $\kappa_{j,s}(\gamma) = h_{j,s}(\gamma)\left[1+\frac{1}{j}\frac{\gamma(1-\gamma)}{2}+O\left(\frac{1}{j^2}\right)\right]$, $h_{j,s}(\gamma) \coloneqq \frac{\gamma}{s^\gamma} j^{\gamma-1}$.
\end{enumerate}    
\end{lemma}

The following remark collects a few observations and notational conventions:

\renewcommand{\theremark}{A.0}
\begin{remark}\textcolor{white}{.}
\begin{itemize}
    \item Clearly, the initial values $a_{0,s} = \mathfrak{a}_s$ do not enter the recursion.
    \item  Because $\sum_{j= \floor{\gamma}}^{s}\kappa_{j,s}(\gamma) = 1$, one obtains for the centred recursion
\[
a^\st_{t,t-s}(\gamma) \coloneqq  a_{t,t-s}(\gamma)   -  \Ex[y_1] = \sum_{j= \floor{\gamma}}^{s}\kappa_{j,s}(\gamma)y^\st_{t-s+j}, \quad y_t^\st \coloneqq y_t-\Ex[y_1].
\]
\item Define
 \begin{align}\normalfont\label{rreq}
\frac{\textsf{d}^k}{\textsf{d}\gamma^k}r_{t,t-s}(\gamma) \eqqcolon r_{t,t-s}^{(k)}(\gamma) = \sum_{j = 1}^{s}h^{(k)}_{j,s}(\gamma)y^\st_{t-s+j},
\end{align} 
where $r_{t,s}(\gamma) \coloneqq r^{(0)}_{t,t-s}(\gamma)$, $k \in \{0,1,\dots\}$, and
\[\normalfont
h^{(k)}_{j,s}(\gamma)\coloneqq\frac{\textsf{d}^k}{\textsf{d}\gamma^k} h_{j,s}(\gamma) = h_{j,s}(\gamma)\lln(j/s)^{k-1}(\lln(j/s)+k/\gamma).
\]
Note that $a_{t,s}(1) = r_{t,s}(1)$ as $\kappa_{j,s}(1) = h_{j,s}(1)$. 
\item Because of the cohort demeaning $\tilde a_{t,s}(\gamma) = a_{t,s}(\gamma)-\bar a_t(\gamma) = a^\st_{t,s}(\gamma)-\bar a^\st_t(\gamma)$, to simplify notation, we re-define in what follows $a_{t,s}(\gamma) = a^\st_{t,s}(\gamma)$ and $y_t = y_t^\st$.
\item Moreover, we set $a_{t,s} \coloneqq a_{t,s}(\gamma_0)$ and $r^{(k)}_{t,s} \coloneqq r^{(k)}_{t,s}(\gamma_0)$.
\end{itemize}
 \end{remark}

The following four auxiliary lemmata are derived under the asymptotic regime of Assumption \ref{ass:A2}. Throughout, we set $\nu_n \coloneqq n\lln(n).$

\renewcommand{\thelemma}{A.1}
\begin{lemma} \label{lem:A1}For any  \(\gamma,\gamma_1,\gamma_2 \in  \Gamma\), and any \(k \in \mathbb{N}_+\)
\begin{align}
s\Ex[r^{(k)}_{t,t-s}(\gamma_1)r^{(k)}_{t,t-s}(\gamma_2)] =\omega^2\Upsilon_k(\gamma_1,\gamma_2)(1+o(1)),  \tag{$i$}
\end{align}
with
\[
\Upsilon_k(\gamma_1,\gamma_2) \coloneqq 
k\Gamma(2k-1)\frac{k(1-(\gamma_1-\gamma_2)^2)+(\gamma_1-1)\gamma_1+(\gamma_2-1)\gamma_2}{(\gamma_1+\gamma_2-1)^{2k+1}}, 
\]
if \(k>1\) and \(\Upsilon_0(\gamma_1,\gamma_2)\coloneqq \varphi(\gamma_1,\gamma_2)\) if \(k = 0\). Moreover,
\begin{align}
s\Ex[r^{(1)}_{t,t-s}(\gamma)r_{t,s}(\gamma)] = \omega^2\frac{\gamma(\gamma-1)}{(2\gamma-1)^2}(1+o(1)), \tag{$ii$}
\end{align}
while
\begin{align}
s^2\Ex[(r^{(k)}_{t,t-s}(\gamma))^4] = \,& O(1)\tag{$iii$}.
\end{align}
\end{lemma}
 
\renewcommand{\thelemma}{A.2}
\begin{lemma}\label{lem:A2}
For any \(m,k \in \mathbb{N}_{+}\)
\begin{align}\normalfont
s\Ex[\ssup\limits_{\gamma \in  \Gamma} (r^{(k)}_{t,t-s}(\gamma))^2] = O(1), \tag{$i$}\label{lem:A2:sup}
\end{align}
and
\begin{align}
  \ssup\limits_{\gamma \in \Gamma} \frac1{\nu_n}\bigg\vert \sum_{t = u+1}^n\sum_{s = t-u}^{t-l} r^{(m)}_{t,s}(\gamma)r^{(k)}_{t,s}(\gamma)\bigg\vert= \,&O_p(1), \tag{$ii$} \\
    \ssup\limits_{\gamma \in \Gamma} \frac{m}{\nu_n}\bigg\vert\sum_{t = u+1}^n \bar r^{(m)}_{t}(\gamma) \bar r^{(k)}_{t}(\gamma)\bigg\vert = \,&O_p({\lln}^{-1}(n)). \tag{$iii$} 
\end{align}
\end{lemma}
\renewcommand{\thelemma}{A.3}
\begin{lemma}\label{lem:A3} Uniformly \(\gamma,\gamma_1,\gamma_2 \inn [\ubar{\gamma},\bar{\gamma}]\) and for $k \in \mathbb{N}_+$
\begin{align}
\frac{1}{\nu_n}	 \sum_{t = u+1}^n\sum_{s = t-u}^{t-l} r^{(k)}_{t,s}(\gamma_1)r^{(k)}_{t,s}(\gamma_2)= \omega^2\lambda^2 \Upsilon_k(\gamma_1,\gamma_2) +  \,& o_p(1)\tag{$i$}\label{lemA:a}\\
\frac{1}{\nu_n}	 \sum_{t = u+1}^n\sum_{s = t-u}^{t-l} r^{(1)}_{t,s}(\gamma)r_{t,s}(\gamma)= \omega^2\lambda^2 \frac{\gamma(\gamma-1)}{(2\gamma-1)^2} + \,& o_p(1)\tag{$ii$}
\end{align}
where  \(\Upsilon_k(\gamma_1,\gamma_2)\), has been defined in Lemma \ref{lem:A1}. Moreover,
\begin{align}
 \frac{1}{\sqrt{\nu_n}}\sum_{t = u+1}^n\sum_{s = t-u}^{t-l} r^{(k)}_{t,s}(\gamma)\eps_{t,s}  \Rightarrow \,&\omega\lambda\mathbb{S}(\gamma;k) \tag{$iii$}
\end{align}
where \(\mathbb{S}(\gamma;k)\) is a mean zero Gaussian process on \(\Gamma\) with covariance kernel \(\Upsilon_k(\gamma_1,\gamma_2) = \cov[\mathbb{S}(\gamma_1;k),\mathbb{S}(\gamma_2;k)]\), \(\gamma_1,\gamma_2 \in \Gamma\).
\end{lemma}

\renewcommand{\thelemma}{A.4}
\begin{lemma}\label{lem:A4} Let \(D_n(\theta) = L_{1,n}(\theta)-2L_{2,n}(\theta)\), where 
\begin{align*}
L_{1,n}(\theta) \coloneqq \sum_{t = u+1}^n\sum_{s = t-u}^{t-l}(\tilde g_{t,s}(\theta)-\tilde g_{t,s}(\theta_0))^2,\;\; 
L_{2,n}(\theta) \coloneqq \sum_{t = u+1}^n\sum_{s = t-u}^{t-l} \eps_{t,s}(\tilde g_{t,s}(\theta)-\tilde g_{t,s}(\theta_0)),
\end{align*}
with $\tilde g_{t,s}(\theta) \coloneqq \beta \tilde a_{t,s}(\gamma)$,  and define similarly \( D^\dagger_{n}(\theta) = L^\dagger_{1,n}(\theta)-2 L^\dagger_{2,n}(\theta)\), where
\begin{align*}
L^\dagger_{1,n}(\theta) \coloneqq  \sum_{t = u+1}^n\sum_{s = t-u}^{t-l}(\tilde g^\dagger_{t,s}(\theta)-\tilde g^\dagger_{t,s}(\theta_0))^2, \;\;
L^\dagger_{2,n}(\theta) \coloneqq  \sum_{t = u+1}^n\sum_{s = t-u}^{t-l}\eps_{t,s}(\tilde g^\dagger_{t,s}(\theta)-\tilde g^\dagger_{t,s}(\theta_0)),
\end{align*}
with $\tilde g^\dagger_{t,s}(\theta) \coloneqq \beta \tilde r_{t,s}(\gamma)$. Analogously to $D^\dagger_n$, define $D^\ddagger_n$ with $\tilde g^\dagger_{t,s}(\theta)$ replaced by $g^\dagger_{t,s}(\theta) \coloneqq \beta  r_{t,s}(\gamma)$. Then, for any $\theta \in \Theta$
\[
(i)\; \frac{L^\dagger_{1,n}(\theta)-L_{1,n}(\theta)}{\nu_n} =  O_p\left(\frac{\lVert \theta-\theta_0\rVert^2}{a_n^2}\right),\; (ii)\;\frac{L^\dagger_{2,n}(\theta)-L_{2,n}(\theta)}{\sqrt{\nu_n}} =  O_p\left(\frac{\lVert \theta-\theta_0\rVert}{a_n}\right)
\]
for $a^{-2}_n = {\lln}^3(u){\lln(n)}^{-1}u^{1-2\ubar\gamma}$, and
\[
(iii)\; \frac{L^\dagger_{1,n}(\theta)-L^\ddagger_{1,n}(\theta)}{\nu_n} =  O_p\left(\frac{\lVert \theta-\theta_0\rVert^2}{\lln(n)}\right),\;(iv)\;\frac{L^\dagger_{2,n}(\theta)-L^\ddagger_{2,n}(\theta)}{\sqrt{\nu_n}} =  O_p\left(\frac{\lVert \theta-\theta_0\rVert}{\sqrt{\lln}(n)}\right)
\]
so that
\[\normalfont
\nu_n^{-1}|D_n(\theta)-D^\dagger_n(\theta)| = o_p(1), \quad \nu_n^{-1}|D^\dagger_n(\theta)-D_n^\ddagger(\theta)| = o_p(1), 
\]
uniformly in \(\theta \in \Theta\)
\end{lemma}

\subsection{Proof of the main results}

\textbf{Proof of Proposition \ref{prop:1}.} 
\underline{Part 1.} Set 
\(
L_{1,n}(\theta) \coloneqq \sum_{t = u+1}^n\sum_{s = u}^{l}(\tilde g_{t,t-s}(\theta)-\tilde g_{t,t-s}(\theta_0))^2,
\)
and
\(
L_{2,n}(\theta) 
= \sum_{t = u+1}^n\sum_{s = l}^{u}\eps_{t,t-s}(\tilde g_{t,t-s}(\theta)-\tilde g_{t,t-s}(\theta_0)),
\)
with $\tilde g_{t,t-s}(\theta) \coloneqq \beta \tilde a_{t,t-s}(\gamma)$, so that $D_n(\theta) = L_{1,n}(\theta)-2L_{2,n}(\theta).$ Clearly, $\Ex[L_{2,n}(\theta)] = 0$, while ${\sf D}_m(\theta) = \Ex[L_{1,n}(\theta)/n].$ To this end, we will verify ($a$) a uniform law of large number (ULLN) for $L_{1,n}(\theta)/n$ and ($b$) show that uniformly $n^{-1/2}L_{2,n}(\theta)=O_p(1)$. Begin with ($a$) and note that the claim follows if we can show that, 
\begin{align}\label{eq:ALLN}
\frac{1}{n}	 \sum_{t = u+1}^nA_t(\gamma_1,\gamma_2)= \varphi_m(\gamma_1,\gamma_2)+  o_p(1) \quad \text{uniformly \(\gamma_1,\gamma_2 \inn \Gamma\)}
\end{align}
for $A_{t}(\gamma_1,\gamma_2)\coloneqq \sum_{s=l}^u \tilde a_{t,t-s}(\gamma_1) \tilde a_{t,t-s}(\gamma_2)$ and 
\[
\varphi_m(\gamma_1,\gamma_2) = \sum_{s,k=l}^u\left[1\{s=k\}\left(1-\frac1{m}\right)-\frac{1}{m}\right]\sum_{i=\floor{\gamma_1}}^s\sum_{j=\floor{\gamma_2}}^kc(k-s+i-j) \kappa_{j,s}(\gamma_1)\kappa_{j,k}(\gamma_1),
\]
where we point out that the final expectation $\Ex[A_{t}(\gamma_1,\gamma_2)] =\varphi_m(\gamma_1,\gamma_2)$ is, by stationarity of $A_{t}(\cdot,\cdot)$ for a fixed $u$, independent of the time index $t$. If it holds that $\var[\sum_{t=u+1}^n A_{t}(\gamma_1,\gamma_2)] = o(n^2)$, then, by Chebychev's inequality, Eq. \eqref{eq:ALLN} holds pointwise. To verify this, let \(A_{t}(\gamma_1,\gamma_2)=A_{t,1}(\gamma_1,\gamma_2)-mA_{t,2}(\gamma_1,\gamma_2),\) where $A_{1,t}\coloneqq \sum_{s=l}^u a_{t,t-s}(\gamma_1) a_{t,t-s}(\gamma_2)$ $A_{2,t}\coloneqq \bar a_t(\gamma_1)\bar a_t(\gamma_2)$, so that
\[
\var[\sum_{t=u+1}^n A_{t}(\gamma_1,\gamma_2)]\leq 2(\var[\sum_{t=u+1}^n A_{1,t}(\gamma_1,\gamma_2)]+m^2\var[\sum_{t=u+1}^n A_{2,t}(\gamma_1,\gamma_2)]).
\]
Next,  using $\kappa_{j,s}(\gamma) \leq 1$ and the triangle inequality, one gets
\begin{align}
\var[\sum_{t=u+1}^n A_{t,1}(\gamma_1,\gamma_2)] \leq & \sum_{t,\tau=u+1}^n  |\cov[A_{t,1}(\gamma_1,\gamma_2),A_{\tau,1}(\gamma_1,\gamma_2)]| \nonumber\\
\leq  & \,\sum_{t,\tau=u+1}^n \sum_{s,r=l}^u \nonumber\\
&\quad\times\sum_{i,j=\floor{\gamma_1}}^s\sum_{k,l=\floor{\gamma_2}}^r|\cov[y_{t-s+i}y_{t-s+j},y_{\tau-r+k}y_{\tau-r+l}]|. \label{A:cov}
\end{align}
Note that (see, e.g., \citealp[Eq. (5.1)]{hannan:70})
\begin{equation}\label{cume4}
\begin{split}
\Ex[y_iy_jy_ky_l] = &c(j-i,k-i,l-i) \\
&\;+c(j-i)c(l-k)+c(k-i)c(l-j)+c(l-i)c(k-j)
\end{split}
\end{equation}
and $\cov[y_iy_j,y_ky_l] = \Ex[y_iy_jy_ky_l]-c(j-i)c(l-k)$. Because $u$ is a finite constant under Assumption \ref{ass:A1}, it follows from Assumption \ref{ass:B} that the right-hand side of Eq. \eqref{A:cov} is of order $O(n)$. Similar arguments apply to $\var[\sum_{t=u+1}^n A_{2,t}(\gamma_1,\gamma_2)].$ This proves a pointwise LLN. Following \citet[Lemma 1]{andrews:92}, to establish uniform convergence, we verify a stochastic Lipschitz condition. Because, by Lemma \ref{lem:A0} ($iv$), $|a_{t,t-s}(\gamma_1)-a_{t,t-s}(\gamma_2)| \leq c \sum_{i = 1}^si|y_{t-i}||\gamma_1-\gamma_2|$, $c \in (0,\infty)$, and $\Ex[|y_1|^2]<\infty$, the claim follows. Turning to ($b$), it suffices to show that
\begin{align}
\frac1{\sqrt{n}}\sum_{t=u+1}^nS_t(\gamma) \Rightarrow \sigma \mathbb{S}_m(\gamma), \label{eq:Sweak}
\end{align}
where $S_t(\gamma) = \sum_{s=l}^u\eps_{t,t-s}\tilde a_{t,t-s}(\gamma)$ and $\mathbb{S}_m(\gamma)$ is a Gaussian process with covariance kernel $\varphi_m(\cdot,\cdot)$.  Weak convergence follows from the convergence of the finite-dimensional (`fidi') marginal distributions and stochastic equicontinuity. To show `fidi'-convergence, we resort to the Cramèr-Wold device; i.e, set $\gamma \coloneqq (\gamma_1,\dots,\gamma_G)^\Tr$, where $\gamma_j \in \Gamma$ are distinct, and consider
\[
\frac1{\sqrt n}\sum_{t=u+1}^n\iota^\Tr (S_t(\gamma_1),\dots,S_t(\gamma_G)) = \frac1{\sqrt n}\sum_{t=u+1}^n\sum_{g=1}^G \iota_rS_t(\gamma_g), \quad \iota \in \mathbb{R}^N.
\]
Because $\sum_{g=1}^G \iota_rS_t(\gamma_g)$ is a homoskedastic martingale difference sequence with finite homokurtosis and $
\var[\sum_{g=1}^G \iota_gS_t(\gamma_r)] = \iota^\Tr \{\varphi(\gamma_g,\gamma_f)\}_{1\leq g,f\leq G} \iota > 0,
$ 
with variance-covariance matrix $\{\varphi_m(\gamma_g,\gamma_f)\}_{1\leq g,f\leq G}$ generated by the kernel $\varphi_m(\cdot,\cdot)$.
the CLT for homoskedastic martingale difference sequences yields the desired result. Stochastic equicontinuity follows from the stochastic Lipschitz continuity of $S_t(\gamma)$ and \citet[Theorem 2]{hansen:1996b}. 

\underline{Part 2.}
By Lemma \ref{lem:A4}, it remains to be shown that \(\nu_n^{-1}|D^\ddagger_n(\theta)-\textsf{D}(\theta)|= o_p(1)\) uniformly in \(\theta \in \Theta\), which, in turn, follows by Lemma \ref{lem:A3} recalling the convention $\Upsilon_k = \varphi$ for $k=0$.

This proves the proposition. \hfill$\square$

\textbf{Remark on Example \ref{example}.}  First, using simple algebra, note that $\frac1{\nu_n}\sum_{t=u+1}^n\sum_{s=l}^u\Ex[\tilde a_{t,t-s}^2]$ equals $  \frac1{\nu_n}\sum_{t=u+1}^n\sum_{s=l}^u\Ex[ a_{t,t-s}^2] - \frac{m}{\nu_n} \sum_{t=u+1}^n\Ex[\bar a_t^2]$. Next, by assumption of Example \ref{example}, we get for $\nu_n = n\lln(n)$
\begin{align*}
\frac1{\nu_n}\sum_{t=u+1}^n\sum_{s=l}^u\Ex[\tilde a_{t,t-s}^2]
\overset{(1)}{=} \,& \frac1{n}\sum_{t=u+1}^n\frac1{\lln(n)}\sum_{s=l}^u \frac1{s}\bigg[s\sum_{j=1}^s\kappa_{j,s}^2\bigg] \\
\,& - \frac1{m}\frac1{n}\sum_{t=u+1}^n\frac1{\lln(n)}\sum_{s=l}^u \frac1{s}\bigg[s\sum_{j=1}^s\kappa_{j,s}^2\bigg] \\
\,& - \frac2{m}\frac1{n}\sum_{s=1}^{u-l}\frac1{\lln(n)}\sum_{k=s+l}^u\frac1{k}\bigg[k\sum_{j=1}^{k-s}\kappa_{j+s,k}\kappa_{j,k-s}\bigg] \\
\overset{(2)}{=} \,& \left(1-\frac{u}{n}\right)\bigg[\left(1-\frac1{m}\right)\frac{\psi(u+1)-\psi(l)}{\lln(n)}\\
\,& - 2\left(1-\frac1{m}\right)\frac1{\lln(n)}\\
\,& + 2\frac{l}{m}\frac{\psi(u+1)-\psi(l+1)}{\lln(n)}\bigg]  \overset{(3)}{=}   \left(1-\frac{u}{n}\right)\frac{\lln(u/l)}{\lln(n)} + O\left(\frac1{\lln(n)}\right). 
\end{align*}
Explanations: \textnormal{(1)} is due to $\cov[y_t,y_s] = 1\{t=s\}$ and Eq. \eqref{amean}, \textnormal{(2)} is due to the fact that the expressions in brackets are equal to one, and \textnormal{(3)} uses that $\psi(x) \sim \lln(x)+1/x$ as $x \rightarrow \infty$ for the digamma-function $\psi(\cdot)$.

\textbf{Proof of Corollary \ref{cor:0}.} 
\underline{Part (1): $\beta_0 \neq 0$.} We just consider the case of the asymptotic regime of Assumption \ref{ass:A1}. The conclusion under Assumption \ref{ass:A2} follows from Proposition \ref{prop:2}. As argued already in the proof of Proposition \ref{prop:1}, from Lemma \ref{lem:A0} ($iv.$) it follows, for a fixed $u$, that $A_t(\gamma) = \sum_{s=l}^u\tilde a_{t,t-s}(\gamma)$ satisfies a stochastic Lipschitz condition such that $|A_t(\gamma_1)-A_t(\gamma_2)| \leq \dot A_t|\gamma_1-\gamma_2|$, with $\Ex[\dot A_t^2] < \infty$. Moreover, as shown in the Supplementary Material $-{\sf D}_m(\theta) \leq -c \Vert \theta-\theta_0 \Vert^2 $, $c \in (0,\infty)$. The claim is then due to second-order stationarity of $A_t$ for a fixed $u$ in conjunction with \citet[Theorem 3.2.5]{van:1996} and \citet[Corollary 5.53]{van:2000}.

\underline{Part (2): $\beta_0 = 0$}. Adapting the discussion in \citet[Section 5.3]{saikkonen:95}, let $\tilde\eps_{1,t,t-s}(\theta) \coloneqq -(\beta-\beta_0)\tilde a_{t,t-s}(\gamma)$, $\tilde\eps_{2,t,t-s}(\theta) \coloneqq \tilde\eps_{t,t-s}-\beta_0(\tilde a_{t,t-s}(\gamma)-\tilde a_{t,t-s})$,
to decompose $Q_n(\theta) = Q_{1,n}(\theta)+Q_{2,n}(\gamma)$ for
\[
Q_{1,n}(\theta) \coloneqq 2\sum_{t = u+1}^n\sum_{s = u}^{l} \tilde\eps_{1,t,t-s}(\theta)\tilde\eps_{2,t,t-s}(\gamma)+\sum_{t = u+1}^n\sum_{s = u}^{l} \tilde\eps^2_{1,t,t-s}(\theta),
\]
and
$Q_{2,n}(\gamma) \coloneqq \sum_{t = u+1}^n\sum_{s = u}^{l} \tilde\eps^2_{2,t,t-s}(\gamma).$
Observe that $Q_{1,n}(\beta_0,\gamma) = 0$, $Q_n(\beta_0,\gamma) = Q_{2,n}(\gamma)$, and, therefore, also $Q_n(\theta_0) = Q_{2,n}(\gamma_0).$ Hence, $D_n(\theta) = Q_{1,n}(\theta)+Q_{2,n}(\gamma)-Q_{2,n}(\gamma_0)$. Thus, for $\beta_0 = 0$, $Q_{2,n}(\gamma)$ is independent of $\gamma$ and equals $Q_{2,n}(\gamma_0)$. Evidently, for any value of $\gamma$, the unique minimum of $\beta \mapsto D_n(\beta,\gamma) = Q_{1,n}(\beta,\gamma)$ is reached at $\beta = 0$. If follows that $\sqrt{\nu_n}\theta_{\beta,n}=O_p(1)$ if, for any  $\delta > 0$ and $r_n \rightarrow \infty$ with $r_n = o(\sqrt{\nu_n})$, $D_n(\theta)$ is uniformly bounded away from zero on $\Theta_n \coloneqq \bar B_{n,\delta} \times \Gamma$, with $\bar B_{n,\delta} \coloneqq \{\beta \in \Xi: |\beta| > \delta/r_n\}$. Similar to \citet[Proof of Theorem 1]{seo:11}, to see that this is the case, we note that $\iinf\limits_{\theta \in \Theta_n}D_n(\theta) > 0 \Leftrightarrow \iinf\limits_{\theta \in \Theta_n}Q_{1,n}(\theta)/\eta_n > 0$
for $\eta_n \coloneqq\sqrt{n}|\beta|$,
\begin{align*}
  \frac1{\eta_n}Q_{1,n}(\beta,\gamma) 
  = \,& \eta_n \left[\frac1{\nu_n}\sum_{t = u+1}^n\sum_{s = u}^{l} \tilde a^2_{t,t-s}(\gamma)\right] - \frac{2\textsf{sign}(\beta)}{\sqrt\nu_n}\sum_{t = u+1}^n\sum_{s = u}^{l} \tilde a_{t,t-s}(\gamma)\eps_{t,t-s}.
\end{align*}
Since ($a$) $\eta_n \rightarrow \infty$ for any $\beta \in \bar B_{n,\delta}$, ($b$) the first term in square brackets on the right-hand side converges uniformly to a positive function (by Eq. \eqref{eq:ALLN} if Assumption \ref{ass:A1} holds, and by Lemma \ref{lem:A3} and \ref{lem:A4} if Assumption \ref{ass:A2} holds) and ($c$) the second term is stochastically bounded uniformly in $\theta \in \Theta$ (by Assumption \ref{ass:C} $|\beta| \leq |\ubar\beta| \vee \bar\beta < \infty$ and Eq. \eqref{eq:Sweak} if Assumption \ref{ass:A1} holds, and by Lemma \ref{lem:A3} and \ref{lem:A4} if Assumption \ref{ass:A2} holds), $Q_n(\theta)/\eta_n \rightarrow \infty$ uniformly on $\bar B_{n,\delta} \times \Gamma$. The latter follows from the proof of Proposition \ref{prop:1}.

This proves the corollary.  \hfill$\square$

\textbf{Proof of Proposition \ref{prop:2}.} 
The proof is an application of \citet[Thm. 7.1]{newmc:94}: Because $\theta_0$ minimizes $\theta \mapsto \textsf{D}(\theta)$, $\theta_0 \in \textsf{int}(\Theta)$, and 
\[
\frac{\partial^2}{\partial\theta\partial\theta^\Tr}\textsf{D}(\theta) \coloneqq 2(\omega\lambda)^2\varphi(\gamma,\gamma) \begin{bmatrix} 1 & \displaystyle\frac{\beta}{\gamma}\frac{\gamma-1}{(2\gamma-1)}\\   \displaystyle\frac{\beta}{\gamma}\frac{\gamma-1}{(2\gamma-1)} & \displaystyle\frac{\beta^2}{\gamma}\frac{1/\gamma+2(\gamma-1)}{(2\gamma-1)^2}\end{bmatrix}
\]
is continuous, viewed as a function of $\theta$, and positive definite when evaluated at $\theta_0$, their conditions ($i$), ($ii$), ($iii$) are satisfied. Next, their final two conditions ($iv$) and ($v$) require the existence of some $2 \times 1$ random vector $C_n \rightarrow_d \mathcal{N}_2(0,\Omega)$, with $\Omega$ positive definite, such that the following \textit{stochastic differentiablity} condition holds: 
\[
\ssup\limits_{\lVert \theta-\theta_0 \rVert \leq \delta_n} \left\vert\frac{R_n(\theta)}{1+\sqrt{\nu_n}\lVert \theta-\theta_0 \rVert}  \right\vert = o_p(1), \quad R_n(\theta) \coloneqq \sqrt{\nu_n}\frac{\nu_n^{-1}D_n(\theta) - \textsf{D}(\theta) - (\theta-\theta_0)^\Tr C_n}{\lVert \theta-\theta_0 \rVert}
\] 
for any sequence of constants $\delta_n \rightarrow 0$. To this end, set
\[
C_n(\theta) \coloneqq \frac1{\sqrt \nu_n} \frac{\partial }{\partial \theta} D^{\ddagger}_n(\theta)
\]
and note that by the the Cram\`{e}r Wold device and part ($iii$) of Lemma \ref{lem:A3},  
\[
C_n \coloneqq C_n(\theta_0) = -\frac{2}{\sqrt{\nu_n}}\sum_{t = u+1}^n\sum_{s = t-u}^{t-l}(r_{t,s},\beta_0 r^{(1)}_{t,s})^\Tr\eps_{t,s} \rightarrow_d \mathcal{N}_2\left(0,2\frac{\partial^2}{\partial\theta\partial\theta^\Tr}\textsf{D}(\theta_0)\right). 
\]
 It remains to be shown that the remainder term $R_n(\theta)$ is stochastically differentiable. Note that $R_n(\theta) = R_{1,n}(\theta)+R_{2,n}(\theta)$, with
\[
R_{1,n}(\theta) = \sqrt{\nu_n}\frac{\nu_n^{-1}D^\ddagger_n(\theta)- \textsf{D}(\theta) - C_n(\theta-\theta_0)}{\lVert \theta-\theta_0 \rVert},\quad R_{2,n}(\theta) = \sqrt{\nu_n}\frac{\nu_n^{-1}(D_n(\theta)- D^\ddagger_n(\theta))}{\lVert \theta-\theta_0 \rVert},
\]
where $D^\ddagger_n(\theta)$ has been defined in Lemma \ref{lem:A4}. Begin with $R_{1,n}$ and notice that
\[
\nu_n^{-1}(D^\ddagger_n(\theta)- \textsf{D}(\theta)) = C_n(\theta-\theta_0) + \frac{1}{2}(\theta-\theta_0)^\Tr W_n(\bar\theta)(\theta-\theta_0),
\]
with
\[
 W_n(\theta)\coloneqq \nu_n^{-1}\left(\frac{\partial^2}{\partial\theta\partial\theta^\Tr} D_n^\ddagger(\theta)-\frac{\partial^2}{\partial\theta\partial\theta^\Tr}\textsf{D}(\theta)\right)
\]
for some  $\bar\theta$ on the line segment connecting $\theta$ and $\theta_0$. Because, by Lemma \ref{lem:A3} ($i$), $\lvert W_n(\bar\theta) \rvert = o_p(1)$ for any $\lvert\theta-\theta_0\rvert = o(1)$ it follows that $R_{1,n}(\theta) = o_p(\sqrt{\nu_n}\lVert \theta-\theta_0 \rVert)$ so that
\[
\ssup\limits_{\lVert \theta-\theta_0 \rVert \leq \delta_n} \left\vert\frac{R_{1,n}(\theta)}{1+\sqrt{\nu_n}\lVert \theta-\theta_0 \rVert}  \right\vert =  o_p(1),
\]
using that $ab/(1+ab) \leq a/b$ for any positive constants $a$ and $b$.
Moreover, by construction of $D_n(\cdot)$ and $D^\ddagger_n(\cdot)$ defined in Lemma \ref{lem:A4}, one obtains
\[
\nu_n^{-1}(D_n(\theta)-D^\ddagger_n(\theta)) = \nu_n^{-1}(L_{1,n}(\theta)-L^\ddagger_{1,n}(\theta))-2\nu_n^{-1}(L_{2,n}(\theta)-L_{2,n}^\ddagger(\theta)),
\]
where, as shown in Lemma \ref{lem:A4},
\[
\nu_n^{-1}(L_{1,n}(\theta)-L_{1,n}^\ddagger(\theta)) =  o_p(\lVert\theta-\theta_0\rVert^2), \quad \nu_n^{-1/2}(L_{2,n}(\theta)-L_{2,n}^\ddagger(\theta)) =  o_p(\lVert\theta-\theta_0\rVert).
\]
Hence, by the triangle inequality
\begin{align}
\ssup\limits_{\lVert \theta-\theta_0 \rVert \leq \delta_n} \left\vert\frac{R_{2,n}(\theta)}{1+\sqrt{\nu_n}\lVert \theta-\theta_0 \rVert}  \right\vert \leq \,&
\ssup\limits_{\lVert \theta-\theta_0 \rVert \leq \delta_n} \left\vert\frac{\nu^{-1/2}_n(L_{1,n}(\theta)-L^\ddagger_{1,n}(\theta))}{\lVert \theta-\theta_0 \rVert(1+\sqrt{\nu_n}\lVert \theta-\theta_0 \rVert)}  \right\vert \nonumber \\
 \,&\quad + 2\ssup\limits_{\lVert \theta-\theta_0 \rVert \leq \delta_n}\left\vert\frac{\nu_n^{-1/2}(L_{2,n}(\theta)-L^\ddagger_{2,n}(\theta))}{\lVert \theta-\theta_0 \rVert(1+\sqrt{\nu_n}\lVert \theta-\theta_0 \rVert)}  \right\vert. 
\end{align}
But
\[
\ssup\limits_{\lVert \theta-\theta_0 \rVert \leq \delta_n} \left\vert\frac{\nu^{-1/2}(L_{1,n}(\theta)-L^\ddagger_{1,n}(\theta))}{\lVert \theta-\theta_0 \rVert(1+\sqrt{\nu_n}\lVert \theta-\theta_0 \rVert)}  \right\vert \leq \ssup\limits_{\lVert \theta-\theta_0 \rVert \leq \delta_n} \left\vert\frac{\nu^{-1}(L_{1,n}(\theta)-L^\ddagger_{1,n}(\theta))}{\lVert \theta-\theta_0 \rVert^2}  \right\vert = o_p(1)
\]
and
\[
\ssup\limits_{\lVert \theta-\theta_0 \rVert \leq \delta_n}\left\vert\frac{\nu_n^{-1/2}(L_{2,n}(\theta)-L^\ddagger_{2,n}(\theta))}{\lVert \theta-\theta_0 \rVert(1+\sqrt{\nu_n}\lVert \theta-\theta_0 \rVert)}  \right\vert \leq \ssup\limits_{\lVert \theta-\theta_0 \rVert \leq \delta_n}\left\vert\frac{\nu_n^{-1/2}(L_{2,n}(\theta)-L^\ddagger_{2,n}(\theta))}{\lVert \theta-\theta_0 \rVert}  \right\vert = o_p(1),
\]
using that $ca/(b(1+ca)) \leq ca/b$ for any positive $a$, $b$, and $c$. This proves the proposition. \hfill$\square$


\textbf{Proof of Corollary \ref{cor:1}.}
Next, turn to the general case. First, let us show that $\left\Vert \Ex[H_n(\theta_0)]-(\omega\lambda)^2{\sf H}\right\Vert = O({\lln}^{-1}(n))$. Too see this, consider
\[
\frac1{\nu_n}\sum_{t=u+1}^n\sum_{s=l}^u\Ex[\tilde r^{(h)}_{t,t-s}\tilde r^{(k)}_{t,t-s}] = \frac1{\nu_n}\sum_{t=u+1}^n\sum_{s=l}^u\Ex[ r^{(h)}_{t,t-s} r^{(k)}_{t,t-s}]-\frac m{\nu_n}\sum_{t=u+1}^n\Ex[\bar r^{(h)}_{t} \bar r^{(k)}_{t}]
\]
for $h,k \in \{0,1\}$, which, up a scaling by $\beta_0$, constitute the elements of $\Ex[H_n(\theta_0)]$. By Lemma \ref{lem:A2}, the second term on the right-hand side is $O({\lln}^{-1}(n))$, while the first summand simplifies by Lemma \ref{lem:A1}.
\[
\omega^2\left[\gamma 1\{h \neq k\}\frac{\gamma-1}{(2\gamma-1)^2}+1\{h = k\}\Upsilon_k(\gamma_0,\gamma_0)\right] \left[\frac{1-u/n}{\lln(n)}\sum_{s=l}^u\frac1{s}(1+o(1))\right]. 
\]
Inspection of ${\sf H}$ in Proposition \ref{prop:2} yields the claim.

\underline{Part ($i$).} By the mean-value theorem, $\Vert H_n(\theta_0)-H_{1,n} \Vert \leq \frac1{\sqrt{\nu_n}}(\sqrt{\nu_n}\Vert \theta_n-\theta_0 \Vert) \Vert \frac{\partial}{\partial \theta} H_{n}(\bar\theta_n) \Vert$ for some $\bar\theta_n$ between $\theta_0$ and $\theta_n$. Because, by Lemma \ref{lem:A2}, $\ssup_\theta\Vert \frac{\partial}{\partial\theta} H_n(\theta) \Vert = O_p(1)$, it follows from Proposition \ref{prop:2}, $\sqrt{\nu_n}\Vert H_n(\theta_0)-H_{1,n} \Vert = O_p(1)$. 

\underline{Part ($ii$).} Let $H_{2,n}^c$ denote $H_{2,n}$ in Eq. \eqref{eq:H2n} with the true autocovariance function $c(\cdot)$ replacing $c_n(\cdot)$. By the triangle inequality,
$\left\Vert \Ex[H_n(\theta_0)]-H_{2,n}\right\Vert \leq \Vert H_{2,n}^c-H_{2,n}\Vert+\Vert \Ex[H_n(\theta_0)]-H_{2,n}^c\Vert$. It will be shown that both summands on the right-hand side are $o_p(1)$. To begin with, another application of Cauchy-Schwarz yields
\[
\left\Vert H_{2,n}^c-H_{2,n}\right\Vert \leq  \sqrt{\frac{m^2}{{\lln}^{2}(n)n}} \left(\mmax\limits_{1 \leq  \tau \leq u}    \sqrt{n}|c_n(\tau)-c(\tau)| \right)\left[\frac1{m}\sum_{s=l}^u\sum_{i=1}^s\Vert e_{i,s}(\theta_n)\Vert\right]^2,
\]
which is $o(1)$ as explained as follows: By assumption of the corollary, the first scaling on the right-hand side of the inequality is $o(1)$, while the scaled maximum divergence of autocovariances in parantheses is $O_p(1)$. Next, recall that $e_{i,s}(\theta) = (h_{i,s},\beta \dot h_{i,s}(\gamma))^\Tr$, where $h_{i,s}(\gamma) \coloneqq \gamma/s^\gamma i^{\gamma-1}$, $\dot h_{i,s}(\gamma) = h_{i,s}(\gamma)(\lln(i/s)+1/\gamma)$. As $|h_{i,s}(\gamma)| \leq 1$, $|\dot h_{i,s}(\gamma)| \leq 1$, $\Vert e_{i,s}(\theta)\Vert^2 \leq 1+\bar\beta^2$ uniformly in $\theta$ and for any $1 \leq i \leq s$. Hence, by Proposition \ref{prop:1}, the dominated convergence theorem in conjunction with $\int_i^s |h_{i,s}(\gamma)| {\sf d}i  = 1+o(1)$ and $\int_i^s |\dot h_{i,s}(\gamma) | {\sf d}i  = 2/(\gamma_0 e)+o(1)$, it follows that the term in square brackets is $O(1)$. Next, define the $2\times 2$ Jacobian matrix of $e_{i,s}(\theta)$,
\[
\nabla e_{i,s}(\theta) \coloneqq \begin{bmatrix}
    0 & h_{i,s}(\gamma) \\
    \dot h_{i,s}(\gamma) & \beta \ddot h_{i,s}(\gamma)
\end{bmatrix},
\]
where $\ddot h_{i,s}(\gamma) = h_{i,s}(\gamma)\lln(i/s)(\lln(i/s)+2/\gamma)$ with $|\ddot h_{i,s}(\gamma)| \leq 1$ and $\int_i^s |\ddot h_{i,s}(\gamma) | {\sf d}i  = 8/(\gamma_0 e)^2+o(1)$. Hence $\Vert \nabla e_{i,s}(\theta) \Vert \leq K$, $K \in (0,\infty)$. Then, by the mean-value theorem and $\mmax_{\tau \leq u}|c(\tau)| < \infty$, we get for some $C \in (0,\infty)$ and $\bar\theta_n$ between $\theta_0$ and $\theta_n$
\begin{align*}
\Vert &\Ex[H_n(\theta_0)]-H_{2,n}^c\Vert \\
    \,& \leq  \frac{Cm}{\lln(n)\sqrt{\nu_n}} \bigg(\left[ \frac1{m}\sum_{s=l}^u\sum_{i=1}^s\Vert e_{i,s}(\bar\theta_n)\Vert \right] \left[ \frac1{m}\sum_{s=l}^u\sum_{i=1}^s\Vert \nabla e_{i,s}(\bar\theta_n)\Vert \right] \sqrt{\nu_n}\Vert \theta_n-\theta_0 \Vert \\
     \,& \qquad + \frac{1}{\sqrt{\nu_n}} \left[ \frac1{m}\sum_{s=l}^u\sum_{i=1}^s\Vert \nabla e_{i,s}(\bar\theta_n)\Vert \right] \left[ \frac1{m}\sum_{s=l}^u\sum_{i=1}^s\Vert \nabla e_{i,s}(\bar\theta_n)\Vert \right] (\sqrt{\nu_n}\Vert \theta_n-\theta_0 \Vert)^2\bigg),
\end{align*}
which is, invoking similar arguments as before, $o(1)$.

\underline{Part ($iii$).} Follows directly from the consistency of $\omega_n^2$ and Proposition \ref{prop:2}.

\underline{Part ($iv$).} The consistency of the numerical Hessian is due to \citet[Thm. 7.4]{newmc:94}. 

This proves the claim. \hfill$\square$

\textbf{Proof of Corollary \ref{cor:2}.}
First, let us show that $s_n^2  = \sigma^2+o_p(1)$. Because, by the LLN for martingale difference arrays (see, e.g., \citealp[Thm. 19.7]{davidson:94}), one gets
\[
\frac{Q_n(\theta_0)}{(n-u)m}-\sigma^2= \frac{1}{(n-u)m}\sum_{t=u+1}^n\sum_{s=t-u}^{t-l}(\eps^2_{t,s}-\sigma^2) = o_p(1),
\]
consistency of $s_n^2$ then follows from Proposition \ref{prop:1}. 
The proof follows now by standard arguments using Corollary \ref{cor:1} in conjunction with Proposition \ref{prop:2}. \hfill$\square$

\textbf{Proof of Corollary \ref{cor:3}.} 
We show only part (1) as part (2) follows analogously. Some algebra reveals that under a sequence of local alternatives $\beta_{0,n} = \Delta_\beta/\sqrt{\nu_n}$, $\nu_n = n$,
\begin{align*}
\frac{N(\tilde\sigma_n^2-\sigma^2_n(\gamma))}{\sigma^2_n(\gamma)} = &\frac{\sum_{t=u+1}^n (\beta_{0,n}A_t(\gamma_0,\gamma) + S_t(\gamma))^2}{\sum_{t=u+1}^n A_t(\gamma,\gamma)} \Big/ \\
 &\quad\frac1{N}\sum_{t=u+1}^n\sum_{s=l}^u\left[\tilde \eps_{t,s}+\beta_{0,n}\tilde a_{t,s}-\frac{\sum_t (\beta_{0,n}A_t(\gamma_0,\gamma)+ S_t(\gamma))}{\sum_t A_t(\gamma,\gamma)}\tilde a_{t,s}(\gamma)\right]^2.
\end{align*}
Thus, by Eqs. \eqref{eq:ALLN}, \eqref{eq:Sweak}, and Slutzky's theorem, it follows for the numerator
\begin{align*}
\frac{\sum_{t=u+1}^n (\beta_{0,n}A_t(\gamma_0,\gamma) + S_t(\gamma))^2}{\sum_{t=u+1}^n A_t(\gamma,\gamma)} \,& = \frac{(\frac{\Delta_\beta}{\nu_n} \sum_{t=u+1}^n A_t(\gamma_0,\gamma) + \frac1{\sqrt{\nu_n}}\sum_{t=u+1}^nS_t(\gamma))^2}{\frac1{\nu_n}\sum_{t=u+1}^n A_t(\gamma,\gamma)} \\
\,& \Rightarrow \frac{(\varphi_m(\gamma_0,\gamma)+\sigma {\mathbb S}_m(\gamma))^2}{\varphi_m(\gamma,\gamma)},
\end{align*}
so that the denominator converges to $\sigma^2$ in probability uniformly in $\gamma$. Application of the continuous mapping theorem completes the proof.  \hfill$\square$

{\bf Proof of Corollary \ref{cor:5}.} Again we focus for brevity on the asymptotic regime under Assumption \ref{ass:A1}. It follows readily, that under the null $\frac1{\sqrt \nu_n}\sum_{t=u+1}^n S_{t,b}(\gamma) \Rightarrow \mathbb{S}_m(\gamma)$, with $S_{t,b}(\gamma) \coloneqq \sum_{s=l}^{u}\tilde a_{t,t-s}(\gamma)z_{t,t-s,b}$, in probability conditionally on the original sample $\mathcal S_n$. By Eq. \eqref{eq:ALLN} in conjunction with Slutzky's theorem and continuous mapping theorem we conclude $p_n \rightarrow_d 1-{\sf G}(T)$. \hfill$\square$






\newpage

\begin{appendices}

\setcounter{page}{1}
    

\renewcommand{\theequation}{S.\arabic{equation}}
\renewcommand{\thesection}{S.\arabic{section}}
\renewcommand{\thetable}{S.\arabic{table}}
\setcounter{equation}{0}
\setcounter{section}{0}
\setcounter{table}{0}

\section{Proofs of auxiliary results}

\textbf{Proof of Lemma \ref{lem:A0}.} 
\underline{Part ($i$)}. Observing that, for all individuals of age $t-s>0$, one gets for any $j<t$: \begin{align}\label{a_rec_deriv0}
a_{t-j,t-s}(\gamma) = a_{t-j-1,t-s}(\gamma)(1-\gamma_{t-j,t-s}(\gamma))+\gamma_{t-j,t-s}(\gamma)y_{t-j},
\end{align}
where $\gamma_{t-j,t-s}=\gamma/(s-j)$ if $s-j>\gamma$ and $\gamma_{t-j,t-s}=1$ otherwise. Thus, by Eq.\ \eqref{a_rec_deriv0}, we have $a_{t-j,t-s}(\gamma)= y_{t-j}$ if $j \geq s-\floor{\gamma}>0$. Hence,  
\begin{align*}
    a_{t,t-s}(\gamma) = \,& a_{t-1,t-s}(\gamma)(1-\gamma_{t,t-s}(\gamma))+\gamma_{t,t-s}(\gamma)y_{t} \\
    = \,& a_{t-2,t-s}(\gamma)(1-\gamma_{t,t-s}(\gamma))(1-\gamma_{t-1,t-s}(\gamma)) \\
    \,& \qquad\qquad\qquad  + (1-\gamma_{t,t-s}(\gamma))\gamma_{t-1,t-s}(\gamma)y_{t-1}+\gamma_{t,t-s}y_t \\
    = \,& a_{t-j,t-s}(\gamma) \prod_{i=0}^{j-1}\left(1-\frac{\gamma}{s-i}\right)\\
    \,&\qquad\qquad\qquad + \sum_{i=1}^{j-1}\frac{\gamma}{s-j+i}\prod_{l=0}^{j-i-1}\left(1-\frac{\gamma}{s-l}\right)y_{t-j+i} + \frac{\gamma}{s}y_t \\
       \stackrel{(a)}{=} \,& a_{t-s+\floor{\gamma},t-s}(\gamma) \prod_{i=0}^{s-\floor{\gamma}-1}\left(1-\frac{\gamma}{s-i}\right)\\
    \,&\qquad\qquad\qquad + \sum_{i=1}^{s-\floor{\gamma}-1}\frac{\gamma}{\floor{\gamma}+i}\prod_{l=0}^{s-\floor{\gamma}-i-1}\left(1-\frac{\gamma}{s-l}\right)y_{t-s+\floor{\gamma}+i} + \frac{\gamma}{s}y_t \\
    \stackrel{(b)}{=} \,& y_{t-s+\floor{\gamma}}\prod_{i=\floor{\gamma}+1}^{s}\left(1-\frac{\gamma}{i}\right)\\
    \,&\qquad\qquad\qquad + \sum_{j=\floor{\gamma}+1}^{s-1}\frac{\gamma}{j}\prod_{i=j+1}^{s}\left(1-\frac{\gamma}{i}\right)y_{t-s+j} + \frac{\gamma}{s}y_t \\
    \stackrel{(c)}{=} \,& \sum_{j= \floor{\gamma}}^{s}\kappa_{j,s}(\gamma)y_{t-s+j}, \quad s \in \{l,l+1,\dots,u-1,u\},
\end{align*}
where the last three equations obtain as follows: Eq.\ ($a$) obtains by recursively applying Eq.\ \eqref{a_rec_deriv0} up to $j = s-\floor{\gamma}$; Eq.\ ($b$) uses $a_{t-j,t-s}(\gamma)= y_{t-j}$ for $j = s-\floor{\gamma}$; Eq.\ ($c$) re-arranges terms using the definition of $\kappa_{j,s}(\gamma)$ as introduced in Eq.\ \eqref{amean}.
 
\underline{Part ($ii$).} Although not immediately obvious, the updating scheme Eq. \eqref{amean}, viewed as a function of $\gamma$, is continuous $a.s.$. To see this, introduce for the sake of the argument the weights
\[
\bar{\kappa}_{j,s}(\gamma)\coloneqq \begin{cases}
\displaystyle \prod_{i = 1}^{s}\left(1-\frac{\gamma}{i}\right)  & \text{ if }\,j = 0\\[3ex] 
 \displaystyle\frac{\gamma}{s} & \,\text{ if }\, j=s\\[2ex]
\displaystyle\frac{\gamma}{j}\prod_{i = j+1}^{s}\left(1-\frac{\gamma}{i}\right)  & \text{ otherwise},
 \end{cases}
\]
and note that that \(\bar\kappa_{j,s}(\gamma) = 0\) on \(j<\gamma,\) \(\gamma \in \Gamma \cap \mathbb{Z}\). Hence, we can express $a_{t,t-s}(\gamma)$ equivalently as\footnote{Under the alternative gain sequence
\[
 \gamma_{t,s}(\gamma) = \begin{cases}
\displaystyle\frac{\gamma}{t-s}  & \text{ if }\,t-s > 0\\
 1 & \text{ otherwise}, \end{cases}
 \]
 the forecast of an individual is $a^0_{t,t-s}(\gamma) \coloneqq \sum_{j = 0}^{s}\bar\kappa_{j,s}(\gamma)y_{t-s+j}$, thereby coinciding with the first term on the right hand side of Eq.\ \eqref{eq:b_recursion}.}
\begin{align}\label{eq:b_recursion}
a_{t,t-s}(\gamma) = \sum_{j = 0}^{s}\bar\kappa_{j,s}(\gamma)y_{t-s+j}+b_{t,t-s}(\gamma),
\end{align}
 where
\begin{align}
b_{t,t-s}(\gamma) \coloneqq y_{t-s+ \floor{\gamma}}\prod_{i = \floor{\gamma}+1}^{s}\left(1-\frac{\gamma}{i}\right) - \sum_{j = 0}^{ \floor{\gamma}}\bar\kappa_{j,s}(\gamma)y_{t-s+j}.
 \end{align}
Since $a_{t,t-s}(\gamma)$ is continuous $a.s.$ on $\Gamma \setminus \mathbb{Z}$, it suffices to show that, conditionally on $\{y_t\}_t$, it holds
\[
\llim\limits_{\gamma \rightarrow \gamma_0^{-}} a_{t,t-s}(\gamma) = a_{t,t-s}(\gamma_0), \quad  \forall\gamma_0 \in \Gamma \cap \mathbb{Z}.
\]
First, we note that $\sum_{j = 0}^{s}\bar\kappa_{j,s}(\gamma)y_{t-s+j}$ is continuous in $\gamma$ $a.s.$. Next, because $\bar\kappa_{j,s}(\gamma_0)=0$, $ \forall\gamma_0 \in \Gamma \cap \mathbb{Z}$, \rd for all $j < \gamma_0$, \bk it follows 
$$b_{t,t-s}(\gamma_0) = \sum_{j = 0}^{ \floor{\gamma_0}-1}\bar\kappa_{j,s}(\gamma_0)y_{t-s+j} = 0, \qquad \forall\gamma_0 \in \Gamma \cap \mathbb{Z}.$$ Hence, the claim follows from Eq. \eqref{eq:b_recursion} because, by the same argument,
\[
 \llim\limits_{\gamma \rightarrow \gamma_0^{-}} b_{t,t-s}(\gamma) = y_{t-s+ \floor{\gamma}}\prod_{i = \floor{\gamma}+1}^{s}\left(1-\frac{\gamma_0}{i}\right) - \sum_{j = 0}^{ \floor{\gamma}}\bar\kappa_{j,s}(\gamma_0)y_{t-s+j} = 0
\]
for any $\gamma\leq\gamma_0$, with $\gamma_0 \in \mathbb{Z}$. This shows that 
$$\llim\limits_{\gamma \rightarrow \gamma_0^{-}} b_{t,t-s}(\gamma) = b_{t,t-s}(\gamma_0)=0 \Leftrightarrow \llim\limits_{\gamma \rightarrow \gamma_0^{-}} a_{t,t-s}(\gamma) = a_{t,t-s}(\gamma_0).$$

\underline{Part ($iii$).} Suppose, without loss of generality, that $\gamma_1 > \gamma_2$.  Moreover, recall the representation in Eq. \eqref{eq:b_recursion} and note that,  $b_{t,t-s}(\gamma)= 0$ for $\gamma \in \mathbb{Z}$. \textit{Case 1:} $\gamma_1,\gamma_2 \in \mathbb{Z}$. By Eq. \eqref{eq:b_recursion} and the triangle inequality
\[
|a_{t,t-s}(\gamma_1)-a_{t,t-s}(\gamma_2)| \leq \sum_{j=0}^s|\bar\kappa_{j,s}(\gamma_1)-\bar\kappa_{j,s}(\gamma_2)||y_{t-s+j}|.
\]
Now, for $0<j<s$ (the cases $j=0$ or $j=s$ are trivial),
\begin{align*}
|\bar\kappa_{j,s}(\gamma_1)-&\bar\kappa_{j,s}(\gamma_1)| \\ = \,& \left\vert\frac{\gamma_2-\gamma_1}{j}\prod_{i=j+1}^s\left(1-\frac{\gamma_2}{i}\right)+\frac{\gamma_1}{j}\left[\prod_{i=j+1}^s\left(1-\frac{\gamma_2}{i}\right)-\prod_{i=j+1}^s\left(1-\frac{\gamma_1}{i}\right)\right]\right\vert \\
\leq \,& |\gamma_2-\gamma_1| + \gamma_1|\gamma_2-\gamma_1|(s-j),
\end{align*}
where the final inequality uses the triangle inequality and $|\prod a_i-\prod b_i| \leq \sum_i |a_i-b_i|$ for constants $a_i$ and $b_i$. It follows
\[
|a_{t,t-s}(\gamma_1)-a_{t,t-s}(\gamma_2)| \leq |\gamma_1-\gamma_2| \ 2\bar\gamma \sum_{j=1}^sj|y_{t-j}| \leq c\eta_{t,s}|\gamma_1-\gamma_2|, \quad \eta_{t,s} \coloneqq \gamma \sum_{j=1}^sj|y_{t-j}|, 
\]
with $c \coloneqq 2\bar\gamma \in (0,\infty).$
\textit{Case 2:} $\gamma_2 \notin \mathbb{Z}$, $\gamma_1 \in \mathbb{Z}$. Because $b_{t,t-s}(\gamma_2) = 0$ and (see item $iii.$)
\[
 y_{t-s+ \floor{\gamma_2}}\prod_{i = \floor{\gamma_2}+1}^{s}\left(1-\frac{\gamma_1}{i}\right) - \sum_{j = 0}^{ \floor{\gamma_2}}\bar\kappa_{j,s}(\gamma_2)y_{t-s+j} = 0
\]
we get
\begin{align*}
a_{t,t-s}(\gamma_2)-a_{t,t-s}(\gamma_1) = \,& \sum_{j=\floor{\gamma_2}+1}^s(\bar\kappa_{j,s}(\gamma_2)-\bar\kappa_{j,s}(\gamma_1))y_{t-s+j} \\
\,&  +y_{t-s+ \floor{\gamma_2}}\left[\prod_{i = \floor{\gamma_2}+1}^{s}\left(1-\frac{\gamma_2}{i}\right)  -\prod_{i = \floor{\gamma_2}+1}^{s}\left(1-\frac{\gamma_1}{i}\right)\right] 
\end{align*}
Thus $|a_{t,t-s}(\gamma)-a_{t,t-s}(\gamma_0)| \leq c\eta_{t,s}|\gamma-\gamma_0|,$ $c \in (0,\infty)$. \textit{Case 3:} $\gamma,\gamma_0 \notin \mathbb{Z}$, $k < \gamma_2 < \gamma_1 < k+1$, $k \in \mathbb{Z}$. Because $\floor{\gamma_1}=\floor{\gamma_2}$, we have 
\[
b_{t,t-s}(\gamma_1) = y_{t-s+ \floor{\gamma_2}}\prod_{i = \floor{\gamma_2}+1}^{s}\left(1-\frac{\gamma_1}{i}\right) - \sum_{j = 0}^{ \floor{\gamma_2}}\bar\kappa_{j,s}(\gamma_1)y_{t-s+j}
\]
and we are in the same situation as {\it Case 2}, i.e. $|a_{t,t-s}(\gamma_1)-a_{t,t-s}(\gamma_2)| \leq c\eta_{t,s} |\gamma_1-\gamma_2|.$ \textit{Case 4:} $\gamma_1,\gamma_2 \notin \mathbb{Z}$, $k < \gamma_2  < m < \gamma_1$, $k < m$, $k,m\in \mathbb{Z}$. Consider
\begin{align*}
    a_{t,t-s}(\gamma_2)\ -&\ a_{t,t-s}(\gamma_1) \\ = \,& \sum_{j=0}^s(\bar\kappa_{j,s}(\gamma_2)-\bar\kappa_{j,s}(\gamma_1))y_{t-s+j} \\
    \,& +  y_{t-s+ \floor{\gamma_2}}\prod_{i = \floor{\gamma_2}+1}^{s}\left(1-\frac{\gamma_2}{i}\right)-y_{t-s+ \floor{\gamma_1}}\prod_{i = \floor{\gamma_1}+1}^{s}\left(1-\frac{\gamma_1}{i}\right) \\ 
    \,& + \sum_{j = 0}^{ \floor{\gamma_1}}\bar\kappa_{j,s}(\gamma_1)y_{t-s+j}-\sum_{j = 0}^{ \floor{\gamma_2}}\bar\kappa_{j,s}(\gamma_2)y_{t-s+j} = A+B+C,
\end{align*}
say. $A \leq \eta_{t,s}c|\gamma_1-\gamma_2|$ has already been treated. Turning to $B$, decompose $B =  B_1+B_2$ via
\[
B_1 = y_{t-s+ \floor{\gamma_2}}\left[\prod_{i = \floor{\gamma_2}+1}^{s}\left(1-\frac{\gamma_2}{i}\right)-\prod_{i = \floor{\gamma_2}+1}^{s}\left(1-\frac{\gamma_1}{i}\right)\right], 
\]
and
\[
B_2 = y_{t-s+ \floor{\gamma_2}}\prod_{i = \floor{\gamma_2}+1}^{s}\left(1-\frac{\gamma_1}{i}\right)- y_{t-s+ \floor{\gamma_1}}\prod_{i = \floor{\gamma_1}+1}^{s}\left(1-\frac{\gamma_1}{i}\right).
\]
As before, $|B_1| \leq c\eta_{t,s} |\gamma-\gamma_0|$, $c \in (0,\infty)$. For $B_2$, note that $\gamma_2/\gamma_2 > 1$ and
\[
0<(\gamma_1/\floor{\gamma_1}-1)^m\eqqcolon\kappa<\prod_{i=\floor{\gamma_2}+1}^{\floor{\gamma_1}}\left(\frac{\gamma_1}{i}-1 \right) \leq |\gamma_1-\gamma_2|.
\]
Using that
$$\prod_{i = \floor{\gamma_2}+1}^{s}\left(1-\frac{\gamma_1}{i}\right) = \prod_{i=\floor{\gamma_2}+1}^{\floor{\gamma_1}}\left(1-\frac{\gamma_1}{i}\right)\prod_{i=\floor{\gamma_1}+1}^{s}\left(1-\frac{\gamma_1}{i}\right)$$
and $\prod_{i>\gamma}(1-\gamma/i)\leq 1$ we obtain the bound
\begin{align*}
|B_2| \leq \eta_{t,s} &\prod_{i=\floor{\gamma_2}+1}^{\floor{\gamma_1}}\left(\frac{\gamma_1}{i}-1 \right) \\
&\times  \left(\prod_{i=\floor{\gamma_1}+1}^{s}\left(1-\frac{\gamma_1}{i}\right)/\prod_{i=\floor{\gamma_2}+1}^{\floor{\gamma_1}}\left(\frac{\gamma_1}{i}-1\right) +\prod_{i = \floor{\gamma_1}+1}^{s}\left(1-\frac{\gamma_1}{i}\right)\right) \\
& \leq \eta_{t,s} |\gamma_1-\gamma_2|(1+1/\kappa).
\end{align*}

\underline{Part ($iv$).} Define the Gamma function
\[
\Gamma(x) =  \int_0^\infty t^{x-1}e^{-t}\textsf{d}t, \quad x > 0,
\]
which is extended by analytic continuation to all real numbers $x\in \mathbb{R}$ except for simple poles at $x \in \{-1, -2\dots\}$. Thus, we note that for $\floor{\gamma} \leq j \leq s$
\begin{align} 
\kappa_{j,s}(\gamma) \stackrel{(1)}{=} \,&\frac{\gamma}{j}\frac{\Gamma(j+1)}{\Gamma(s+1)}\frac{\Gamma(s+1-\gamma)}{\Gamma(j+1-\gamma)}\label{eq:kappa}\\
                  \stackrel{(2)}{=} \,& h_{j,s}(\gamma)\left[1+\frac{1}{j}\frac{\gamma(1-\gamma)}{2}+O\left(\frac{1}{j^2}\right)\right],\nonumber \quad h_{j,s}(\gamma) \coloneqq \frac{\gamma}{s^\gamma} j^{\gamma-1}, \nonumber
\end{align}
where equality (1) is due to the definition of the gamma function (see, e.g., \citealp[Ch. 12]{apostol:1976}) and (2) uses \citet[Eq. (1)]{erdtric:1951}.

\textbf{Proof of Lemma \ref{lem:A1}.} 
\underline{Part ($i$).} We first verify the case \(k=0\); as discussed below, the case \(k > 0\) follow analogously. Consider
\(
\Ex[r_{t,t-s}(\gamma)r_{t,t-s}(\gamma')] = A_s+B_s+C_s,
\)
where
\[
A_s \coloneqq c(0)\sum_{j=1}^sh_{j,s}(\gamma)h_{j,s}(\gamma'),
\]
and
\[
B_s\coloneqq \sum_{j=2}^s\sum_{i=1}^{j-1}h_{j,s}(\gamma)h_{i,s}(\gamma')c(j-i), \quad C_s \coloneqq \sum_{j=2}^s\sum_{i=1}^{j-1}h_{j,s}(\gamma')h_{i,s}(\gamma)c(j-i).
\]
Now, one gets
\[
s\sum_{j=1}^sh_{j,s}(\gamma)h_{j,s}(\gamma') = \frac{\gamma\gamma'}{s^{\gamma+\gamma'-1}}\sum_{j=1}^{s}\frac{1}{j^{2-\gamma-\gamma'}} = \varphi(\gamma,\gamma')(1+o(1)), \quad \text{as }\;s \rightarrow \infty,
\]
where \(\Upsilon_0(\gamma,\gamma') = \varphi(\gamma,\gamma')\); see, e.g., \citet[Ch. 3]{apostol:1976}. Turning to \(B_s\), note that, by Toeplitz's lemma and Assumption \ref{ass:B},
\[
\frac{1}{j^{\gamma'-1}}\sum_{i=1}^{j-1} \frac{c(j-i)}{i^{1-\gamma'}} = \sum_{i=1}^{\infty}c(i)(1+o(1))
\]
as \(j \rightarrow \infty\). Moreover,
\[
sB_s =  \frac{\gamma\gamma'}{s^{\gamma+\gamma'-1}}\sum_{j=2}^{s}\frac{1}{j^{2-\gamma-\gamma'}}\left[\frac{1}{j^{\gamma'-1}}\sum_{i=1}^{j-1} \frac{c(j-i)}{i^{1-\gamma'}}\right] = \sum_{i=1}^{\infty}c(i)\varphi(\gamma,\gamma')+o(1),
\]
as \(s \rightarrow \infty\). By the same arguments, \(sC_s = \sum_{i=1}^{\infty}c(i)\varphi(\gamma,\gamma')+o(1)\). The claim is proven upon collecting terms. Similarly, if \(k > 0\), note first that 
\[
\sum_{j=1}^sh_{j,s}^{(k)}(\gamma)h^{(k)}_{j,s}(\gamma')=\frac{1}{s^{\gamma+\gamma'-1}}\sum_{j=1}^s\frac{1}{j^{2-\gamma-\gamma'}}{\lln}^{2(k-1)}(j/s)(\gamma\lln(j/s)+k)(\gamma'\lln(j/s)+k).
\]
Use the same arguments employed to prove the case \(k=0\) and note
\[
\frac{1}{s^{\gamma+\gamma'-1}}\sum_{j=1}^s\frac{1}{j^{2-\gamma-\gamma'}}{\lln}^{k}(j/s) = (-1)^kk\Gamma(k)(\gamma+\gamma'-1)^{-k-1}(1+o(1)), \quad k > 0,
\]
to finish the proof.

\underline{Part ($ii$).} This follows directly from part ($i$) noting that $s \sum_{j=1}^s (h_{j,s}(\gamma))^2\lln(j/s) = -\gamma^2/(2\gamma-1)^2(1+o(1)).$

\underline{Part ($iii$).} Recall from Eq. \eqref{cume4} that
\begin{align*}
\Ex[y_iy_jy_ky_l] = &c(j-i,k-i,l-i) \\
&\qquad+c(j-i)c(l-k)+c(k-i)c(l-j)+c(l-i)c(k-j)
\end{align*}
what, in conjunction with the triangle inequality, yields  \(\Ex[|r^{(r)}_{t,t-s}|^4] \leq A + B\), where
\begin{align*}
A \coloneqq \,& \sum_{i,j,k,l=1}^s|h^{(r)}_{i,s}(\gamma)||h^{(r)}_{j,s}(\gamma)||h^{(r)}_{k,s}(\gamma)||h^{(r)}_{l,s}(\gamma)|\\
\,& \hspace*{1.5cm}\times(|c(j-i)c(l-k)|+|c(k-i)c(l-j)|+|c(l-i)c(k-j)|).\\
B \coloneqq \,& \sum_{i,j,k,l=1}^s|h^{(r)}_{i,s}(\gamma)||h^{(r)}_{j,s}(\gamma)||h^{(r)}_{k,s}(\gamma)||h^{(r)}_{l,s}(\gamma)||c(j-i,k-i,l-i)|.
\end{align*}
To this end, we will show that $s^2A$ and $s^2B$ are bounded as $s \rightarrow \infty$.

Begin with \(A\) and note that by Assumption \ref{ass:B} and construction of $h^{(r)}_{i,s}(\gamma)$, there exists a constant $C_1 \in (0,\infty)$ such that $$(1+|\tau|)^2|c(\tau)| \leq C_1, \quad |h^{(r)}_{i,s}(\gamma)| \leq C_1h_{i,s}(\gamma){\lln}^{r}(s/i),$$ so that 
\begin{align*}
s^2A \leq \,&  3C_1 s^2\sum_{i,j,k,l=1}^s\frac{h_{i,s}(\gamma)\lln^r(s/i)h_{j,s}(\gamma)\lln^r(s/j)h_{k,s}(\gamma)\lln^r(s/k)h_{l,s}(\gamma)\lln^r(s/l)}{(1+|j-i|)^2(1+|l-k|)^2} \\
= \,& 3C_1 \gamma^4 s^{2-4\gamma}\sum_{i=1}^s i^{\gamma-1}{\lln}^r(s/i)\\
\,& \times \sum_{p,q,r=-(i-1)}^{s-i}\frac{(p+i)^{\gamma-1}\lln^r(s/(p+i))(q+i)^{\gamma-1}\lln^r(s/(q+i))(r+i)^{\gamma-1}\lln^r(s/(r+i))}{(1+|p|)^2(1+|r-q|)^2} \\
\leq \,& 3C_1 \gamma^4 s^{2-4\gamma}\sum_{i=1}^s i^{\gamma-1}{\lln}^r(s/i)\\
\,& \times \sum_{p,q,r=-(i-1)}^{s-i}\frac{(p+i)^{\gamma-1}\lln^r(s/(p+i))(q+i)^{\gamma-1}\lln^r(s/(q+i))(r+i)^{\gamma-1}\lln^r(s/(r+i))}{(1+|p|)^2(1+|r|)^2} \\
= \,&  3C_1 \gamma^4 s^{2-4\gamma}\sum_{q=1}^{s}q^{\gamma-1}{\lln}^r(s/q)\sum_{i=1}^s i^{\gamma-1}{\lln}^r(s/i)\bigg(\sum_{p=-(i-1)}^{s-i}\frac{(p+i)^{\gamma-1}{\lln}^r(s/(p+i)}{(1+|p|)^2}\bigg)^2 \\
\leq \,& C_2 s^{2-3\gamma}\sum_{i=1}^s i^{\gamma-1}{\lln}^r(s/i)\bigg(\sum_{p=-(i-1)}^{s-i}\frac{(p+i)^{\gamma-1}\lln^r(s/(p+i))}{(1+|p|)^2}\bigg)^2,
\end{align*}
where $C_2 \in (0,\infty)$ and the final inequality uses $\sum_{q=1}^s{\lln}^r(s/q)q^{\gamma-1} = O(s^\gamma)$. Next, some elementary manipulations reveal
\begin{align*}
\sum_{p=-(i-1)}^{s-i}\frac{(p+i)^{\gamma-1}\lln^r(s/(p+i))}{(1+|p|)^2} = \,& \sum_{p=1}^{i}\frac{p^{\gamma-1}\lln^r(s/p)}{(i-p+1)^2}\\
\,&+\sum_{p=1}^{s-i}\frac{(p+i)^{\gamma-1}\lln^r(s/(p+i))}{(1+p)^2} \coloneqq A_1+A_2,
\end{align*}
say. By the $c_r$-inequality,
\[
A_1 \leq 2^{r-1}\left(\sum_{p=1}^{i}\frac{p^{\gamma-1}\lln^r(s/i)}{(i-p+1)^2}+\sum_{p=1}^{i}\frac{p^{\gamma-1}\lln^r(i/p)}{(i-p+1)^2}\right),
\]
while, by Toeplitz's lemma,  $$i^{1-\gamma}\sum_{p=1}^{i}\frac{p^{\gamma-1}}{(i-p+1)^2}=\frac{\pi^2}{6}(1+o(1))$$ so that $A_1 = O(i^{\gamma-1}{\lln}^r(s/i)).$ Next, consider $A_2$ and note that
\[
A_2 \leq (1\{\gamma \leq 1\}i^{\gamma-1}+1\{\gamma >1\}s^{\gamma-1}){\lln}^r(s/i)\sum_{q\geq 1}q^{-2}.
\]
 Thus, if $\gamma \leq 1$, then $A_1+A_2 = O(i^{\gamma-1}{\lln}^r(s/i))$ while $A_1+A_2 = O(s^{\gamma-1}{\lln}^r(s/i))$ if $\gamma > 1$. Therefore, there exists a constant $C \in (0,\infty)$ such that $s^2A$ is bounded from above by
\[
 s^{2-3\gamma}\sum_{t=1}^s t^{3(\gamma-1)}{\lln}^{3r}(s/t) \leq  Cs^{2-3\gamma} \int^s_{1}t^{3(\gamma-1)} {\lln}^{3r}(s/t)\textsf{d}t \rightarrow  C \Gamma(3r+1)(3\gamma-2)^{-3r-1}, 
\]
if $\gamma \in (2/3,1]$ and
\[
Cs^{-\gamma}\sum_{t=1}^s t^{(\gamma-1)}{\lln}^{3r}(s/t) \leq C\int^s_{1}t^{\gamma-1}{\lln}^{3r}(s/t) \textsf{d}t \rightarrow  C\Gamma(3r+1)\gamma^{-3r-1},
\]
if $\gamma \in (1,\bar\gamma]$. This shows that $s^2A = O(1)$. 

It thus remains to be shown that $s^2B = O(1).$ Because the cumulant is absolutely summable, one has for any \(a,b,c \in \mathbb{R}\)
\[
|c(a,b,c)| \leq C\frac{1}{(1+|a|)(1+|b|)(1+|c|)}, \quad C \in (0,\infty),
\]
see \citet[Lemma 3]{dem:2008}. Therefore, 
\begin{align*}
 s^2A \leq 
 \,&C s^{2-4\gamma}\sum_{i=1}^s i^{\gamma-1}{\lln}^r(s/i)\left(\sum_{p=-(i-1)}^{s-i}\frac{(p+i)^{\gamma-1}{\lln}^{r}(s/(p+i))}{(1+|p|)}\right)^3.
\end{align*}
The following mimics the treatment of $A$. First, we get 
\begin{align*}
\sum_{p=-(i-1)}^{s-i}\frac{(p+i)^{\gamma-1}\lln^r(s/(p+i))}{(1+|p|)^2} = \,& \sum_{p=1}^{i}\frac{p^{\gamma-1}\lln^r(s/p)}{(i-p+1)}\\
\,&+\sum_{p=1}^{s-i}\frac{(p+i)^{\gamma-1}\lln^r(s/(p+i))}{(1+p)} \coloneqq A_1+A_2,
\end{align*}
say. Application of the $c_r$ inequality reveals that
\[
A_1 \leq 2^{k-1}\left(\sum_{p=1}^{i}\frac{p^{\gamma-1}\lln^r(s/i)}{i-p+1}+\sum_{p=1}^{i}\frac{p^{\gamma-1}\lln^r(i/p)}{i-p+1}\right).
\]
Because 
\[
\frac{i^{1-\gamma}}{\lln(i)}\sum_{p=1}^{i}\frac{p^{\gamma-1}}{i-p+1} \rightarrow 1,
\]
one gets $A_1 = O(\lln^r(s/i)\lln(i)i^{\gamma-1}).$ Turning to $A_2$, we get
\[
A_2 \leq (1\{\gamma \leq 1\}i^{\gamma-1}+1\{\gamma >1\}s^{\gamma-1}){\lln}^r(s/i)\lln(s).
\]
Therefore, $A_1+A_2 = O(\lln^r(s/i)\lln(s)i^{\gamma-1})$ if $\gamma \leq 1$ and $A_1+A_2 = O(\lln^r(s/i)\lln(s)s^{\gamma-1})$ if $\gamma > 1$.  
Using similar arguments as above, one gets that $s^2A = o(1).$ This finishes the proof. \hfill$\square$

\textbf{Proof of Lemma \ref{lem:A2}.} 
\underline{Part ($i$).} Recall from Eq. \eqref{rreq}
\(
r^{(m)}_{t,t-s}(\gamma) = \sum_{j=1}^sh^{(m)}_{j,s}(\gamma)y_{t-s+j},\) \(h^{(m)}_{j,s}(\gamma) = \frac{\textsf{d}^m}{\textsf{d}\gamma^m}h_{j,s}(\gamma).
\)
Now, \( \ssup\limits_{\gamma \in \Gamma}|r^{(m)}_{t,t-s}(\gamma)|^2 \leq 2(|r^{(m)}_{t,t-s}|^2+B^2)\), where \(B \coloneqq  \ssup\limits_{\gamma \in \Gamma}|r^{(m)}_{t,t-s}(\gamma)-r^{(m)}_{t,t-s}|\).
Similar to \citet[Eq. (3.8)]{lai:1994}, we obtain from Cauchy-Schwarz
\[
\Ex[B^2] \leq (\bar\gamma-\ubar\gamma)\Ex\left[\int_{\Gamma}|r^{(m+1)}_{t,t-s}(\gamma)|^2\textsf{d}\gamma\right] =  (\bar\gamma-\ubar\gamma)\int_{\Gamma}\Ex\left[|r^{(m+1)}_{t,t-s}(\gamma)|^2\right]\textsf{d}\gamma,
\]
where the second equality follows from Tonelli's theorem. Now, by Lemma \ref{lem:A1}, for each \(\gamma \in \Gamma\) and any integer \(m \geq 0\),
\[
s\Ex[|r^{(m)}_{t,t-s}(\gamma)|^2] = s\sum_{i,j=1}^sh^{(m)}_{i,s}(\gamma)h^{(m)}_{j,s}(\gamma)c(i-j) \rightarrow \omega^2 \Upsilon_m(\gamma,\gamma).
\]
Note that the preceding expectation is independent of $t$ and that the convergence is uniform because \(\gamma \mapsto \Upsilon_m(\gamma,\gamma)\) is continuous and \(\gamma \mapsto s\Ex[|r^{(m)}_{t,t-s}(\gamma)|^2]\) is convex for any integer $m \geq 0$ and $s \geq \bar s >0$ for some fixed $\bar s$. As \(\int_{\Gamma}\Upsilon_m(\gamma,\gamma)\textsf{d}\gamma < \infty\), the claim follows.  \underline{Part ($ii$).} Using the triangle inequality and Cauchy-Schwarz we get
\[
\Ex\left[\ssup\limits_{\gamma \in \Gamma} \frac1{\nu_n}\bigg\vert \sum_{t = u+1}^n\sum_{s = t-u}^{t-l} r^{(m)}_{t,s}(\gamma)r^{(k)}_{t,s}(\gamma)\bigg\vert\right] \leq  \frac1{\lln(n)}\sum_{s = l}^{u} \sqrt{\Ex[\ssup\limits_{\gamma \in \Gamma}(r^{(m)}_{t,t-s}(\gamma))^2]\Ex[\ssup\limits_{\gamma \in \Gamma}(r^{(k)}_{t,t-s}(\gamma))^2]},
\]
which is $O(1)$ by part ($i$). Thus, the claim follows by Markov's inequality. \underline{Part ($iii$).} First note
\begin{align*}
\sum_{t = u+1}^n |\bar r_t^{(m)}(\gamma)\bar r_t^{(k)}(\gamma)| \leq \,&\sum_{t = u+1}^n  |\bar r_t^{(m)}(\gamma)-\bar r_t^{(m)}||\bar r_t^{(k)}(\gamma)-\bar r_t^{(k)}| + \sum_{t = u+1}^n |\bar r_t^{(m)}\bar r_t^{(k)}| \\
\,& \qquad + \sum_{t = u+1}^n |\bar r_t^{(m)}(\gamma)-\bar r_t^{(m)}||\bar r_t^{(k)}| +  \sum_{t = u+1}^n |\bar r_t^{(m)}||\bar r_t^{(k)}(\gamma)-\bar r_t^{(k)}|.
\end{align*}
Next, use again repeatedly Cauchy-Schwarz to obtain
\begin{align*}
m&\sqrt{\Ex[\ssup\limits_{\gamma\in \Gamma} |\bar r^{(m)}_t(\gamma)-\bar r^{(m)}_t|^2]\Ex[\ssup\limits_{\gamma\in \Gamma} |\bar r^{(k)}_t(\gamma)-\bar r^{(k)}_t|^2]} \\
\leq \,& \frac1{m}\sqrt{\Ex\left[\ssup\limits_{\gamma\in  \Gamma}  \left[\sum_{s=l}^u(r^{(m)}_{t,t-s}(\gamma)-r^{(m)}_{t,t-s})\right]^2\right]\Ex\left[\ssup\limits_{\gamma\in  \Gamma}  \left[\sum_{s=l}^u(r^{(k)}_{t,t-s}(\gamma)-r^{(k)}_{t,t-s})\right]^2\right]} \\
\leq \,& \frac{\bar\gamma-\ubar\gamma}{m} \sqrt{\int_{ \Gamma}\Ex\left[\big(\sum_{s=l}^u r^{(m)}_{t,t-s}(\gamma)\big)^2\right] \textsf{d}\gamma \int_{ \Gamma}\Ex\left[\sum_{s=l}^u r^{(k)}_{t,t-s}(\gamma)\right]^2 \textsf{d}\gamma} \\
\leq \,& \frac{\bar\gamma-\ubar\gamma}{m} \sqrt{\int_{ \Gamma}\left[\sum_{s=l}^u \Ex^{1/2}[(r^{(m)}_{t,t-s}(\gamma))^2] \right]^2 \textsf{d}\gamma \int_{ \Gamma}\left[\sum_{s=l}^u \Ex^{1/2}[(r^{(k)}_{t,t-s}(\gamma))^2] \right]^2 \textsf{d}\gamma} \\
= \,& (\bar\gamma-\ubar\gamma) \sqrt{\int_{ \Gamma}\left[\frac1{\sqrt{m}}\sum_{s=l}^u \Ex^{1/2}[(r^{(m)}_{t,t-s}(\gamma))^2] \right]^2 \textsf{d}\gamma \int_{ \Gamma}\left[\frac1{\sqrt{m}}\sum_{s=l}^u \Ex^{1/2}[(r^{(k)}_{t,t-s}(\gamma))^2] \right]^2 \textsf{d}\gamma},
\end{align*}
which is $O(1)$ because, by Lemma \ref{lem:A1} and the arguments used in the proof of part ($ii$),
 \begin{align*} 
 \frac{1}{\sqrt{u}} \sum_{s=1}^u \Ex^{1/2}[|r^{(k)}_{t,t-s}(\gamma)|^2] \rightarrow \sqrt{4\omega^2 \Upsilon_k(\gamma,\gamma)} < \infty \quad \text{uniformly in $\gamma \in \Gamma$}.
 \end{align*}
Hence, by Markov's inequality,
\[
\frac{m}{\nu_n}\sum_{t = u+1}^n \ssup\limits_{\gamma\in \Gamma}|\bar r_t^{(m)}(\gamma)-\bar r_t^{(m)}|\ssup\limits_{\gamma\in \Gamma}|\bar r_t^{(k)}(\gamma)-\bar r_t^{(k)}|  = O_p({\lln}^{-1}(n)).
\]
Similarly,
\begin{align*}
  \frac{m}{\nu_n} \Ex\left[ \sum_{t = u+1}^n \vert\bar r^{(m)}_{t} \bar r^{(k)}_{t}  \vert \right] = \,& \frac1{\nu_n}\sum_{t = u+1}^n \left[\frac{1}{m}\sum_{s,k=l}^u \Ex[|r^{(m)}_{t,t-s}r^{(k)}_{t,t-k}|]\right] \\
   \leq \,& \frac{1}{\lln(n)} \bigg[ \frac{1}{\sqrt{m}} \sum_{s=l}^u \Ex^{1/2}[|r^{(m)}_{t,t-s}|^2] \bigg]\bigg[ \frac{1}{\sqrt{m}} \sum_{s=l}^u \Ex^{1/2}[|r^{(k)}_{t,t-s}|^2] \bigg] = O(1),  
\end{align*}
using Cauchy-Schwarz and the triangle inequality so that, by Markov's inequality,
$\frac{m}{\nu_n}\sum_{t = u+1}^n |\bar r_t^{(m)}\bar r_t^{(k)}|  = O_p({\lln}^{-1}(n)).$ The term $\sum_{t = u+1}^n |\bar r_t^{(m)}||\bar r_t^{(k)}(\gamma)-\bar r_t^{(k)}|$ can be now treated analogously. 

This completes the proof. \hfill$\square$

\textbf{Proof of Lemma \ref{lem:A3}.} 
\underline{Part ($i$).} First, define
\[
R_n(\gamma_1,\gamma_2) \coloneqq \frac{1}{\nu_n}\sum_{t = u+1}^n\sum_{s = t-u}^{t-l} \Ex[r^{(m)}_{t,s}(\gamma_1)r^{(m)}_{t,s}(\gamma_2)], \quad \gamma_1,\gamma_2 \in \Gamma.
\]
Next, we deduce from Lemma \ref{lem:A1} and Assumption \ref{ass:A2} that 
\begin{align}\label{Err}
\Ex[R_{n}(\gamma_1,\gamma_2)] = \omega^2\lambda^2\Upsilon_m(\gamma_1,\gamma_2) + o(1).
\end{align}
To see this, note
\begin{align*}
R_{n}(\gamma_1,\gamma_2) = \frac{n-u}{n}\bigg[\frac{\lln(u)}{\lln(n)}\frac{1}{\lln (u)}&\sum_{s = 1}^{u} r^{(m)}_{t,t-s}(\gamma_1)r^{(m)}_{t,t-s}(\gamma_2)\\
&\qquad-\frac{\lln(l)}{\lln (n)}\frac{1}{\lln(l)}\sum_{s = 1}^{l-1} r^{(m)}_{t,t-s}(\gamma_1)r^{(m)}_{t,t-s}(\gamma_2)\bigg]. 
\end{align*}
Thus, Eq. \eqref{Err} follows from Lemma \ref{lem:A1} and Assumption \ref{ass:A2}.  Next, by Markov's inequality, pointwise convergence in probability follows if we can show that 
\[
\var[\sum_{t = u+1}^n \zeta_{nt}(\gamma_1,\gamma_2)] = o(n^2), \quad \zeta_{nt}(\tilde\gamma) \coloneqq  \frac{1}{\lln(u)}\sum_{s = 1}^{u} r^{(m)}_{t,t-s}(\gamma_1)r^{(m)}_{t,t-s}(\gamma_2).
\]
Begin by considering \(\sum_{t = u+1}^n \var[\zeta_{nt}(\gamma_1,\gamma_2)]\), where
\begin{align*}
\var[\zeta_{nt}(\gamma_1,\gamma_2)] = \,& \frac{1}{\lln^2(u)}\sum_{k,s =1}^u \cov[r^{(m)}_{t,t-s}(\gamma_1)r^{(m)}_{t,t-s}(\gamma_2),r^{(m)}_{t,t-k}(\gamma_1)r^{(m)}_{t,t-k}(\gamma_2)] \\
\leq \,&  \left[\frac{1}{\lln(u)}\sum_{s = 1}^u \Ex^{1/4}[|r^{(m)}_{t,t-s}(\gamma_1)|^4]\Ex^{1/4}[|r^{(m)}_{t,t-s}(\gamma_2)|^4]\right]^2 = O(1),
\end{align*}
using repeatedly Cauchy-Schwarz's inequality and Lemma \ref{lem:A1}. Hence, \[\sum_{t = u+1}^n \var[\zeta_{nt}(\tilde\gamma)] = O(n).\]
Moreover, as
\begin{align*}
\cov[\zeta_{t}(\gamma_1,\gamma_2),\zeta_{t+\tau}(\gamma_1,\gamma_2)] =  \,& \frac{1}{\lln^2(u)}\sum_{s,s' =1}^u\cov[r^{(m)}_{t,t-s}(\gamma_1)r^{(m)}_{t,t-s}(\gamma_2),r^{(m)}_{t+\tau,t+\tau-s'}(\gamma_1)r^{(m)}_{t+\tau,t+\tau-s'}(\gamma_2)] 
\end{align*}
one gets, by arguments similar to those used to verify part ($iv$) of Lemma \ref{lem:A1}, 
\[
\sum_{t = u+1}^{n-1}\sum_{\tau = 1}^{n-t}\cov[\zeta_{t}(\gamma_1,\gamma_2),\zeta_{t+\tau}(\gamma_1,\gamma_2)] = o(n^2).
\]
Uniform convergence follows from \citet[Lemma 1]{andrews:92} because
\[
|R_n(\gamma_1,\gamma_2)-R_n(\gamma_1',\gamma_2')| \leq (|\gamma_1-\gamma_1'|+|\gamma_2-\gamma_2'|)\dot R_n,
\]
with
\[
\dot R_n \coloneqq \ssup\limits_{\gamma_1,\gamma_2 \in \Gamma}\frac{1}{\nu_n}\left\Vert\sum_{t = u+1}^n\sum_{s = 1}^{u}(r^{(m)}_{t,t-s}(\gamma_1) r^{(m+1)}_{t,t-s}(\gamma_2), r^{(m+1)}_{t,t-s}(\gamma_1)r^{(m)}_{t,t-s}(\gamma_2))^\Tr\right\Vert,
\]
where \(\dot R_n = O_p(1)\) follows from Lemma \ref{lem:A2} and Cauchy-Schwarz. This verifies the claim. 

\underline{Part ($ii$).} Follows by similar arguments. 

\underline{Part ($iii$).} First, we show that for any \(J \geq  1\)
\[
S_n(\gamma_1),\dots,S_n(\gamma_J) \rightarrow_d \omega \lambda\,\mathcal{N}_J(0,\textsf{M}_J), \quad \textsf{M}_J \coloneqq \big[\Upsilon_k(\gamma_j,\gamma_h)\big]_{1\leq j,h \leq J}.
\]
By the Cram\`{e}r Wold device, it suffices to consider, for some \(\iota \coloneqq (\iota_1,\dots,\iota_J)^\Tr \in \mathbb{R}^J\), \(\Vert\iota \Vert = 1\), the partial sum
\[
Z_n(\gamma) \coloneqq \frac{1}{\sqrt{n}}	\sum_{t = u+1}^n z_{nt}(\gamma), \quad \gamma \coloneqq (\gamma_1,\dots,\gamma_J)^\Tr,
\]
with
\[
z_{nt}(\gamma) \coloneqq \frac{1}{\sqrt{\lln(n)}}\sum_{s=t-u}^{t-l} \iota^\Tr R^{(k)}_{t,s}(\gamma)\eps_{t,s}, \quad R^{(k)}_{t,s}(\gamma) \coloneqq (r^{(k)}_{t,s}(\gamma_1),\dots,r^{(k)}_{t,s}(\gamma_J))^\Tr,  
\]
 By Assumption \ref{ass:B}, \(\{z_{nt}(\gamma)\}_t\) forms a martingale difference sequence with respect to \(\mathcal{F}_t\) so that
 \begin{align*}
\Ex[z_{nt}(\gamma)^2\mid \mathcal{F}_{t-1}] = \,&\frac{1}{\lln(n)}\sum_{s,k = t-u}^{t-l} \iota^\Tr R^{(k)}_{t,s}(\gamma) \iota^\Tr R^{(k)}_{t,k}(\gamma)\Ex[\eps_{t,s}\eps_{t,k}\mid \mathcal{F}_{t-1}]\\
= \,& \frac{\sigma^2}{\lln(n)}\sum_{s = t-u}^{t-l} (\iota^\Tr R^{(k)}_{t,s}(\gamma))^2.
\end{align*}
Therefore,
\begin{align}\label{eq:EZ2}
\Ex[Z^2_n(\gamma)] = \frac{\sigma^2}{\nu_n}\sum_{t = u+1}^n\sum_{s = t-u}^{t-l} \Ex[(\iota^\Tr R^{(k)}_{t,s}(\gamma))^2] \rightarrow \sigma^2\iota^\Tr \textsf{M}_J \iota \eqqcolon \tau_J > 0.
\end{align}
Thus, by \citet[Cor. 5.26]{white:2000},  if (1) \(\Ex[|z_{nt}(\gamma)|^{2+\delta}] < \infty\) for some \(\delta > 0\) and (2) 
\[
\frac{1}{n} \sum_{t = u+1}^n z^2_{nt}(\gamma) \rightarrow_p \tau_J,
\]
then the claim follows from the Cram\'{e}r-Wold device. Condition  (1) is due to Lemma \ref{lem:A1} ($iii$) while condition (2) follows from part ($ii$) of this Lemma. Stochastic equicontinuity is a direct result of the proof of part ($i$) of this Lemma and \citet[Thm. 2]{hansen:1996b}. This finishes the proof. \hfill$\square$

\textbf{Proof of Lemma \ref{lem:A4}.} 
 Because the objective function is quadratic in $\beta$, the claim follows if we can show
\begin{align}\label{eq:ardiff_mean}
\frac1{\nu_n}\sum_{t=u+1}^n\sum_{s=l}^u (\tilde a_{t,t-s}(\gamma)-\tilde r_{t,t-s}(\gamma)-\tilde a_{t,t-s}+ \tilde r_{t,t-s})^2 =  O_p((\gamma-\gamma_0)^2a_n^2)
\end{align}
and
\begin{align}\label{eq:ardiff_eps}
\frac1{\sqrt{\nu_n}}\sum_{t=u+1}^n\sum_{s=l}^u \eps_{t,t-s}(\tilde a_{t,t-s}(\gamma)-\tilde r_{t,t-s}(\gamma)-\tilde a_{t,t-s}+ \tilde r_{t,t-s}) =  O_p((\gamma-\gamma_0)a_n)
\end{align}
for $a^2_n = {\lln}^3(u){\lln(n)}^{-1}u^{1-2\ubar\gamma}$ (Note that, by Assumption \ref{ass:A2} \(\ubar\gamma > 1/2\), so that $a_n = o(1)$.), as well as 
\begin{align}\label{eq:rrdiff_mean}
\frac1{\nu_n}\sum_{t=u+1}^n\sum_{s=l}^u (\tilde r_{t,t-s}(\gamma)- r_{t,t-s}(\gamma)-\tilde r_{t,t-s}+  r_{t,t-s})^2 = O_p\left(\frac{(\gamma-\gamma_0)^2}{{\lln}(n)}\right)
\end{align}
and
\begin{align}\label{eq:rrdiff_eps}
\frac1{\sqrt{\nu_n}}\sum_{t=u+1}^n\sum_{s=l}^u \eps_{t,t-s} (\tilde r_{t,t-s}(\gamma)- r_{t,t-s}(\gamma)-\tilde r_{t,t-s}+  r_{t,t-s}) = O_p\left(\frac{\gamma-\gamma_0}{\sqrt{\lln}(n)}\right).
\end{align}

\underline{Proof of Eq. \eqref{eq:ardiff_mean}.} From Lemma \ref{lem:A0} ($iv.$) we get 
\begin{align}\label{eq:ardiff}
a_{t,t-s}(\gamma)-r_{t,t-s}(\gamma) =  \frac{\gamma(1-\gamma)}{2}\sum_{j=1}^s  \frac1{j}h_{j,s}(\gamma)y_{t-s+j}  + R_{t,t-s}(\gamma),
\end{align}
where 
\[
R_{t,t-s}(\gamma) \coloneqq \sum_{j=1}^s  O\left(\frac1{j^2}\right)h_{j,s}(\gamma)y_{t-s+j} 
\]
Thus, by the mean-value theorem and Eq. \eqref{eq:ardiff}
\begin{align*}
|\tilde a_{t,t-s}(\gamma)-\tilde r_{t,t-s}(\gamma)-\tilde a_{t,t-s}+ \tilde r_{t,t-s}|\leq  C|\gamma-\gamma_0| \ssup\limits_{\gamma \in \Gamma} |\tilde\Delta_{t,t-s}(\gamma)|+o_p(1),
\end{align*}
where $\Delta_{t,t-s}(\gamma) \coloneqq \sum_{j=1}^s \frac1{j}(h^{(1)}_{j,s}(\gamma))y_{t-s+j}$ for some constant $C \in (0,\infty).$ By Cauchy-Schwarz,
\(
(\Delta_{t,t-s}(\gamma))^2 \leq \sum_{j=1}^s\frac1{j}y^2_{t-s+j}\sum_{j=1}^s\frac1{j}(h^{(1)}_{j,s}(\gamma))^2.
\)
Next, we note that the map $\gamma \mapsto \frac1{\gamma}\sum_{j=1}^s\frac1{j}(h^{(1)}_{j,s}(\gamma))^2$ is a decreasing. Hence, 
assuming $w.l.o.g.$ that $\ubar\gamma \in (2/3,1)$, one gets    
\[
\ssup\limits_{\gamma \in  \Gamma}\sum_{j=1}^s\frac1{j}(h^{(1)}_{j,s}(\gamma))^2 \leq \frac{\bar\gamma}{\ubar\gamma}\sum_{j=1}^s\frac1{j}(h^{(1)}_{j,s}(\ubar\gamma))^2 = O\left(\int_1^s \frac{1}{j}(h^{(1)}_{j,s}(\ubar\gamma))^2 {\sf d}j\right) = O\left(\frac{{\lln}^2(s)}{s^{2\ubar\gamma}}\right).
\]
Since, by Assumption \ref{ass:B}, $\Ex[\ssup_{i\leq t}|y_i|^2] < \infty$, it follows, by Markov's inequality, $\sum_{j=1}^s\frac1{j}y^2_{t-s+j} \leq \ssup_{i\leq t}|y_i|^2\sum_{j=1}^s\frac1{j}=O_p(\lln(n))$, and we obtain
\[
\sum_{s=l}^u\Ex[\ssup\limits_{\gamma \in  \Gamma} |\Delta_{t,t-s}(\gamma)|^2] = \sum_{s=l}^uO({\lln}^3(s)/s^{2\ubar\gamma}) = O({\lln}^{3}(u)/u^{2\ubar\gamma-1})
\]
and, therefore, by Markov's inequality,
\[
\frac{1}{n\lln(n)}\sum_{t = u+1}^n\sum_{s = l}^{u}\ssup\limits_{\gamma \in  \Gamma} |\Delta_{t,t-s}(\gamma)|^2 = O_p\left(\frac{{\lln}^3(u)}{\lln(n) u^{2\ubar\gamma-1}}\right).
\]
\underline{Proof of Eq. \eqref{eq:ardiff_eps}.} Set
$X \coloneqq \frac1{\sqrt{\nu_n}}\sum_{t=u+1}^n\sum_{s=l}^u \eps_{t,t-s} \Delta_{t,t-s}(\gamma)$. The claim follows if  $X = O_p(\sqrt{a_n})$. Note that $\Ex[X] = 0$ and $\var[X] \leq \frac{\sigma^2}{\nu_n}\sum_t\sum_s \Ex[\ssup_\gamma |\Delta_{t,t-s}(\gamma)|^2]$, using by Assumption \ref{ass:B}. Because, by Chebychev's inequality, $X = O_p(\sqrt{\var[X]})$, the claim follows from the proof of Eq. \eqref{eq:ardiff_mean}

\underline{Proof of Eq. \eqref{eq:rrdiff_mean}.} Clearly, $\tilde r_{t,t-s}(\gamma)- r_{t,t-s}(\gamma)-\tilde r_{t,t-s}+r_{t,t-s} = \bar r_{t} - \bar r_t(\gamma)$. Hence, by the mean-value theorem, adding a zero, and Cauchy-Schwarz, we get $|\bar r_t - \bar r_t(\gamma)| \leq |\gamma-\gamma_0| \ssup_{\gamma \in \Gamma}(|\bar r^{(1)}_t(\gamma)-\bar r^{(1)}_t|+|\bar r^{(1)}_t|)$. Next, notice that
  \begin{align*}
  \frac{m}{\nu_n}\sum_{t = u+1}^n \Ex[(\bar r^{(1)}_t)^2] = \,& \frac1{\lln(n)} \frac{1}{n}\sum_{t = u+1}^n \left[\frac{1}{m}\sum_{s,k=l}^u \Ex[r^{(1)}_{t,t-s}r^{(1)}_{t,t-k}]\right] \\
  = \,& \frac1{\lln(n)} (1-u/n)\left[\frac{1}{m}\sum_{s,k=l}^u \Ex[r^{(1)}_{t,t-s}r^{(1)}_{t,t-k}]\right] \\
  \leq \,&  \frac{1}{\lln(n)}(1-u/n)\bigg[ \frac{1}{\sqrt{m}} \sum_{s=l}^u \Ex^{1/2}[|r^{(1)}_{t,t-s}|^2] \bigg]^2 = O\left(\frac1{\lln(n)}\right),
\end{align*}
 where the first equality is due to the definition of $\bar r^{(1)}_t$, the second equality uses the second-order stationarity of $r^{(1)}_{t,t-s}$, the inequality follows from Cauchy-Schwarz, and the order of magnitude is due to the fact that the term in square brackets is $O(1)$ because, by Lemma \ref{lem:A1},
 \begin{align}\label{A4:rksq}
 \frac{1}{\sqrt{u}} \sum_{s=1}^u \Ex^{1/2}[|r^{(k)}_{t,t-s}(\gamma)|^2] \rightarrow \sqrt{4\omega^2 \Upsilon_k(\gamma,\gamma)} < \infty.
 \end{align}
 The claim then follows by Markov's inequality because $m = O(u)$.  Next, use again Cauchy-Schwarz to obtain
\begin{align*}
\frac{m}{\nu_n}\sum_{t = u+1}^n\ssup\limits_{\gamma\in \Gamma} |\bar r^{(1)}_t(\gamma)-\bar r^{(1)}_t|^2 \leq \,& \frac{1}{m\nu_n}\sum_{t = u+1}^n \ssup\limits_{\gamma\in  \Gamma}  \left[\sum_{s=l}^u(r^{(1)}_{t,t-s}(\gamma)-r^{(1)}_{t,t-s})\right]^2 \\
\leq \,& \frac{\bar\gamma-\ubar\gamma}{m\nu_n}\sum_{t = u+1}^n\int_{ \Gamma}\left[\sum_{s=l}^u r^{(2)}_{t,t-s}(\gamma)\right]^2 \textsf{d}\gamma.
\end{align*}
Taking expectations, using Eq. \eqref{A4:rksq}, and $\int_{ \Gamma}\Upsilon_2(\gamma,\gamma) \textsf{d}\gamma < \infty$, the claim follows by Cauchy-Schwarz and Markov's inequality.

\textbf{Additional proofs of Corollary \ref{cor:0}.} 
It remains to be verified that $-{\sf D}_m(\theta) \leq -c\Vert \theta-\theta_0\Vert^2$, $c\in (0,\infty)$. Since the crucial parameter is $\gamma$ that causes the "kink" points, we will assume for the sake of brevity $\beta = \beta_0$ and also assume that no mean-adjustment is used, i.e., $\tilde a_{t,s}= a_{t,s}$. The general case follows readily. In addition, assume, without loss of generality, $\gamma < \gamma_0$. Consider
 \begin{align}
  {\sf D}_m(\beta_0,\gamma) = \beta_0^2\sum_{t=u+1}^n\sum_{s=l}^u\Ex[(a_{t,t-s}(\gamma)-a_{t,t-s}(\gamma_0))^2] 
\end{align}
where the expectation is independent of $t$ and given by
\begin{align*}
    \Ex[(a_{t,t-s}(\gamma)-a_{t,t-s}(\gamma_0))^2] = \,& \sum_{i,j=\floor{\gamma}}^s\kappa_{j,s}(\gamma)\kappa_{i,s}(\gamma)c(i-j)\\
&+\sum_{i,j=\floor{\gamma_0}}^s\kappa_{j,s}(\gamma_0)\kappa_{i,s}(\gamma_0)c(i-j)\\
&-2\sum_{i=\floor{\gamma_0}}^s\sum_{j=\floor{\gamma}}^s\kappa_{i,s}(\gamma_0)\kappa_{j,s}(\gamma)c(i-j) \\
= \,& (\kappa_s(\gamma)-\tilde\kappa_s(\gamma_0))^\Tr T(s)(\kappa_s(\gamma)-\tilde\kappa_s(\gamma_0)
\end{align*}
 where $\kappa_s(\gamma) \coloneqq (\kappa_{\floor{\gamma},s}(\gamma),\kappa_{\floor{\gamma}+1,s}(\gamma).\dots,\kappa_{s,s}(\gamma))^\Tr$ and $\tilde\kappa_s(\gamma_0) = (0_{\floor{\gamma}-\floor{\gamma_0}}^\Tr,\kappa_s(\gamma_0)^\Tr)^\Tr$
 are $(s-\floor{\gamma}+1)\times 1$ vectors while $ T(s) = [c(i-j)]_{0 \leq i,j \leq s-\floor{\gamma}}$  is a $(s-\floor{\gamma}+1)\times (s-\floor{\gamma}+1)$ positive definite Toeplitz matrix. Hence, using a well-known inequality for Rayleight quotientes, we get
\begin{align}\nonumber
    {\sf D}_m(\gamma) \geq  \mmin\limits_{1\leq i \leq u}\mu_1(T(i))(1-u/n)\sum_{s=l}^u \Vert\kappa_s(\gamma)-\tilde\kappa_s(\gamma_0)\Vert^2,
\end{align}
where, by Assumption \ref{ass:B}, the minimum eigenvalue $\mu_1(T(i))$ is bounded away from zero for any $i \geq 1$ and
\begin{align*}
\Vert\kappa_s(\gamma)-\tilde\kappa_s(\gamma_0)\Vert^2 = \,&
 \sum_{j=\floor{\gamma}}^s\kappa^2_{j,s}(\gamma)+\sum_{j=\floor{\gamma_0}}^s\kappa^2_{j,s}(\gamma_0)-2\sum_{j=\floor{\gamma_0}}^s\kappa_{j,s}(\gamma)\kappa_{j,s} (\gamma_0) \\
 \geq \,& \sum_{j=\floor{\gamma_0}}^s\kappa^2_{j,s}(\gamma)+\sum_{j=\floor{\gamma_0}}^s\kappa^2_{j,s}(\gamma_0)-2\sum_{j=\floor{\gamma_0}}^s\kappa_{j,s}(\gamma)\kappa_{j,s} (\gamma_0) \\
 = \,& \sum_{j=\floor{\gamma_0}}^s(\kappa_{j,s}(\gamma)-\kappa_{j,s}(\gamma_0))^2 \\
 \geq \,& \sum_{j=\floor{\gamma_0}+1}^s(\kappa_{j,s}(\gamma)-\kappa_{j,s}(\gamma_0))^2 = \sum_{j=\floor{\gamma_0}+1}^s(\bar\kappa_{j,s}(\gamma)-\bar\kappa_{j,s}(\gamma_0))^2,
\end{align*}
where 
the inequalities are immediate and the final equality follows from the definition of $\bar\kappa_{j,s}(\gamma) = \kappa_{j,s}(\gamma)$, $j>\floor{\gamma}$.
Now, for any $\floor{\gamma}\leq j \leq s$, define the first derivative
\[
 \dot{\bar\kappa}_{j,s}(\gamma)\coloneqq \frac{\sf d}{{\sf d}\gamma}\bar\kappa_{j,s}(\gamma) = \bar\kappa_{j,s}(\gamma) [\psi(j+1-\gamma)-\psi(s+1-\gamma)+1/\gamma].
\]
By the mean-value theorem, for some $\bar\gamma$ that lies on the line segment connecting $\gamma$ and $\gamma_0$, we get
\begin{align*}
\Vert\kappa_s(\gamma)- \tilde\kappa_s(\gamma_0)&\Vert^2 \\
\geq \ & (\gamma-\gamma_0)^2\sum_{j=\floor{\gamma_0}+1}^s( \kappa_{j,s}(\bar\gamma)[\psi(j+1-\bar\gamma)-\psi(s+1-\bar\gamma)+1/\bar\gamma])^2 \\
\geq \ & c (\gamma-\gamma_0)^2,
\end{align*}
with
\[
c \coloneq \mmax\limits_{1 \leq s \leq u} \iinf\limits_{\gamma \inn \Gamma} \sum_{j=\floor{\gamma_0}+1}^s( \kappa_{j,s}(\gamma)[\psi(j+1-\gamma)-\psi(s+1-\gamma)+1/\gamma])^2.
\]

\newpage

\section{Additional finite sample results}

{\bf This section provides additional Monte Carlo (Section \ref{sec:SMC}) and empirical (Section \ref{sec:Semp}) results that, among others, entertain a generalisation of the recursion used by agents to update their beliefs.}

As emphasised before, our model specification is a special case of the one considered by \cite{MadeiraZafar15}, \cite{mn:2016}, \cite{Gwak22}, and \cite{nagel:24}. Although, strictly speaking, the aforementioned specifications are not covered by our theory in Sections
\ref{sec:estimation}--\ref{sec:inf}, it may still be instructive to explore the limits of our theory in finite samples by moving towards these models. In particular, we outlined in Section \ref{sec:model} that agents in our setup are assumed to forecast inflation by recursively estimating the
level of inflation. This amounts to a perceived law of motion (PLM) that comprises as sole regressor a constant term:
\begin{equation} \label{eq:constantPLM}
y_t = \phi  + \eta_t,
\end{equation}
where $\eta_t$ is some error term and the level $\phi$ is recursively estimated using Eq.\ \eqref{recursion}.
We believe that this PLM offers a plausible approximation to agents' boundedly rational behaviour, since computing a
weighted average is arguably intuitive and reasonably straightforward, even for agents without much statistical training. Indeed, the PLM in \eqref{eq:constantPLM} is the one considered by \cite{naknun:15} as part of their
application in finance. As opposed to that, \cite{MadeiraZafar15}, \cite{mn:2016}, \cite{Gwak22}, and \cite{nagel:24} equip agents with more elaborate skills. In particular, they assume individuals to employ an AR(1) model with intercept as their
PLM of macro-level inflation $y_t$:
\begin{align}
  y_{t} = \phi_0 + \phi_1 y_{t-1} + \eta_{t}.  \label{eq:AR1PLM}
\end{align}
This requires agents to recursively obtain a generalised least-squares estimate of the $2 \times 1$ parameter vector $\phi \coloneqq (\phi_0,\phi_1)^\Tr$. In particular, individuals born in period $s$ use the regressor $ x_t \coloneqq (1,y_{t-1})^\Tr$ to update in period $t$ their beliefs about inflation following the stochastic recursive algorithm
\begin{equation}\label{eq:recursion_general}
\begin{split}
  r_{t,s}                    & = r_{t-1,s} + \gamma_{t,s} (x_t x_t^\Tr - r_{t-1,s}) \\
  \phi_{t,s}               & = \phi_{t-1,s} + \gamma_{t,s} r^{-1}_{t,s} x_t (y_t - \phi_{t-1,s}^\Tr x_t), 
  \end{split}
\end{equation}
with \(\gamma_{t,s} = \gamma_{t,s}(\gamma_0)\) as in Eq.\ \eqref{gain_mn}. 
The recursion in Eq.\ \eqref{eq:recursion_general} is a multivariate generalisation of the learning rule in Eq.\ \eqref{recursion}, which obtains with \(x_t=1\). Given the recursive estimate of $\phi_{t-1,s}$, the learnt expectation is defined as
\(
  a_{t,s} \coloneqq \phi_{t-1,s}^\Tr x_t
\)
so that the data generating process of the dependent variable $z_{t,s}$ is assumed to be
\begin{equation} \label{eq:Sdgp}
z_{t,s} = \alpha_t + \beta a_{t,s} + \eps_{t,s},
\end{equation}
cf.\ the nonlinear cohort panel model in Eq.\ \eqref{surveyexp}.

\subsection{Monte Carlo simulation}\label{sec:SMC}

The data generating process consists of Eqs.\ \eqref{eq:recursion_general} and \eqref{eq:Sdgp}, where, as in the main text, we generate \(y_t\) as an AR(1) process
\[
y_t = \varphi_y y_{t-1} + v_t,\quad v_t \stackrel{\textsf{IID}}{\sim} \mathcal{N}(0,1-\varphi_y^2).
\]
The PLM used by the individuals to predict $y_{t+1}$ is given by a linear projection of $y_t$ on a set of regressors. More specifically, using some observed regressor $x_t$ further specified below, an individual born in period \(s\) estimates in each period \(t\) the parameter of a linear regression $\phi$, say, following the stochastic recursive algorithm \eqref{eq:recursion_general}.
It allows for more than one predictor
variable in the individual's PLM and thus enables a direct comparison with more elaborate learning rules such as those investigated in the following Section \ref{sec:Semp} below. 

Here, we consider for the regressor \(x_t\), three scenarios --labelled \textsf{S1}, \textsf{S2}, and \textsf{S3}-- are considered:
\begin{align}
x_t= \,&  1 \tag{\textsf{AR0}}\label{MCS1} \\[1pt]
x_t = \,&  \varphi_x x_{t-1} + w_t, \quad w_t  \stackrel{\textsf{IID}}{\sim} \mathcal{N}(0,1-\varphi_x^2) \tag{\textsf{ARX}}\label{MCS2} \\[4pt]
x_t = \,& y_{t-1}. \tag{\textsf{AR1}}\label{MCS3}
\end{align}
Scenarios \ref{MCS1}, \ref{MCS2}, and \ref{MCS3} refer to the case where individuals estimate a linear PLM with a constant, a strictly exogenous or a weakly exogenous regressor, respectively. It is apparent that only \ref{MCS1}
is covered by our theory above, whereas \ref{MCS2} and \ref{MCS3} serve as robustness checks.

The fixed effects in Eq.\ \eqref{eq:Sdgp} are generated by $\alpha_t \stackrel{\textsf{IID}}{\sim} \textsf{UNIF}[0,1]$, while the error term $\varepsilon_{t,s}$ \bk satisfies one of the
following two  scenarios
\begin{align}
\eps_{t,s} \,&\stackrel{\textsf{IID}}{\sim}  \mathcal{N}(0,1) \tag{\sf E1}\label{E1}  \\
\eps_{t,s} = \rho \eps_{t-1,s}+\,&e_{t,s},\, e_{t,s}  \stackrel{\textsf{IID}}{\sim} \mathcal{N}(0,1-\rho^2). \tag{\sf E2}\label{E2} 
\end{align}

With the data thus generated for a particular choice of parameter values $\theta = (\beta, \gamma)^\Tr$ and $(\varphi_y, \varphi_x, \rho)^\Tr$, the model in Eqs.\ \eqref{surveyexp}, \eqref{recursion} and \eqref{gain_mn} is
estimated by NLS. Numerical optimisation over $\theta$ is based on the \textsf{optim} routine from the statistical software \textsf{R} (\citealp{rcovre:21}). More specifically, \(\theta_n\) is the minimizer of the profiled NLS
objective discussed in Section \ref{sec:estimation} based on the BFGS algorithm on \([2/3,10]\) with starting values from an initial grid search.

We report rejection frequencies of two-sided \(t\)-tests for $H_0:$ \(\gamma = \gamma_0\) and for $H_0:$ \(\beta = \beta_0\) based on asymptotic critical values derived from Corollary \ref{cor:2}. We use numerical derivatives for the standard errors in Eq.\ \eqref{eq:Wald}, which are calculated using a tuning parameter \(\ell_n = \delta_{n}(\gamma+\delta_n)\), where $\delta_{n} = \nu_n^{-2/5}$ in accordance with the requirement $\sqrt{\nu_n}\ell_n \rightarrow \infty$ of Corollary \ref{cor:1}. Note that $\delta_n$ is larger by at least one order of magnitude than step-sizes for numerical derivatives typically encountered in statistical software. As an example, the default setting in {\sf STATA}'s {\sf nl} routine (\citealp{stata:23}) is $4e^{-7}$ and the {\sf numDeriv} package in {\sf R} uses $1e^{-4}$. Since standard errors have been derived assuming spherical innovations in both dimensions, error design \ref{E2} is not covered. Following  MN, we therefore also report two-way cluster robust standard errors (see \citealp{cameron:11}). Some comment is warranted here: Taking a look at Table \ref{tab:indices}, one observes that cluster sizes are constant if we cluster at the time period ($t$) level (the clusters correspond to the number of blue cells  per row in Table \ref{tab:indices}). However, cluster sizes vary substantially if we cluster at the birth period level ($s$), i.e. slicing Table \ref{tab:indices} column by column.\footnote{Importantly, one should not compute birth-period clusters from first coercing the $(n-u) \times (n-l)$ Table \ref{tab:indices} by stacking the blue cells and creating a $(n-u) \times m$ matrix. This would distort the panel structure inducing unwanted dependencies.} 
 
\setlength{\tabcolsep}{4.66pt}
\begin{table}[htbp]
\begin{center}\small
\begin{tabular}{lccccccccccccc} \toprule \\[-.55cm]  \hline\\[-.4cm]
\sf (A)&&&\multicolumn{4}{c}{$\gamma = 3.0$} & &  \multicolumn{6}{c}{$\beta = 0.6 $} \\
 \cmidrule(r){4-7} \cmidrule(r){9-14}
& &$k$& \sf mean  &   \sf var   &  \(t\)  &\(t_{\sf CL}\)  &&  \sf mean  &   \sf var   &  \(t\)  & \(t_{\sf CL}\)  &$\sf supF$ &$\sf supF_{\sf CL}$\\  \hline
\sf E1&\sf AR0 &	2	&	3.011	&	0.078	&	0.051	&	0.061	&	&	0.600	&	0.003	&	0.051	&	0.059	&	1.000	&	1.000	\\
&&	3	&	3.003	&	0.028	&	0.046	&	0.045	&	&	0.602	&	0.001	&	0.052	&	0.057	&	1.000	&	1.000	\\
&&	4	&	2.999	&	0.016	&	0.050	&	0.057	&	&	0.598	&	0.001	&	0.040	&	0.042	&	1.000	&	1.000	\\
\cmidrule(l){2-14}
&\sf ARX &	2	&	3.039	&	0.208	&	0.083	&	0.072	&	&	0.608	&	0.009	&	0.089	&	0.079	&	1.000	&	1.000	\\
&&	3	&	3.003	&	0.066	&	0.086	&	0.060	&	&	0.601	&	0.003	&	0.079	&	0.057	&	1.000	&	1.000	\\
&&	4	&	3.004	&	0.027	&	0.044	&	0.057	&	&	0.599	&	0.001	&	0.055	&	0.059	&	1.000	&	1.000	\\
\cmidrule(l){2-14}														
&\sf AR1 &	2	&	3.020	&	0.224	&	0.085	&	0.079	&	&	0.609	&	0.009	&	0.072	&	0.077	&	1.000	&	1.000	\\
&&	3	&	3.040	&	0.088	&	0.081	&	0.071	&	&	0.601	&	0.004	&	0.086	&	0.074	&	1.000	&	1.000	\\
&&	4	&	3.008	&	0.052	&	0.072	&	0.053	&	&	0.602	&	0.002	&	0.082	&	0.070	&	1.000	&	1.000	\\
\hline\\[-.4cm]
\sf E2&\sf AR0 &	2	&	3.033	&	0.264	&	0.298	&	0.068	&	&	0.609	&	0.012	&	0.287	&	0.057	&	1.000	&	1.000	\\
&&	3	&	3.026	&	0.111	&	0.299	&	0.054	&	&	0.602	&	0.004	&	0.293	&	0.053	&	1.000	&	1.000	\\
&&	4	&	3.015	&	0.062	&	0.306	&	0.061	&	&	0.605	&	0.003	&	0.308	&	0.063	&	1.000	&	1.000	\\
\cmidrule(l){2-14}																&\sf ARX &	2	&	3.037	&	0.186	&	0.144	&	0.076	&	&	0.606	&	0.009	&	0.135	&	0.069	&	1.000	&	1.000	\\
&&	3	&	3.014	&	0.078	&	0.144	&	0.058	&	&	0.602	&	0.004	&	0.118	&	0.056	&	1.000	&	1.000	\\
&&	4	&	3.013	&	0.049	&	0.146	&	0.063	&	&	0.600	&	0.002	&	0.121	&	0.055	&	1.000	&	1.000	\\
\cmidrule(l){2-14}														
&\sf AR1 &	2	&	3.036	&	0.268	&	0.146	&	0.086	&	&	0.603	&	0.011	&	0.124	&	0.077	&	1.000	&	1.000	\\
&&	3	&	3.016	&	0.106	&	0.137	&	0.071	&	&	0.603	&	0.005	&	0.153	&	0.071	&	1.000	&	1.000	\\
&&	4	&	3.011	&	0.057	&	0.126	&	0.052	&	&	0.605	&	0.002	&	0.132	&	0.054	&	1.000	&	1.000	\\
\hline \\[-.4cm]
\sf (B) &&&\multicolumn{4}{c}{$\gamma = 3.0$} & &  \multicolumn{6}{c}{$\beta = 0.0 $} \\
 \cmidrule(r){4-7} \cmidrule(r){9-14}
\sf E1&	\sf AR0 &	2	&	5.571	&	10.564	&	0.084	&	0.088	&	&	-0.004	&	0.004	&	0.100	&	0.108	&	0.039	&	0.039	\\
&	&	3	&	5.722	&	10.789	&	0.081	&	0.087	&	&	0.001	&	0.002	&	0.102	&	0.111	&	0.046	&	0.044	\\
&	&	4	&	5.692	&	11.059	&	0.079	&	0.080	&	&	0.000	&	0.001	&	0.110	&	0.110	&	0.055	&	0.048	\\
\cmidrule(l){2-14}
&	\sf ARX &	2	&	5.890	&	9.941	&	0.081	&	0.089	&	&	0.000	&	0.006	&	0.084	&	0.122	&	0.037	&	0.041	\\
&	&	3	&	5.758	&	10.467	&	0.086	&	0.092	&	&	-0.001	&	0.003	&	0.104	&	0.127	&	0.057	&	0.052	\\
&	&	4	&	5.644	&	10.662	&	0.078	&	0.083	&	&	-0.001	&	0.002	&	0.117	&	0.134	&	0.053	&	0.052	\\
\cmidrule(l){2-14}														
&	\sf AR1 &	2	&	5.836	&	10.391	&	0.077	&	0.084	&	&	0.006	&	0.008	&	0.084	&	0.123	&	0.049	&	0.051	\\
&	&	3	&	5.765	&	10.206	&	0.071	&	0.076	&	&	-0.006	&	0.003	&	0.073	&	0.107	&	0.052	&	0.056	\\
&	&	4	&	5.714	&	10.935	&	0.090	&	0.094	&	&	0.000	&	0.002	&	0.084	&	0.095	&	0.043	&	0.046	\\
\hline\\[-.4cm]
\sf E2&	\sf AR0 &	2	&	5.428	&	9.525	&	0.188	&	0.087	&	&	0.006	&	0.016	&	0.495	&	0.125	&	0.397	&	0.038	\\
&	&	3	&	5.674	&	10.319	&	0.166	&	0.067	&	&	-0.001	&	0.006	&	0.501	&	0.100	&	0.405	&	0.043	\\
&	&	4	&	5.473	&	11.006	&	0.204	&	0.098	&	&	-0.002	&	0.004	&	0.507	&	0.120	&	0.417	&	0.040	\\		\cmidrule(l){2-14}											
&	\sf ARX &	2	&	5.454	&	9.562	&	0.124	&	0.102	&	&	-0.001	&	0.011	&	0.213	&	0.128	&	0.138	&	0.044	\\
&	&	3	&	5.496	&	9.770	&	0.114	&	0.087	&	&	0.000	&	0.004	&	0.233	&	0.115	&	0.151	&	0.038	\\
&	&	4	&	5.469	&	10.222	&	0.118	&	0.092	&	&	0.002	&	0.003	&	0.235	&	0.105	&	0.145	&	0.040	\\	\cmidrule(l){2-14}														
&	\sf AR1 &	2	&	5.545	&	9.907	&	0.100	&	0.091	&	&	0.002	&	0.014	&	0.185	&	0.111	&	0.163	&	0.040	\\
&	&	3	&	5.456	&	10.283	&	0.122	&	0.094	&	&	0.003	&	0.005	&	0.191	&	0.089	&	0.136	&	0.039	\\
&	&	4	&	5.442	&	10.558	&	0.139	&	0.116	&	&	0.000	&	0.003	&	0.207	&	0.096	&	0.155	&	0.037	\\
\hline\\[-.50cm]
\bottomrule
\end{tabular}\vspace*{-.25cm}
\caption{Simulation results under error designs \ref{E1} and \ref{E2} based on 1,000 Monte Carlo repetitions. Rejection frequencies of two-sided $t$ tests based on Corollary \ref{cor:2} ($t$) and the two-way clustered standard errors ($t_{\sf CL}$), respectively. The `supF' statistics are based on the wild bootstrap ($\sf supF$) or the cluster wild bootstrap ($\sf supF_{\sf CL}$).}\label{tab:MC1}  
\end{center}
\end{table}

Moreover, rejection frequencies of the `supF' statistic Eq.\ \eqref{supF} for the null hypothesis \(\beta = 0\) are reported, where \(p\) values are obtained using a wild bootstrap (see, e.g., \citealp[Algorithm 1]{hansen:17}) with \(B = 99\) bootstrap repetitions. This resampling scheme will, however, fail under the error design \ref{E2} due to the neglected serial dependence as we do not bootstrap a pivotal statistic. Instead, we equip in these cases the `supF' statistic with the two-way cluster standard errors and adapt the wild bootstrap along the lines of the wild two-way cluster bootstrap of \cite{mackinnon:21}. We stress again, that error design \ref{E2} is not covered by our theory, let alone the cluster bootstrap, the theoretical properties of which constitute an ongoing research field with only a few results available for nonlinear models (see \citealp[Section 4.4]{mackinnon:23}).

All test decisions are executed at a nominal significance level of five per cent. Inspired by the empirical application and the analysis in MN, we consider different sample sizes
\[
n = k\times 150,\quad u = k\times 75, \quad l = 25, \qquad k \in \{2,3,4\}.
\]
The gain parameter is fixed at \(\gamma = 3\), while \(\beta \in \{0,0.6\}\). We employed 1,000 Monte Carlo repetitions\footnote{The computations were parallelised and performed using CHEOPS, the DFG-funded (Funding number: INST 216/512/1FUGG) High Performance Computing (HPC) system of the Regional Computing Center at the University of Cologne (RRZK) using 1,000  iterations.}, where we set for the marginal time series processes \(\varphi_y = \varphi_x = 1/2\). We present our simulation results in Table \ref{tab:MC1}, divided in two panels, {\sf (A)} and {\sf (B)}, pertaining to the cases $\beta = 0.6$ and $\beta = 0.0$, respectively. 
\begin{enumerate}
\item[\sf (A)] In line with Propositions \ref{prop:1} and \ref{prop:2}, estimation precision is observed to increase with sample size. If the error term is uncorrelated (i.e., error design \ref{E1}), then the empirical size of the $t$-statistics, based on Corollary \ref{cor:2}, becomes reasonably close to the nominal size of five per cent. As anticipated, size control is lost if error terms are correlated over time periods (i.e., error design \ref{E2}). In this scenario, we see that the approach of MN of equipping $t$-tests with two-way cluster robust standard errors is doing its job as size is controlled, regardless of the error design. The ``supF'' test, using either the wild bootstrap of \cite{hansen:17} ({\sf supF}) or a wild cluster robust extension  (${\sf supF}_{\sf CL}$) based on \cite{mackinnon:21}, appears to consistently reject the alternative $\beta = 0.6$ of the null $\beta = 0$.
\item[\sf (B)] As our discussion of Proposition \ref{prop:1} indicates, identification breaks down if  $\beta = 0$. This is reflected by the poor performance of $\gamma_n$. Accordingly, we observe that the $t$-statistics for $\beta = 0$ are oversized because of the non-identified gain $\gamma$ under the null. This problem is solved when a corresponding ``supF'' statistic is used. Again, the robust statistic successfully controls size irrespective of the error design.
\end{enumerate}

\subsection{Additional empirical results}\label{sec:Semp}

The following empirical application uses the same sample as in the main text. The main difference concerns, similar to the previous section, the updating scheme used be the agent.

In particular, we report in Table \ref{tab:Semp} parameter estimates and test statistics based on the PLM in \eqref{eq:AR1PLM}, estimated by the agent using the generalised least-squares recursion \eqref{eq:recursion_general} (denoted by {\sf AR1}) using $x_t = (1, y_{t-1})^\Tr$ as regressor.  Comparing the estimates of $\beta$ and $\gamma$ to those based on our {\sf AR0} specification for the
prime sample of data from 1978 to 2023, it is clear that they are very similar. The null hypothesis of `no use of private experiences in forecasting inflation' is again rejected. Here, we use our {\it sup}{\sf F} statistic with the wild bootstrap as discussed in the previous section. Similarly, we can confidently reject the null of `no recency bias'. 

Interestingly, however, the more sophisticated belief updating of the {\sf AR1} specifications in
\eqref{eq:AR1PLM} does not yield a model fit, as measured by the $R^2$, that is superior to that based on our {\sf AR0} specifications with the simple learning rule in \eqref{eq:constantPLM}. The {\sf AR1} parameter estimates
for the extended sample and for the sample used by MN are again markedly lower than for the 1978--2023 period. For instance, for MN's 1953--2009 data, the point estimates of the slope and gain parameters are $\hat \beta = 0.62$
and $\hat \gamma = 3.10$, respectively. On the one hand, this corresponds in the main to the results presented in \citet[Table 1]{mn:2016}. On the other hand, these estimates confirm the aforementioned (see the main text) impression that the estimates may be driven by pre-1978
observations.

\begin{sidewaystable}
\setlength{\tabcolsep}{2.2pt}
\captionsetup{width=.9\textwidth}
\begin{center} \small
\begin{tabular}{ r cc c cc c cc c cc c cc c cc} \toprule                                                                                                                                                                                                                                                                                                                                                                                                                                                                 \\[-.55cm]  \hline \\[-.4cm]
                        & \multicolumn{5}{c}{\it 1978--2023}                                                                                                                                        &  & \multicolumn{5}{c}{\it 1953--2023}                                                                                 &  & \multicolumn{5}{c}{\it 1953--2009}                                                                                                                                                        \\ \cmidrule(l){2-6} \cmidrule(l){8-12} \cmidrule(l){14-18}
                        & \multicolumn{2}{c}{\sf AR0}                                           &                                           & \multicolumn{2}{c}{\sf AR1}                           &  & \multicolumn{2}{c}{\sf AR0}                           &  & \multicolumn{2}{c}{\sf AR1}                             &  & \multicolumn{2}{c}{\sf AR0}                                           &                                           & \multicolumn{2}{c}{\sf AR1}                                           \\ \cmidrule(l){2-3} \cmidrule(l){5-6} \cmidrule(l){8-9} \cmidrule(l){11-12}  \cmidrule(l){14-15} \cmidrule(l){17-18}
                        & $\beta$                                   & $\gamma$                  &                                           & $\beta$                   & $\gamma$                  &  & $\beta$                   & $\gamma$                  &  & $\beta$                   & $\gamma$                    &  & $\beta$                                   & $\gamma$                  &                                           & $\beta$                   & $\gamma$                                  \\ \hline \\[-5pt]
{\sf estimate}          & 0.8338                                    & 3.1551                    &                                           & 0.8667                    & 3.4942                    &  & 0.6987                    & 2.8969                    &  & 0.6189                    & 3.0712                      &  & 0.7311                                    & 2.7472                    &                                           & 0.6199                    & 3.0994                                    \\[-2pt]
                        & {\footnotesize (0.0394) }                 & {\footnotesize (0.2115) } &                                           & {\footnotesize (0.0515) } & {\footnotesize (0.2064) } &  & {\footnotesize (0.0343) } &	{\footnotesize (0.2010) } &  & {\footnotesize (0.0340) } & {\footnotesize (0.2127) }&  & {\footnotesize (0.0402) }                 & {\footnotesize (0.2173) } &                                           & {\footnotesize (0.0385) }	& {\footnotesize (0.2526) }              \\ \cmidrule(l){2-6} \cmidrule(l){8-12}  \cmidrule(l){14-18}  
{\sf $\#$ observations} & \multicolumn{2}{c}{8,800}                                             &                                           & \multicolumn{2}{c}{8,800}                             &  & \multicolumn{2}{c}{10,615}                            &  & \multicolumn{2}{c}{10,615}                              &  & \multicolumn{2}{c}{8,215}                                             &                                           & \multicolumn{2}{c}{8,215}                                             \\  
{$R^2$}                 & \multicolumn{2}{c}{0.5612}                                            &                                           & \multicolumn{2}{c}{0.5583}                            &  & \multicolumn{2}{c}{0.6346}                            &  & \multicolumn{2}{c}{0.6324}                              &  & \multicolumn{2}{c}{0.6373}                                            &                                           & \multicolumn{2}{c}{0.6346}                                            \\ \hline\\[-12pt]
                        & \multicolumn{5}{c}{$H_0$: $\beta = 0$} 	                                                                                                                               &  & \multicolumn{5}{c}{$H_0$: $\beta = 0$}                                                                              &  & \multicolumn{5}{c}{$H_0$: $\beta = 0$} 	                                                                                                                                                \\  \cmidrule(l){2-6} \cmidrule(l){8-12} \cmidrule(l){14-18}
${\it sup}{\sf F}$            & \multicolumn{2}{c}{476.82}                                            &                                           & \multicolumn{2}{c}{411.07}                            &  & \multicolumn{2}{c}{425.60}                            &  & \multicolumn{2}{c}{358.19}                              &  & \multicolumn{2}{c}{370.70}                                            &                                           & \multicolumn{2}{c}{267.38}                                            \\[-2pt]
                        & \multicolumn{2}{c}{\footnotesize (0.00) }	                           &                                           & \multicolumn{2}{c}{\footnotesize (0.00) }             &  & \multicolumn{2}{c}{\footnotesize (0.00) }             &  & \multicolumn{2}{c}{\footnotesize (0.00) }                &  & \multicolumn{2}{c}{\footnotesize (0.00) }                             &                                           & \multicolumn{2}{c}{\footnotesize (0.00) }	                          \\ \hline\\[-.50cm] \bottomrule
\end{tabular}
\caption{NLS estimation results for the sample currently available at MSC (1978Q1--2023Q3), for the current sample extended by MN's preparation of the MSC archive data (1953Q4--2023Q3), and for the sample used by MN
  (1953Q4-2009Q4). Note that the archive data contains missings. Standard errors of the  coefficient estimates and $p$-values of the {\sf supF}-statistics, based on Corollaries \ref{cor:2} and \ref{cor:3}, are given in parentheses underneath the estimates and test statistics, respectively.}\label{tab:Semp}
\end{center}
\end{sidewaystable}

\end{appendices}

\end{document}